\documentclass[preprint2,twocolappendix]{aastex6}

\usepackage[T1]{fontenc}
\usepackage{graphicx}
\usepackage{amsmath}

\begin{document}

\title{Numerical simulation of hot accretion flows (IV): effects of black hole spin and magnetic field strength on the wind and the comparison between wind and jet properties}

\author{Hai Yang\altaffilmark{1,2}, Feng Yuan\altaffilmark{1,2}, Ye-fei Yuan\altaffilmark{3}, and Christopher J. White\altaffilmark{4}}

\affil{\altaffilmark{1} Shanghai Astronomical Observatory, Chinese Academy of Sciences, 80 Nandan Road, Shanghai 200030, China \\
{hyang@shao.ac.cn (HY), fyuan@shao.ac.cn (FY)}\\
\and \altaffilmark{2} University of Chinese Academy of Sciences, 19A Yuquan Road, Beijing 100049, China
\\
\and \altaffilmark{3} Department of Astronomy University of Science and Technology of China Hefei, Anhui, China; yfyuan@ustc.edu.cn
\\
\and \altaffilmark{4} Department of Astrophysical Sciences, Princeton University, Peyton Hall, Princeton, NJ, USA; cjwhite@princeton.edu}

\begin{abstract}

This is the fourth paper of our series of works studying winds from hot accretion flows around black holes. In the first two papers, we have shown the existence of strong winds in hot accretion flows using hydrodynamical and magnetohydrodynamical (MHD) simulations. In the third paper, by using three dimensional general relativity MHD numerical simulation data of hot accretion flows and adopting a ``virtual particle trajectory'' data analysis approach, we have calculated the properties of wind, such as its mass flux and velocity. However, that paper focuses only on a non-spinning black hole and SANE (standard and normal accretion). In the present paper, we extend the third paper by including cases of a rapidly rotating black hole and MAD (magnetically arrested disk). We focus on investigating the effect of spin and magnetic field on the properties of wind and jet. It is found that a larger spin and stronger magnetic field usually enhance the wind and jet. The formulae describing the mass flux, poloidal velocity, and fluxes of momentum, kinetic energy, and total energy of wind and jet are presented. 
One interesting finding, among others, is that even in the case of very rapidly spinning black hole where the jet is supposed to be the strongest, the momentum flux of jet is smaller than that of wind, while the total energy flux of jet is larger than that of wind by at most a factor of 10. This result suggests that  wind potentially plays a more important role than jet at least for some problems in active galactic nuclei feedback.
\end{abstract}


\section{Introduction} \label{sec:intro}
Depending on the value of mass accretion rate, black hole accretion is divided into two modes, namely cold mode above roughly $2\% \dot{M}_{\rm Edd}$ and hot mode below this rate. Here $\dot{M}\equiv 10L_{\rm Edd}/c^2$ is defined as the Eddington accretion rate, with $L_{\rm Edd}$ being the Eddington luminosity. The cold mode includes standard thin disk \citep{Shakura1973,Pringle1981} and super-Eddington accretion \citep{Abra88,Olek2014,Jiang2014}, bounded by the Eddington accretion rate. It corresponds to the cold mode (or quasar mode or radiative mode) in the study of active galactic nuclei (AGN) feedback. The hot mode includes advection-dominated accretion flow \citep{Narayan1994,Narayan1995,Yuan2014Narayan} and luminous hot accretion flows \citep{Yuan2001}, bounded by $\sim 0.1 \alpha^2 \dot{M}_{\rm Edd}$. It corresponds to the hot mode (or radio mode or jet mode or maintenance mode) of AGN feedback. 

In the present paper we continue our study of wind from hot accretion flows. The motivation of such a study is two-fold. First, wind is a fundamental ingredient of black hole accretion flow, affecting both the dynamics and radiation of the accretion flow. Second, active galactic nuclei feedback is now believed to play a crucial role in galaxy formation and evolution \citep{fabian2012,Kormendy2013Ho,ostriker2017}, while wind from hot accretion flow is potentially one of the most important mediums of feedback \citep{weinberger2017,yuan2018,yoon2019}. For example, recent IllustrisTNG cosmological
simulations \citep{weinberger2017} find that to overcome some serious problems in
galaxy formation, e.g., reducing star formation efficiency in the most massive halos, winds launched from hot accretion flows must be invoked to interact with the interstellar medium on the galaxy scale. \citet{yuan2018} recently comprehensively include feedback by wind and
radiation from AGNs in both cold and hot feedback modes and find that wind plays a dominant role in controlling the star formation and black hole growth,
although radiative feedback cannot be neglected.

The numerical study of winds from hot accretion flows was started by \citet{Stone1999}. In this milestone paper they find for the first time that the mass accretion rate of the hot accretion flow decreases inward. Two models have been proposed to explain this surprising result. In the adiabatic inflow-outflow solution (ADIOS;  \citealp{Blandford1999,Blandford2004,Begelman2012}; see also an earlier speculation for the existence of strong wind in hot accretion flows in \citealp{Narayan1994}), it is assumed that the inward decrease of accretion rate is because of mass lost in the wind. In the convection-dominated accretion flow (CDAF) model, it is thought to be because of convective motion in the hot accretion flow \citep{Narayan2000,Quataert2000,Abramowicz2002}. 

To solve this debate, in the first paper of this series of works, \citet{Yuan2012Wu} first significantly extend the  radial dynamical range of \citet{Stone1999} based on  a  ``two-zone'' approach and confirm the \citet{Stone1999} result. In the second paper of this series \citep{Yuan2012Bu}, by systematically comparing the statistical value of some physical quantities such as radial velocity and angular momentum of  ``inflow'' and ``outflow'' and analyzing the convective stability of the MHD accretion flow, they show that strong wind must exist and is the reason for the inward decrease of mass accretion rate (see also \citet{Narayan2012}). In addition to numerical simulations, some analytical solutions of accretion with wind have also been obtained \citep{Gu2015,Amin2016}, with the results broadly consistent with the numerical simulations.  \citet{Bu2016a,Bu2016b} address the question of where the wind can be produced, they find that  winds can only be produced within the Bondi radius. This is likely because of the change of the gravitational potential beyond the Bondi radius due to the contribution of stellar cluster. The theoretical prediction of the existence of strong wind from hot accretion flow has been confirmed by some observations in recent years, including the supermassive black hole in the Galactic center \citep{Wang2013,Ma2019},  low-luminosity AGNs \citep{Tombesi2014,Cheung2016,Park2019}, and black hole X-ray binaries in the hard state \citep{Homan2016}.

The remaining question is what are the main properties of the wind such as mass flux and velocity. This question is difficult to answer because the accretion flow is strongly turbulent; it is hard to discriminate the turbulent outflow and the real wind. The widely adopted solution in literature is to time-average the simulation data first to filter out turbulence and then use the time-averaged velocity and density to calculate the mass flux of wind. However, note that the real wind is often instantaneous. This means that, suppose at one time real wind exists at a certain $\theta$ and $\varphi$ angles, at another time this region may be occupied by inflow. Therefore, although such a simple ``time-average'' procedure will be able to eliminate the turbulent outflow, it will also inevitably  eliminate the real wind. Consequently, the mass flux of wind will be significantly underestimated. 

This difficult is solved  in \citet{Yuan2015}, the third paper of this series. In this work, they propose a ``virtual particle trajectory'' approach, which can faithfully reflect the motion of fluid elements and thus clearly distinguish turbulent outflow and wind. Using this approach, and based on the three dimensional general relativity MHD simulation data of hot accretion flows, they successfully obtain the main properties of wind. They have also compared the mass flux of wind obtained by the trajectory approach with that obtained by the usual time-average streamline approach. The exact difference between the two results depends on the radius and time interval adopted in the time average calculation. The longer the interval, the larger the difference.  Taking $50\,r_{\rm g}$ as an example, it is found that  the flux of wind obtained by the time-average streamline approach is smaller by a factor of 10 compared to that obtained by the trajectory approach (refer to Fig. 5 in \citealp{Yuan2015}). 

The current work is a direct extension of \citet{Yuan2015}. \citet{Yuan2015} only deals with SANE (standard and normal evolution; \citealp{Narayan2012}) around non-spinning black holes. In reality, it is possible that the magnetic field in the flow may be much stronger, i.e., the accretion is in the MAD (magnetically arrested disk; \citealp{Narayan2003}) state, and the black hole spin is more likely nonzero. In the present work, we investigate how the properties of  wind will change  with the strength of magnetic field and black hole spin\footnote{In fact, \citet{Sadowski2013} have studied the wind in the cases of MAD and rapidly spinning black hole. However, as in most works, the time-average streamline approach is adopted. This is why they find the wind very weak.}. Since a rapidly spinning black hole  will power a relativistic jet, it is interesting to compare the properties of winds with the jet such as their momentum and energy fluxes. Such information is also valuable for us to evaluate the respective role of wind and jet in AGN feedback. 

The main structure of the paper is as follows. In Section~\ref{sec:simulations}, we will describe our equations and simulation method (\S\ref{simulationmethod}), the three models we study (\S\ref{threemodels}), and a brief overview to the trajectory approach we use to analyze the simulation data (\S\ref{trajectory}). The results will be presented in Section~\ref{sec:result}. We then summarize our work in Section~\ref{sec:sum}. The implications of our results in the context of AGN feedback are discussed in Section~\ref{discussion}.

\section{Model}
\label{sec:simulations}
\subsection{Equations and numerical method}
\label{simulationmethod}

The equations of ideal MHD describing the evolution of the accretion flow read \citep[e.g.,][]{Gammie2003}, 
\begin{equation}
    \bigtriangledown_{\mu}(\rho{u}^{\mu})=0,
	\label{eq:mhd1}
\end{equation}
\begin{equation}
    \bigtriangledown_{\mu}T^{{\mu}{\nu}}=0,
	\label{eq:mhd2}
\end{equation}
\begin{equation}
   \bigtriangledown_{\mu}{\ast}F^{{\mu}{\nu}}=0.
	\label{eq:mhd3}
\end{equation}
with 
\begin{equation}
    T^{{\mu}{\nu}}=(\rho{h}+b_{\lambda}b^{\lambda})u^{\mu}u^{\nu}+(p_{\mathrm{ gas}}+\frac{1}{2}b_{\lambda}b^{\lambda})g^{{\mu}{\nu}}-b^{\mu}b^{\nu},
	\label{eq:Tmunu}
\end{equation}
\begin{equation}
    {\ast}F^{{\mu}{\nu}}=b^{\mu}u^{\nu}-b^{\nu}u^{\mu}.
	\label{eq:Fmunu}
\end{equation}
Here $T^{{\mu}{\nu}}$ is the stress-energy tensor of the MHD, $h$ is the specific enthalpy of the fluid. We use the Athena++ \citep{White2016,Stone2020} code to solve the above GRMHD equations in the Kerr metric. This code uses a finite-volume Godunov scheme to ensure total energy conservation, with the flux of conserved quantities obtained by solving the Riemann problem at each interface. In our simulation, we use the HLLE Riemann solvers \citep{Einfeldt1988}, which is commonly used in GRMHD simulation. In order to satisfy the divergence-free constraint to prevent spurious production of magnetic monopoles, the staggered-mesh constrained transport (CT) method is applied. For the spatial reconstruction, we use the piecewise linear method \citep[PLM;][]{vanLeer1974}. For the mesh grid, we use the static mesh refinement (SMR). It allows us to easily use higher resolution in areas of interest, and can give a good balance between accuracy and performance, which is very useful for large-scale 3D simulations. 

All our simulations are performed in the Kerr--Schild (horizon penetrating) coordinates ($t$, $r$, $\theta$, $\varphi$). The radius of the black hole horizon\,$r_{\mathrm{H}}$ is $r_{\mathrm{H}}=(1+\sqrt{1-a^2})r_{\rm g}$, with $r_{\mathrm{g}}\equiv GM_{\rm{BH}}/c^2$ being the gravitational radius and $a$ the black hole spin parameter. The determinant of the metric\,$g \equiv \det(g_{\mu\nu})=-\Sigma^2\sin^2\theta$, where $\Sigma \equiv r^2+a^2\cos^2\theta$ \citep{McKinney2004}. The comoving rest-mass density is denoted as $\rho$, $u^{\mu}$ is the component of the coordinate-frame 4-velocity, the equation of the state of the gas is $u=p_{\rm{ gas}}/(\Gamma-1)$, with $p_{\rm gas}$ being the gas pressure of the comoving mass, $u$ being the internal energy of the gas, and $\Gamma$ being the adiabatic index. In our work, $\Gamma$ is taken to be 4/3, and time is measured in unit of $r_{\rm g}/c$. The units we use is Heaviside-Lorentz, both the light speed and gravity constant are set to be unity, and the metric sign convection is $(-, +, +, +)$. The metric of our simulation is stationary, the self-gravity is ignored.


\begin{deluxetable}{cccc}[th!]
	\centering
	\tablecaption{The static mesh refinement grid of MAD00.
	\label{tab:grid00}}
	\tablenum{1}
	\tablewidth{0pt}
	\tablehead{
	\colhead{Level} &
	\colhead{ $r \,/\, r_{\rm g} $} & 
	\colhead{$\theta \,/\, \pi$} & 
	\colhead{$\varphi \,/\, \pi$}
	}
	\startdata
        0 & [1.6, 1200]& [0, 1]&[0, 2]   \\
		\hline
		{1} &  $ [1.6, 200]$& [0.1305, 0.8695]& [0, 2]\\
		&  $ [8.5, 1200]$& [0, 0.1210]& [0, 2]\\
		  &$ [8.5, 1200]$& [0.8790, 1]& [0, 2]\\
	    \hline 
		2 &  $ [1.6, 30]$& [0.1942, 0.8058]& [0, 2]\\
		&  $ [44.4, 1200]$& [0, 0.0573]& [0, 2]\\
		 &  $ [44.4, 1200]$& [0.9427, 1]& [0, 2]\\
          \hline
	\enddata
\end{deluxetable}

\begin{deluxetable}{cccc}[th]
	\centering
	\tablecaption{The static mesh refinement grid of  MAD98 and SANE98.\label{tab:grid98}}
	\tablenum{2}
	\tablehead{
		\colhead{Level} & 
		\colhead{$r \,/\, r_{\rm g}$} & 
		\colhead{$\theta \,/\, \pi$} & 
		\colhead{$\varphi \,/\, \pi$}
		}
		\startdata
		0 & [1.1, 1200]& [0, 1]&[0, 2]   \\
		\hline
		&  $ [1.1, 200]$& [0.1305, 0.8695]& [0, 2]\\
		1&  $ [6.4, 1200]$& [0, 0.1210]& [0, 2]\\
		&  $ [6.4, 1200]$& [0.8790, 1]& [0, 2]\\
	   \hline
		&  $ [1.1, 30]$& [0.1942, 0.8058]& [0, 2]\\
		2 &  $ [36.7, 1200]$& [0, 0.0573]& [0, 2]\\
		 &  $ [36.7, 1200]$& [0.9427, 1]& [0, 2]\\
         \hline
	\enddata
\end{deluxetable}

\begin{deluxetable}{cccccc}[th]
	\centering
	\tablecaption{The setup of three models.\label{tab:para}}
	\tablenum{3}
	\tablehead{		
		\colhead{Model} &
		\colhead{ $a$} & 
		\colhead{$N_{r}$} &
		\colhead{ $N_{\theta}$} &
		\colhead{ $ N_{\varphi}$}&
		\colhead{ Duration}
		}
		\startdata
		MAD00 & 0 & 288 & 128& 64&40 000\\
		 SANE98 & 0.98 & 352 & 128 & 64&80 000\\
		MAD98 & 0.98 & 352 & 128& 64&40 000\\
		\hline
		\enddata	
\end{deluxetable}

\subsection{Three models and their setup}
\label{threemodels}

In this paper, we consider three models, i.e., MAD00, SANE98, and MAD98. They denote the MAD accretion flow around a black hole of spin $a=0$, SANE accretion flow around a black hole of $a=0.98$, and MAD accretion flow around a black hole of $a=0.98$, respectively. We will combine these three models with the SANE around a black hole of $a=0$ presented in \citep{Yuan2015} to analyze the dependence of wind properties on black hole spin and magnetic field strength. 

\begin{figure*}
\plottwo{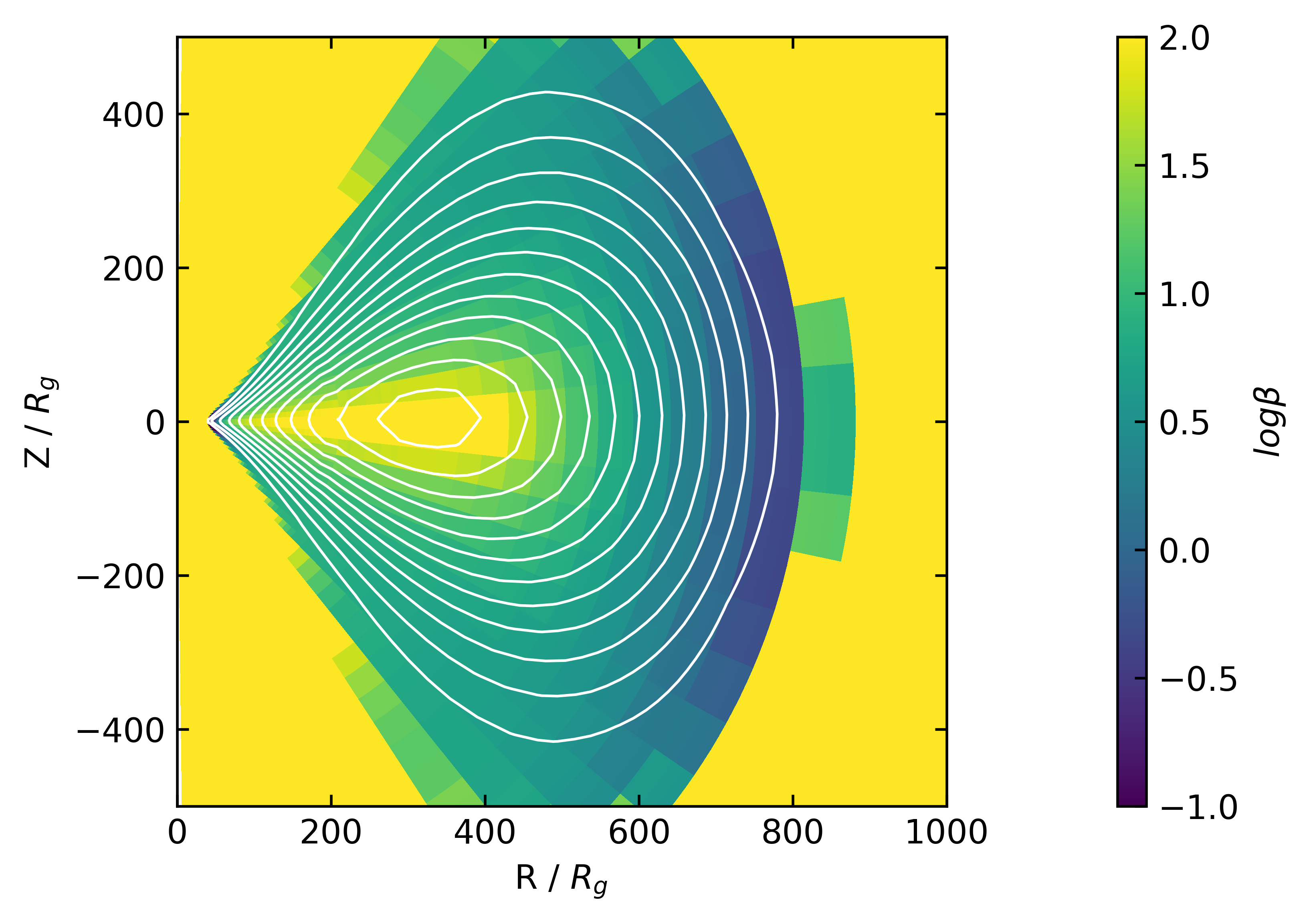}
{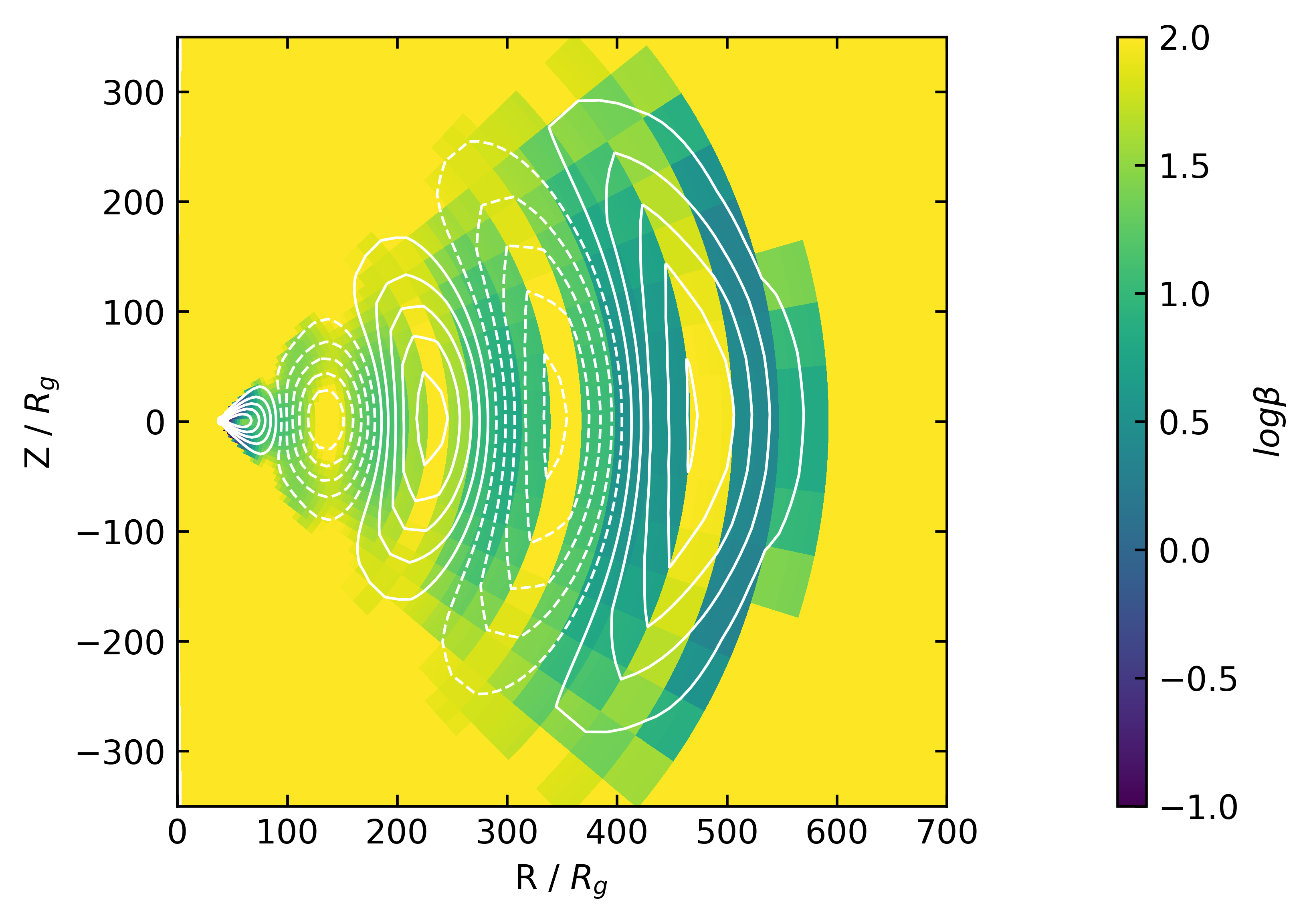}
\caption{The initial magnetic field configuration of MAD (left) and SANE (right). The dashed line in the SANE means that it has a different polarity compared with the solid line. The color is the logarithm of the gas-to-magnetic pressure ratio $\beta $.}
\label{fig:b0fieldline}
\end{figure*}

Our simulations starts with a torus that rotates around a black hole. The torus is initially in hydrostatic equilibrium, described by \citet{Fishbone1976}. The inner edge of the torus is at $r=40.5\,r_{\rm g}$, and the radius of pressure maximum is at $r=80\,r_{\rm g}$. 
The gas torus is thread by a poloidal magnetic field \citep{Penna2013}. In our simulation, we use two different initial magnetic field configurations for MAD and SANE. For the MAD, we set one poloidal loop threading the whole torus. This will result in rapid accretion. The magnetic flux would be accumulated and finally impede the accretion of mass. The magnetic field flux will quickly saturate at a  maximum value on the black hole for a given mass accretion rate, reaching the MAD state  \citep[]{Narayan2003,McKinney2012,Sadowski2013}. For the SANE, we initially set a seed field that consists of  multiple poloidal loops of magnetic field with changing polarity. Such a configuration  makes magnetic reconnection easy to occur, thus the magnetic field will always stay weak, preventing the accumulation of magnetic flux so the accretion flow is in the SANE state \citep{Narayan2012}. Figure~\ref{fig:b0fieldline} shows the initial magnetic field configuration of MAD and SANE.

Three parameters are required to determine the specific magnetic field of the torus, namely $r_{\rm start}$, $r_{\rm end}$, and $\lambda_{B}$.  The first two denote the inner and outer edges of the magnetized region, while the last parameter controls the size of the poloidal loops, or equivalently the number of the loops. The normalization of the magnetic field is determined by the gas-to-magnetic pressure ratio $\beta \equiv p_{\rm gas}/p_{\rm mag}$. It has the minimum value in the equatorial plane for each loop. It  peaks at loop edges and drops to the loop center. For MAD, we set  $r_{\rm start}=25\,r_{\rm g}$, $r_{\rm end}=810\,r_{\rm g}$, $\lambda_{B}=25$ (it just has one loop), and $\beta_{\rm min}=0.1$. For SANE, we set  $r_{\rm start}=25\,r_{\rm g}$, $r_{\rm end}=550\,r_{\rm g}$, $\lambda_{B}=3.75$ (it has five loops), and $\beta_{\rm min}=0.1$.

\begin{figure*}[bh!]
\plottwo{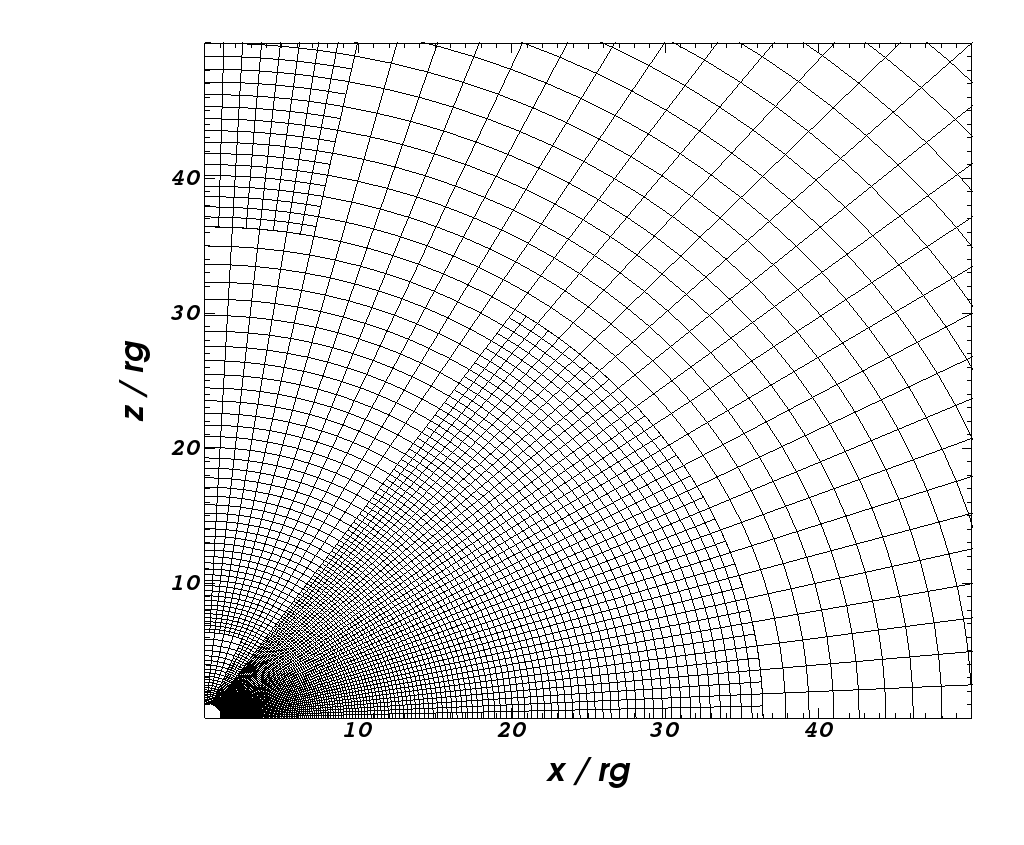}{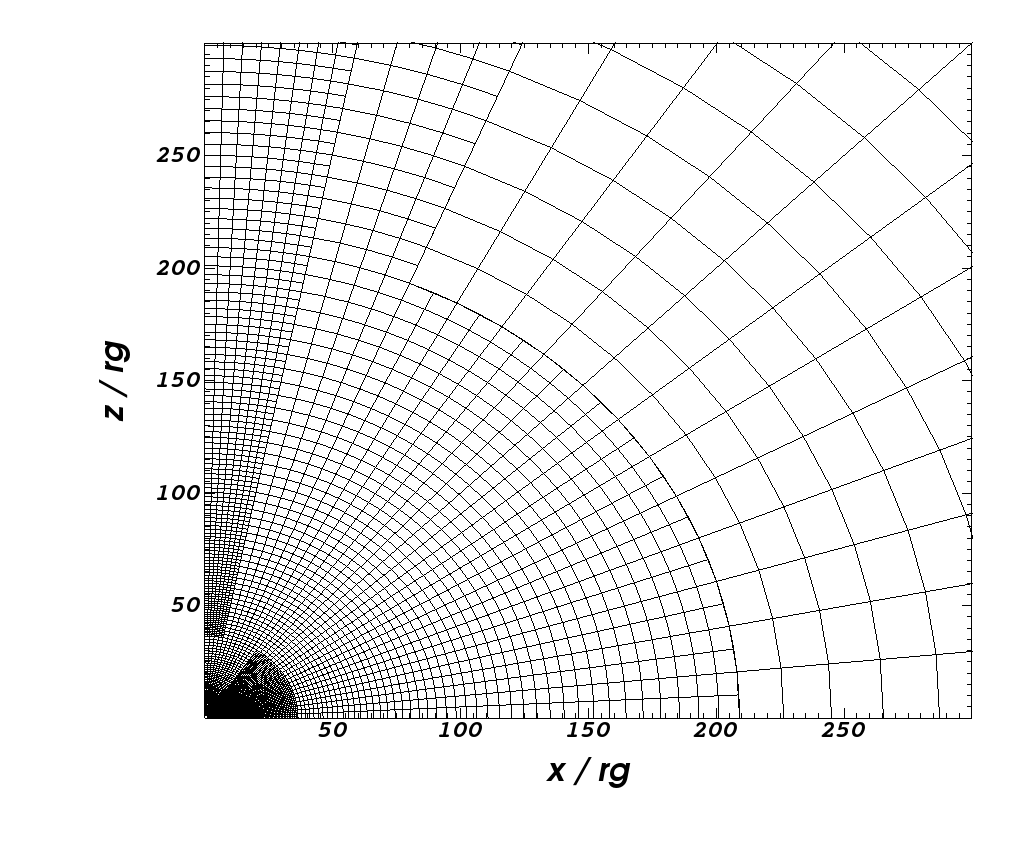}
\caption{Different zoom levels of the mesh grid of the poloidal plane of the MAD98.
\label{fig:grid}}
\end{figure*}

Our grid uses static mesh refinement with a root grid plus a more refined grid. The root grid is $88\times32\times16$ cells in radial, polar, and azimuthal direction. In the $r$ direction, the inner and outer edges are located at 1.1\,${r_{\rm g}}$ and 1200\,${r_{\rm g}}$ respectively. Logarithmic spacing is adopted, with the ratio $r_{i+1}/r_{i}$ being 1.0827. The grid in the polar and azimuthal directions are uniform. Because of the use of Kerr--Schild coordinates, the inner edge of our simulations are within the black hole horizon. For different models we use different static mesh refinement. Table~\ref{tab:grid00} and Table~\ref{tab:grid98} give the details of the three models. For a refinement region, each additional level means that the grid of this area is refined by a factor of 2 for all directions based on the previous level grid. Table~\ref{tab:para} is the final effective grid of the different model. Figure~\ref{fig:grid} shows the different zoom levels of the mesh grid in the poloidal plane of  MAD98.

For the inner and outer boundaries we use the outflow boundary conditions. We use the polar axis boundary condition in the $\theta$ direction and periodic boundary conditions in the $\varphi$ direction.
Since numerical MHD can not deal with vacuum, like all GRMHD numerical simulations, we need to impose the density and gas pressure floors: $\rho_{\rm min}={\rm max}(10^{-2}r^{-1.5},10^{-6})$, $p_{\rm gas,min}={\rm max}(10^{-2}r^{-2.5},10^{-8})$. Following \citet{White2020b}, we also enforce $\sigma<\,100$ and $\gamma<\,50$. Here $\sigma = 2p_{\rm mag}/\rho $ is the magnetization parameter, $\gamma\equiv \alpha u^t$ with $\alpha \equiv (-g^{\rm tt})^{-1/2}$ the lapse. The former is an extra limitation for the density and gas pressure floors. The latter is to  limit the velocity to keep the normal-frame Lorentz factor from becoming too large. 

\begin{figure*}[t]
\gridline{\fig{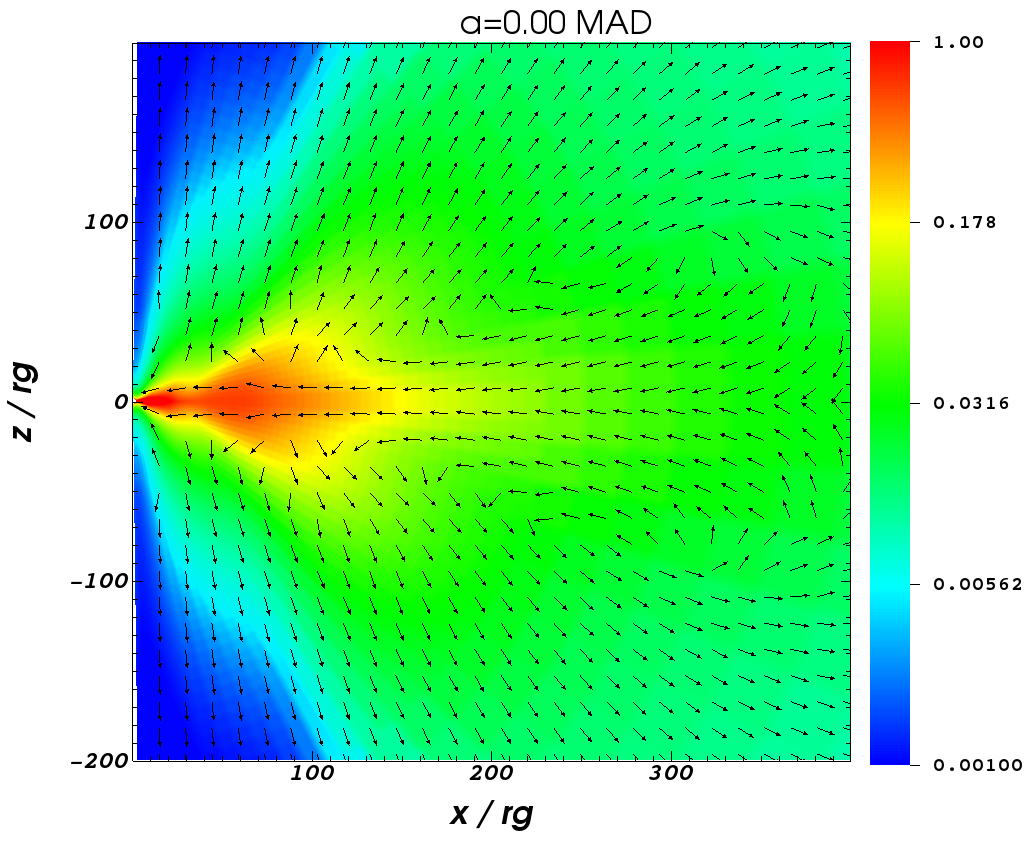}{0.33\textwidth}{}
\fig{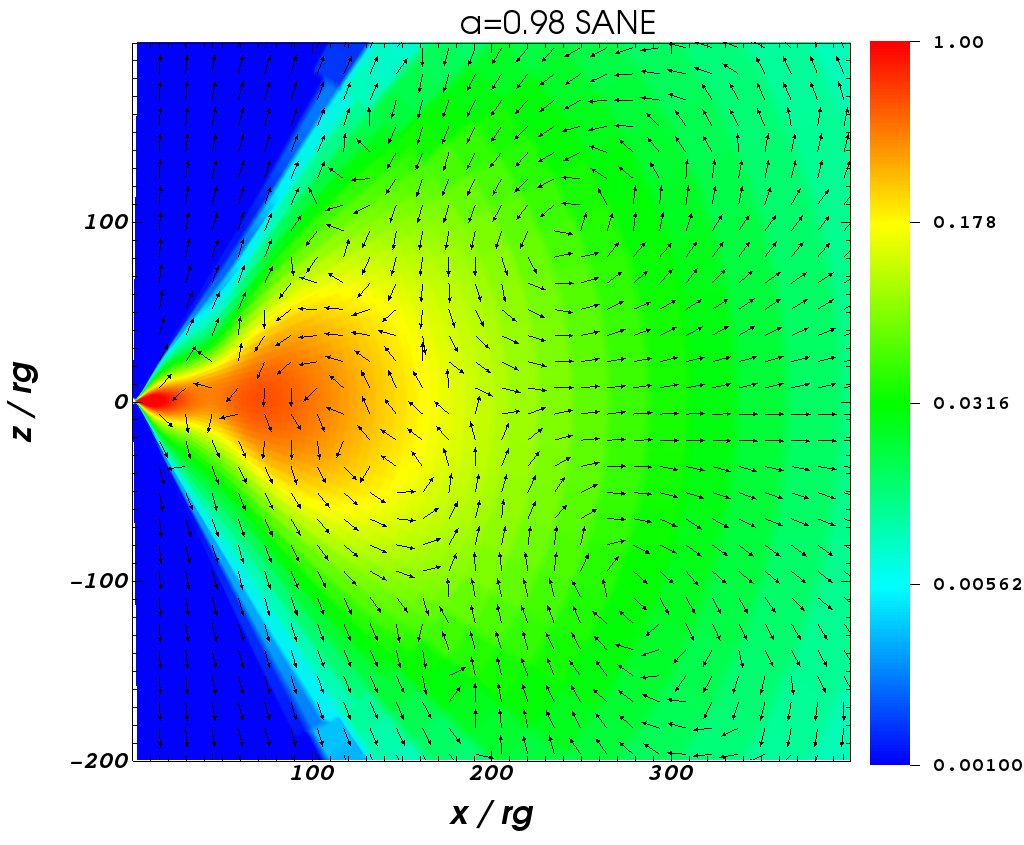}{0.33\textwidth}{}
\fig{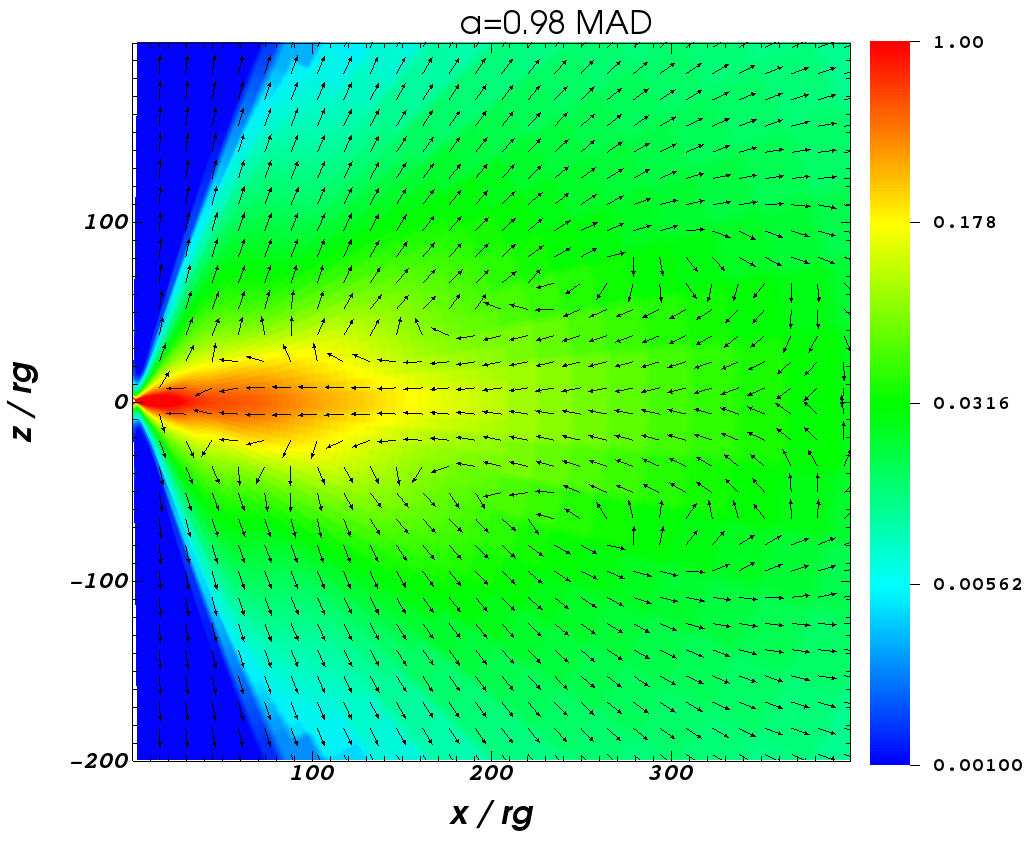}{0.33\textwidth}{}
}
\gridline{\fig
{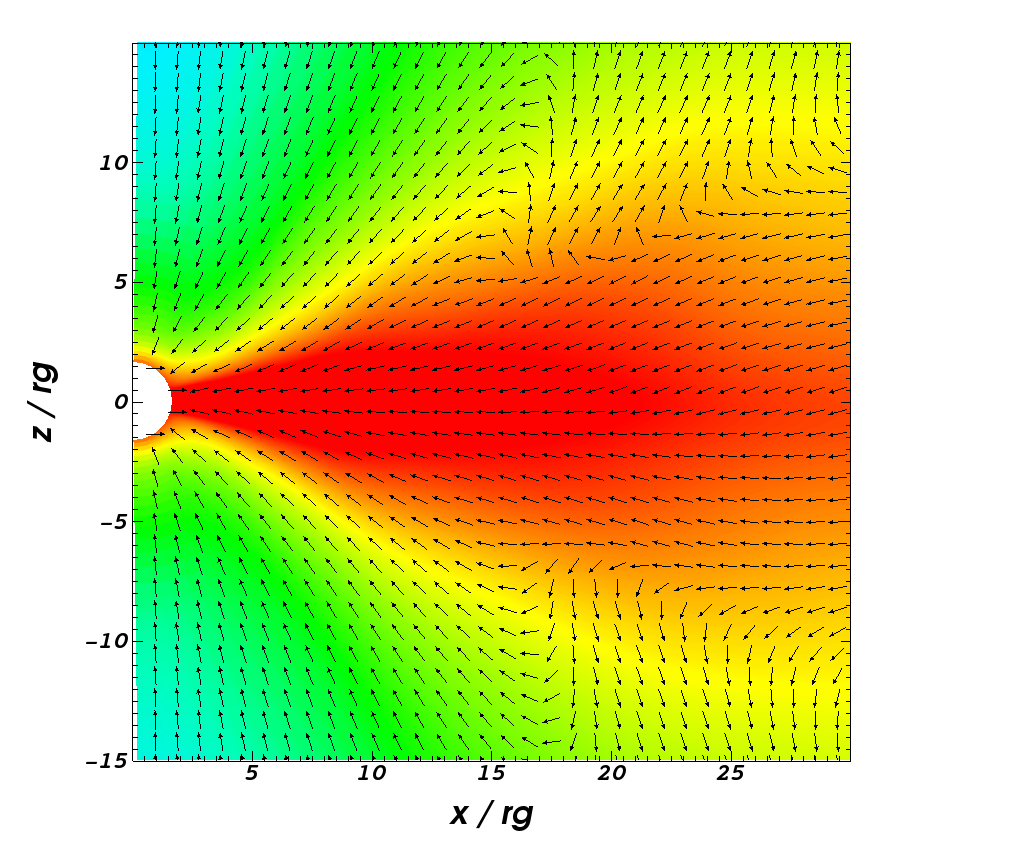}{0.33\textwidth}{}
\fig{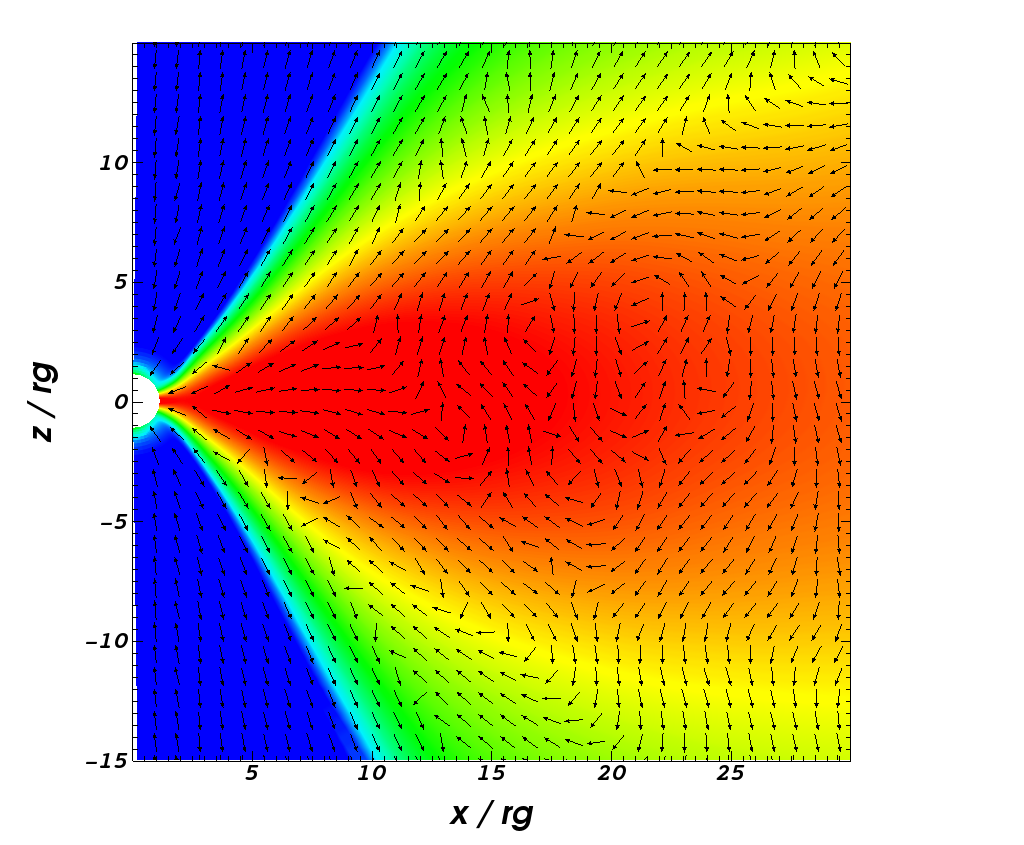}{0.33\textwidth}{}
\fig{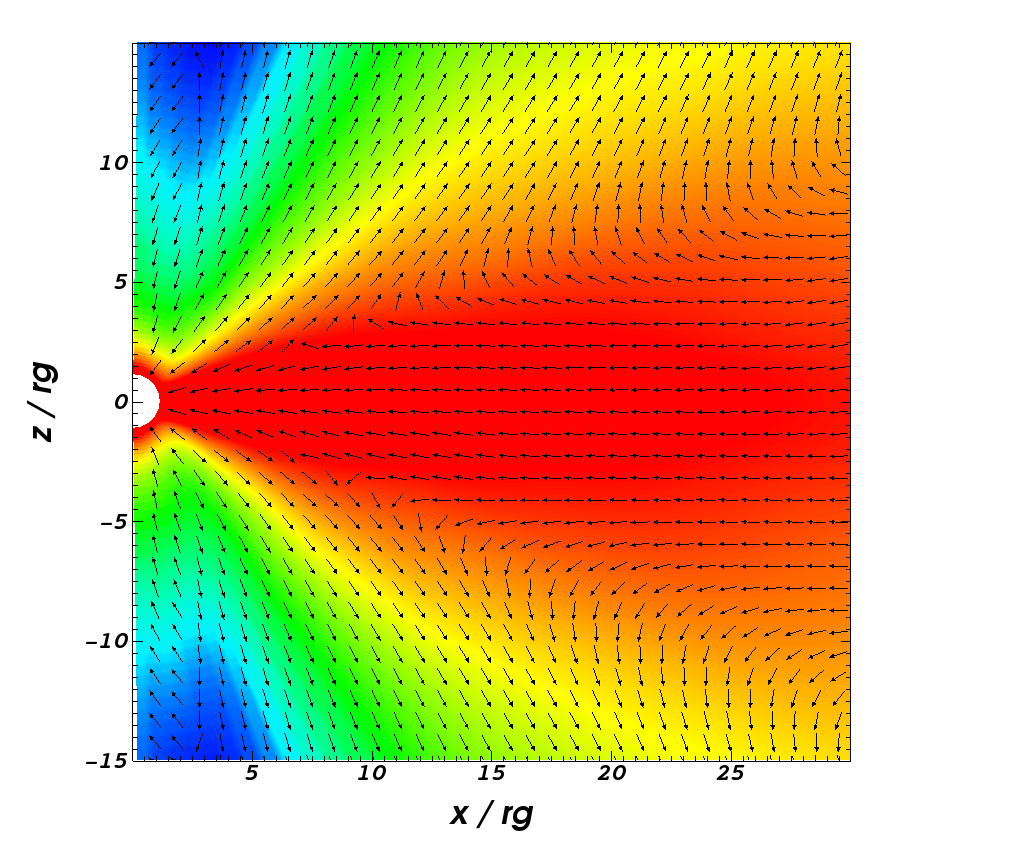}{0.33\textwidth}{}
}
  
    \caption{Two dimensional distribution of density (by color) and velocity vector (by arrows) of the accretion flow averaged in $\varphi$ direction and in time interval of $t=55{,}000\text{--}68{,}000$, $10{,}000\text{--}20{,}000$ and $8000\text{--}23{,}000$ for SANE98, MAD00 and MAD98 respectively.  The bottom panels  are the zoom-in pictures of the top panels.}
    \label{fig:densityvelocity}
\end{figure*}


\begin{figure*}[t]
\gridline{\fig{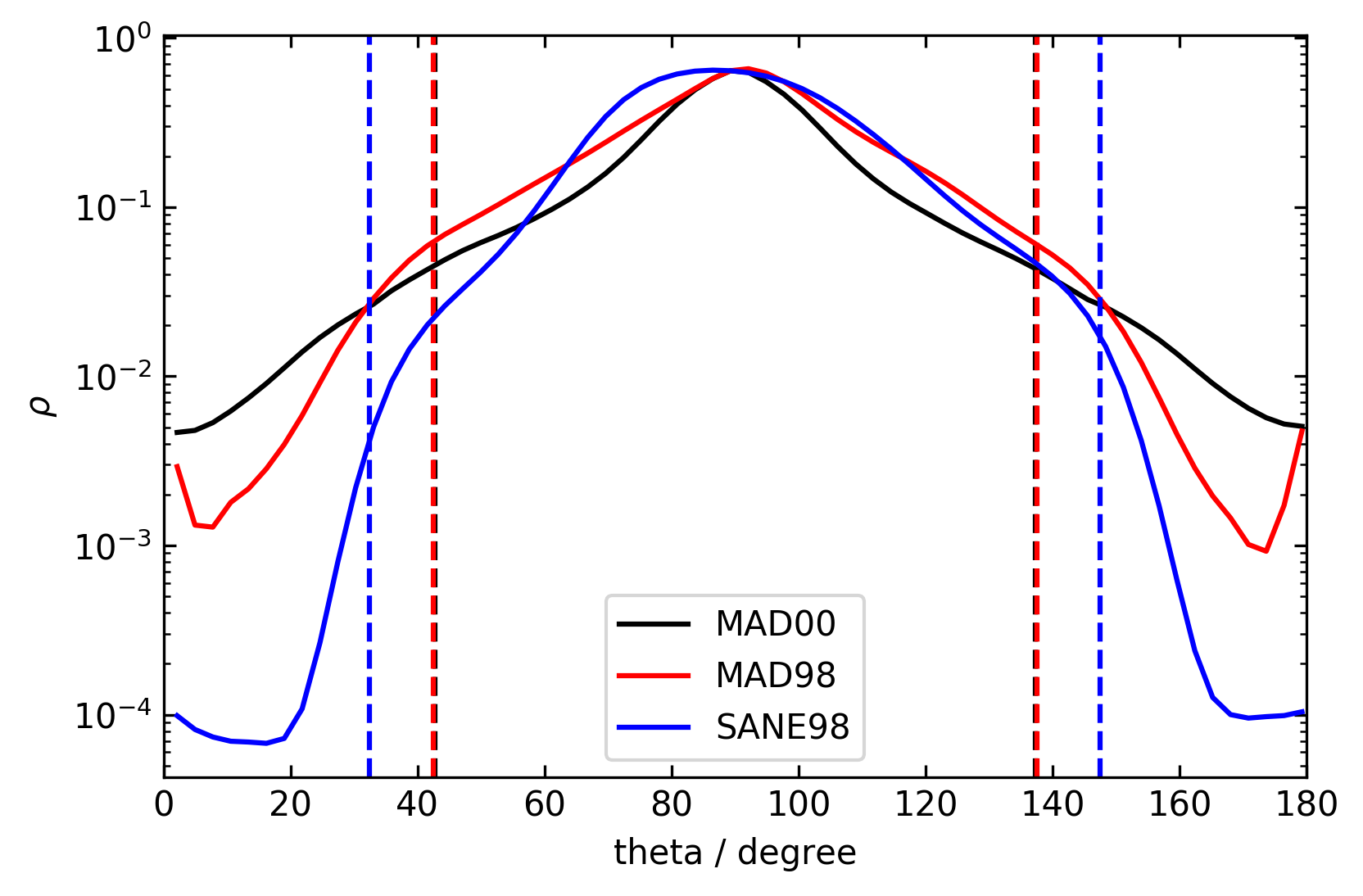}{0.48\textwidth}{(a)}
\fig{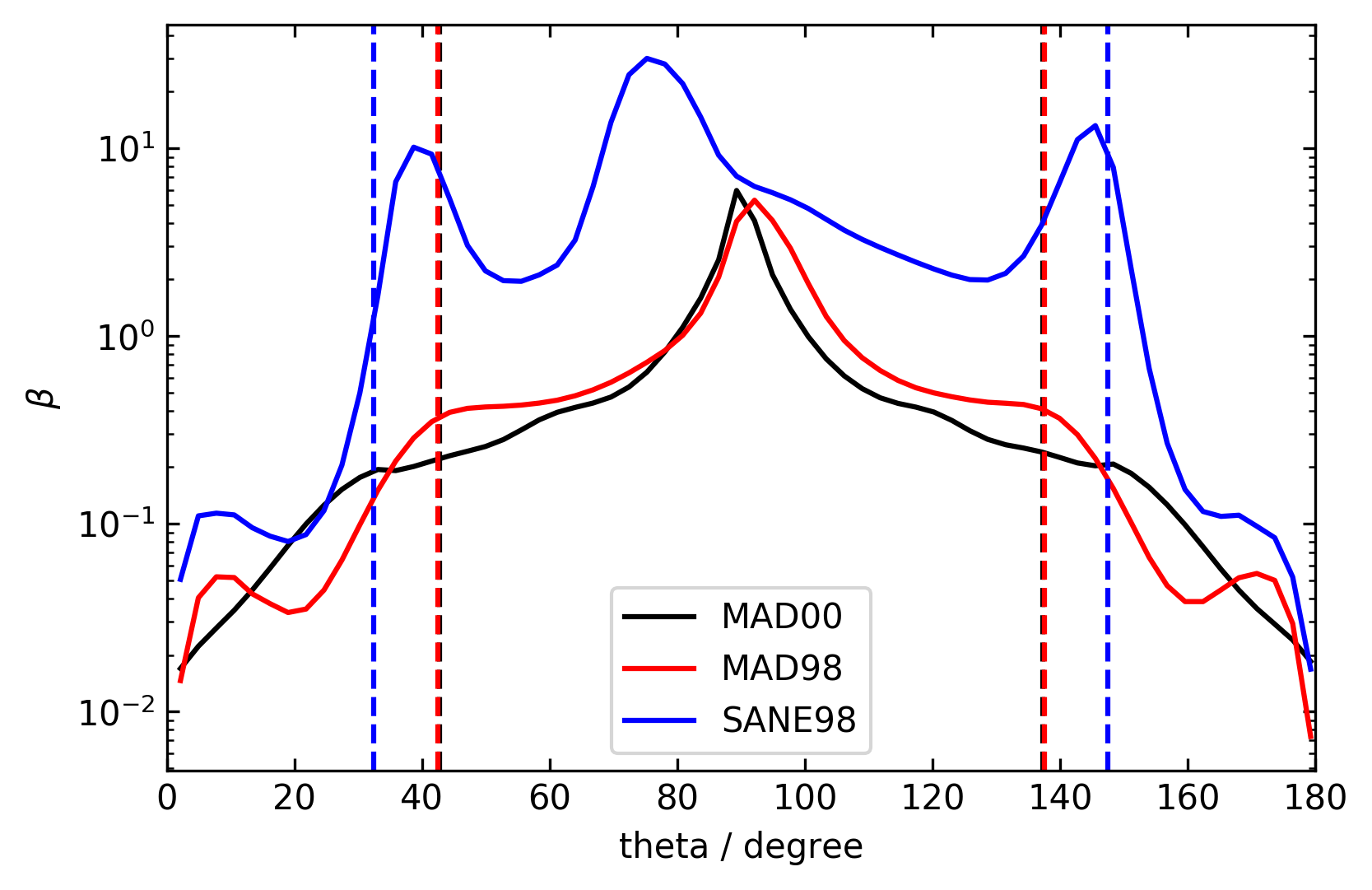}{0.48\textwidth}{b}
/
}
\gridline{\fig
{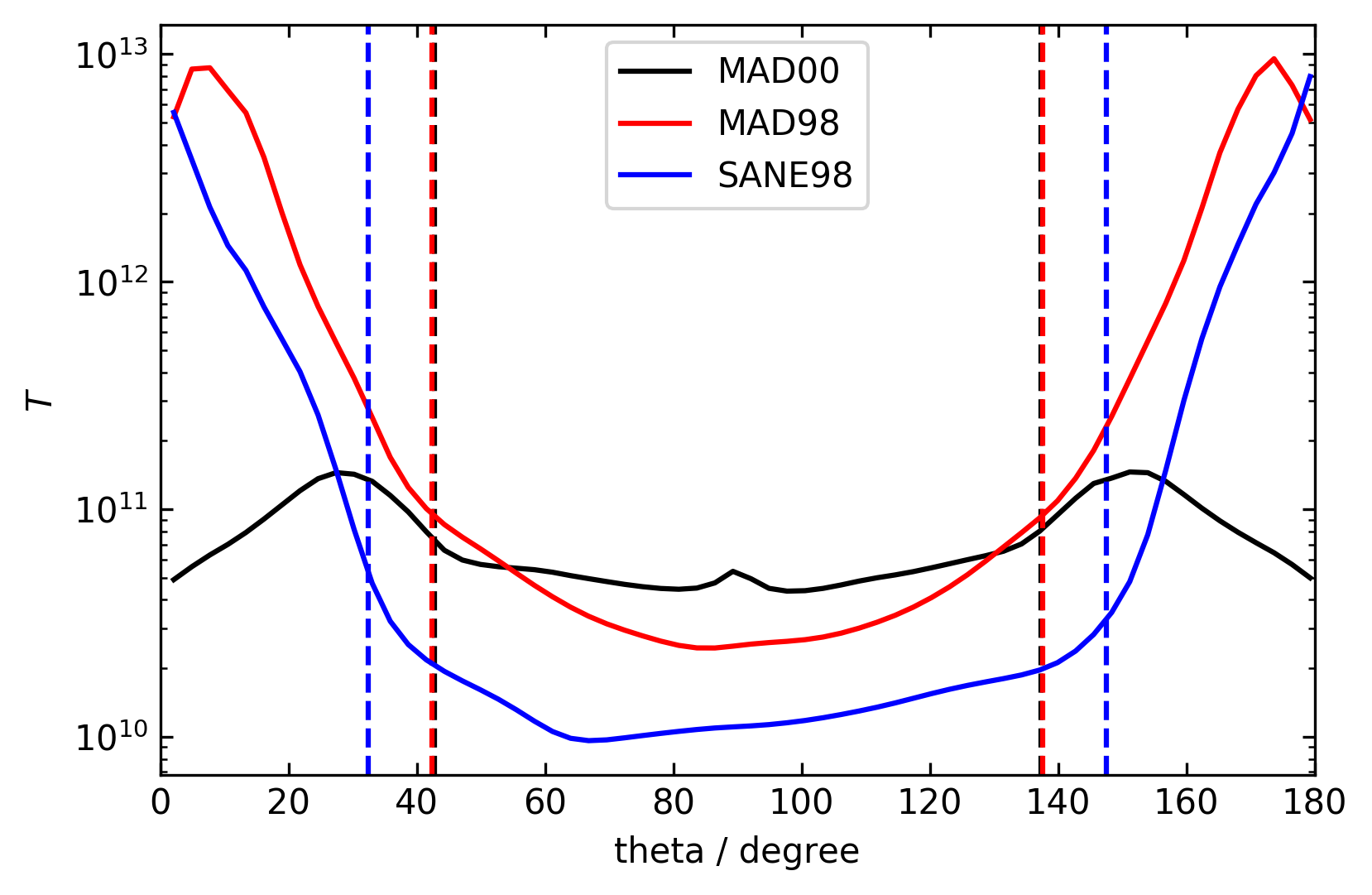}{0.48\textwidth}{(c)}
\fig{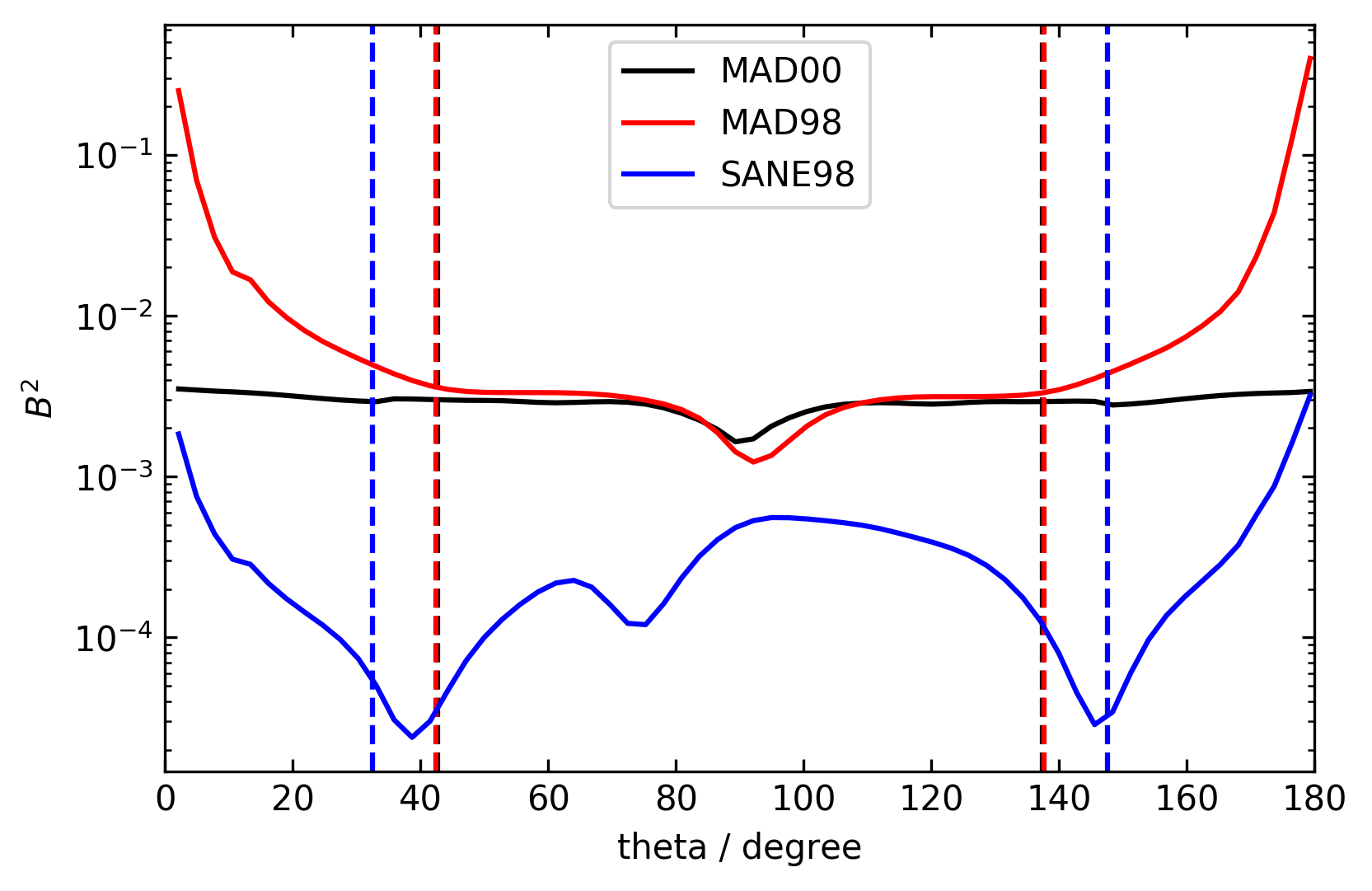}{0.48\textwidth}{(d)}
}

\caption{ Different quantities of the flow averaged over $\varphi$ and $t$ as a function of $\theta$ for  MAD00 (black), SANE98 (blue), and MAD98 (red)  at $r=40\,r_{\rm g}$.  The vertical dashed line represents the boundary between the BZ jet and wind. The density (and therefore also $B^2$) are normalized so that different models have the same value of density at the equatorial plane.}
\label{fig:density}
\end{figure*}
\begin{figure*}
 
   \gridline{
   \fig{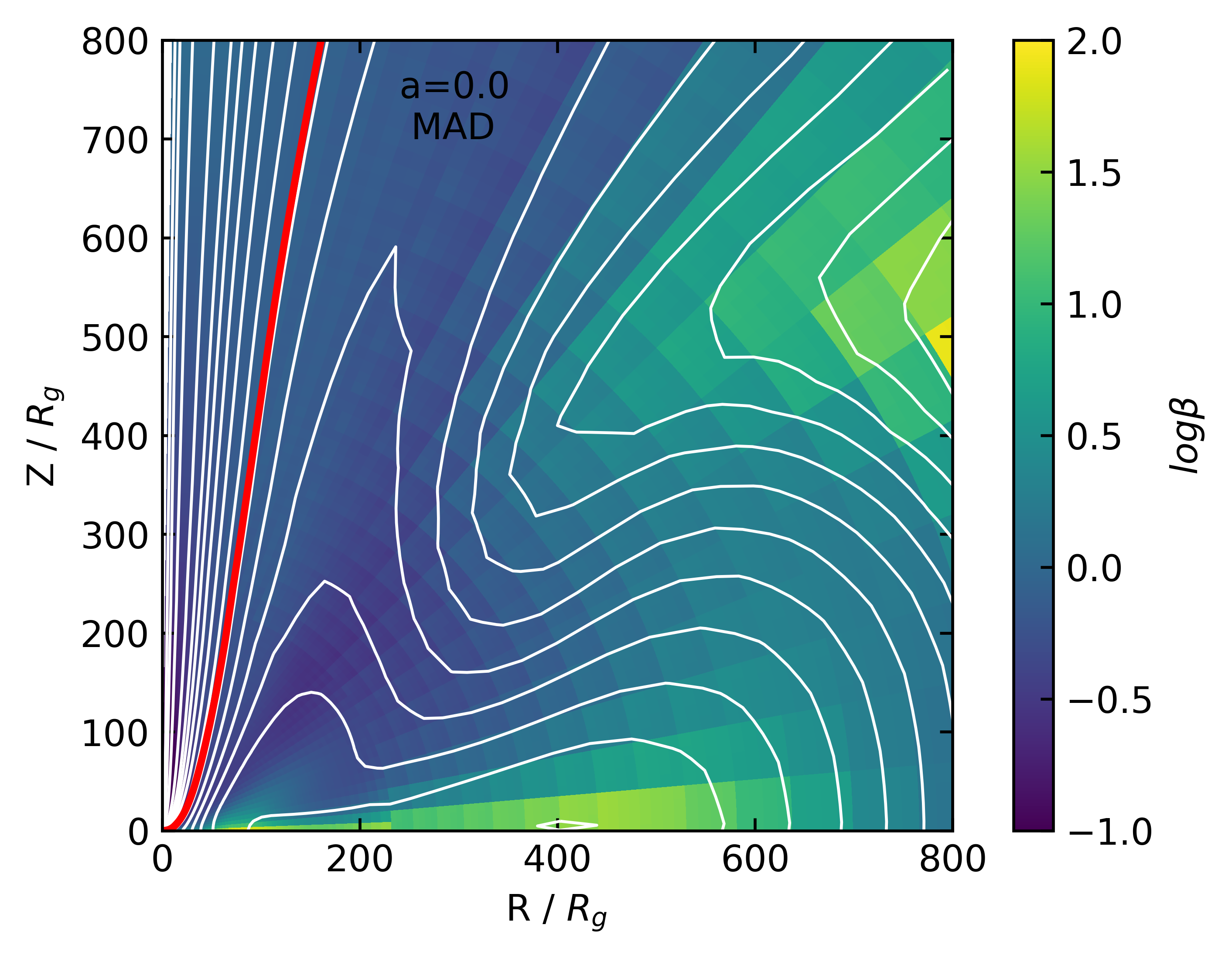}{0.33\textwidth}{}
   \fig{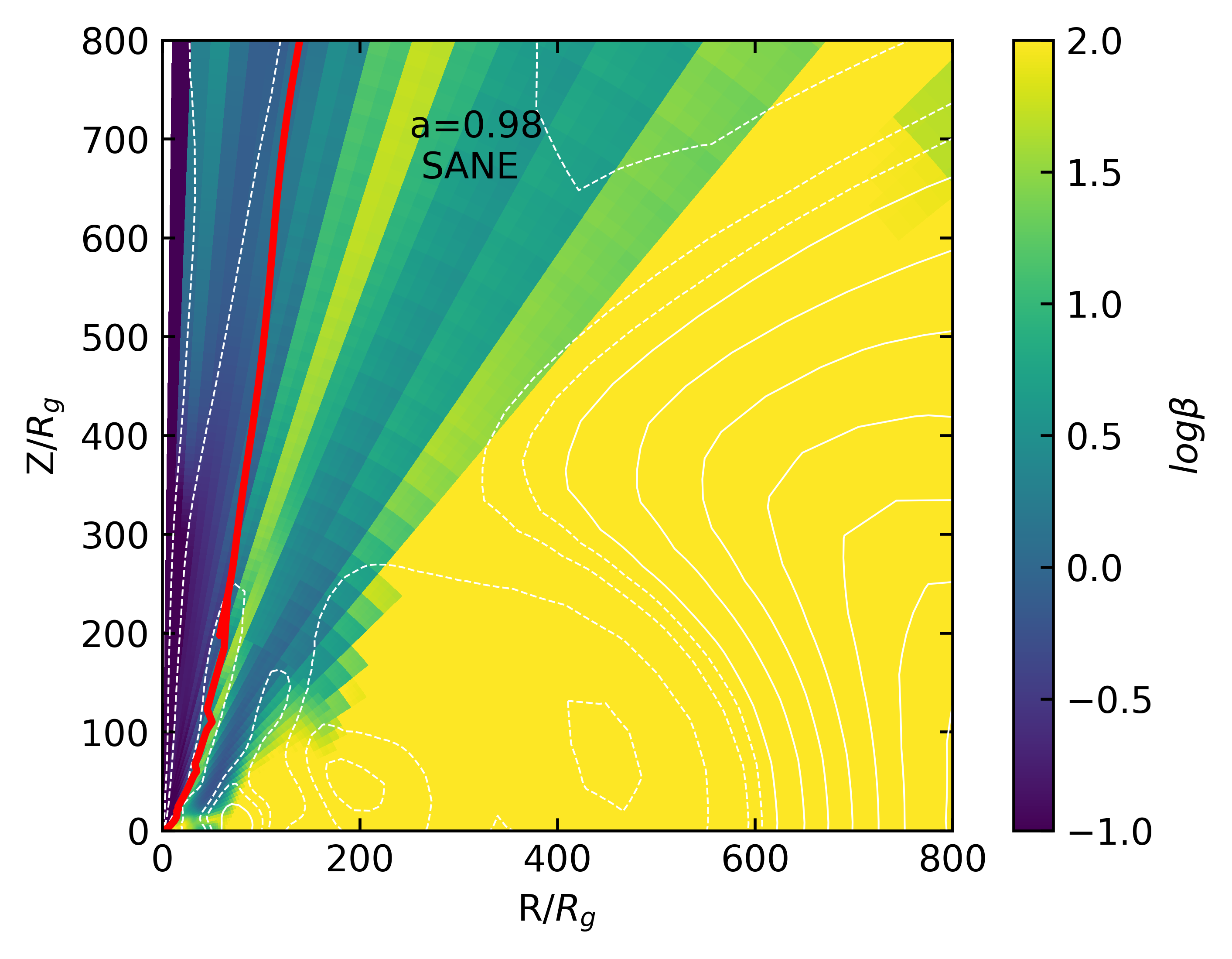}{0.33\textwidth}{}
   \fig{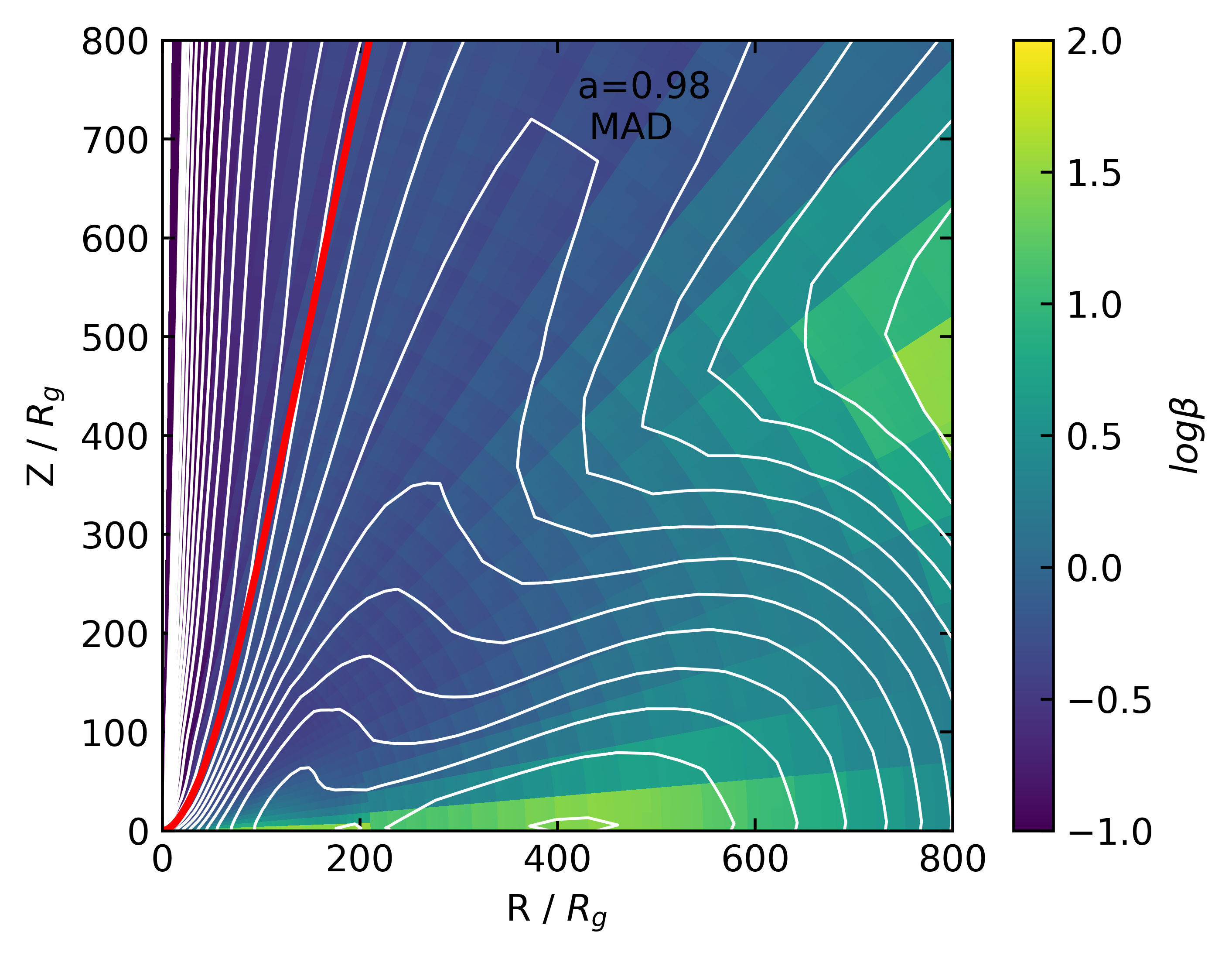}{0.33\textwidth}{}
   }
    \caption{ Magnetic field distribution of  MAD00, SANE98 and MAD98. The white lines denote $t-$ and $\varphi-$averaged magnetic field lines; the colors denote the logarithm of the $\beta$ of the plasma. The red line is the magnetic field line that is rooted at the black hole ergosphere with $\theta=90^{\circ}$, i.e., the boundary between the ergosphere and the accretion flow. It is the boundary between the BZ-jet and wind.}
    \label{fig:mag}
\end{figure*}

\subsection{Virtual particle trajectory approach}
\label{trajectory}

We briefly overview the ``virtual particle trajectory'' approach proposed in \citet{Yuan2015}. Trajectory is related to the Lagrangian description of fluid, obtained by following the motion of fluid elements at consecutive times. Since the motion of fluid is not strictly steady but turbulent, trajectory is very different from streamlines. To get the trajectory, we first need to choose a set of “test particles” in the simulation domain. Note that they are not real particles, but a collection of spatial coordinates as the starting points for the
trajectory calculation.  Usually we put some  “test particles” in different $\theta$ and $\varphi$ at a given initial radius $r_0$ and a given initial time $t_0$. Their velocity can be obtained from the interpolation of the simulation data. With this information, we can obtain their location at time  $t_0+\delta t$. We use the ``VISIT'' software to perform the calculation. Repeating this process,  we can obtain the trajectory of these particles. It is important to choose an appropriate time step $\delta t$ to obtain a convergent trajectory. After some tests, we choose $\delta t=100\,r_{\rm g}/c$ in our work; it corresponds to the Keplerian timescale at $r\approx 6\,r_{\rm g}$. In this way, we can obtain the trajectories of fluid elements or ``virtual test particles'' based on our high time resolution numerical simulation data. The readers can refer to Figures 1 \& 2 in \citet{Yuan2015} for examples of various trajectories we have obtained in the case of SANE00. Once we obtain the trajectories, we can judge whether the motion is turbulence or real wind, and  calculate the wind properties such as mass flux and poloidal velocity, as we will discuss in \S\ref{sec:result}.  Obviously, obtaining the trajectory of fluid elements is much more time consuming than the streamlines approach, but it can faithfully reflect the motion of fluid compared to the latter.

\section{Result}
\label{sec:result}
\subsection{Definitions of wind and jet}
Let us first define some terminology before describing our results. “Outflow” means the flow with a
positive radial velocity $v_r$, including both “turbulent outflow” and “real outflow.”  In the former case the test particle will first move outward but eventually will return after moving outward for some distance. In the latter case the test particle continues to flow outward and eventually escapes the outer boundary of the simulation domain. “Real outflow” consists of two components, i.e., jet and wind. The jet region is defined as the region occupied by the time-averaged magnetic field lines connected to the ergosphere (described by $r_{\rm erg}\equiv r_{\rm g}+\sqrt{1-a^2{\cos}^2\theta}\,r_{\rm g}$; \citealp   {Visser2007}) of the black hole. We note that this definition coincides with other notions of the jet based on velocity and magnetization, as seen even in early two dimensional simulations \citep{McKinney2004}. Moreover, turbulence tends not to have a strong effect on these field lines, so that in both individual snapshots and in a time-averaged sense the field lines connecting to the ergosphere coincide with $\beta < 2$ and $\sigma > 1$ (\citealp{McKinney2012}, see Figures 3 and 6). 
Thus the jet region is bounded by the magnetic field line whose foot point is rooted at the black hole ergosphere  with $\theta=90^{\circ}$, i.e., the boundary between the black hole ergosphere and  the accretion flow (refer to the red lines in Figure~\ref{fig:mag}). In this case all magnetic field lines in the jet region are anchored to the black hole ergosphere and thus can extract the spin energy of the black hole via the Blandford \& Znajek mechanism \citep{Blandford1977Znajek} to power the jet (``BZ-jet''). Real outflows outside of this boundary are powered by the rotation energy of the accretion flow and we call them wind. Note that our definition of wind adopted here is different from that adopted in some literature, where they require that the Bernoulli parameter of wind must satisfy $\mathrm{Be} > 0$.  We do not add this requirement, because we find that for non-steady
accretion flow $\mathrm{Be}$ is not constant along trajectories, but usually increases outward at least until the radius within which turbulence is well developed \citep{Yuan2015}. We find that  even though $\mathrm{Be}$ is negative at a small radius, it can become positive when it propagates outward. 

In \citet{Yuan2015}, although the black hole $a=0$, we still find jet-like outflow in terms of high velocity. This is confirmed by the results of the present paper, as we will show in the left panels of Figure \ref{fig:pv} for MAD00. We can see in this figure that the poloidal speed of the plasma within the red line  is significantly larger than that of the wind. Of course, these outflows must be powered by the rotation energy of the underlying accretion disk rather than by the black hole, so we call them ``disk-jet'' (\citealp{Yuan2014Narayan,Yuan2015}; see also \citealp{Ghosh97,Livio99}). These results suggest that in AGNs and black hole X-ray binaries jets may still be present even though the black hole is non-spinning. In the present work, we focus on the difference between outflows powered by the spinning black hole and by the rotating accretion disk, so we simply call the former ``BZ-jet'' or ``jet'' and the latter  ``wind''. 

\subsection{Overview of the simulation results}

We have run simulations up to $t_f\,=\,40{,}000$, $40{,}000$ and $80{,}000$ for MAD00, MAD98 and SANE98 respectively. They correspond to $8.9$, $8.9$ and $17.8$ orbital periods of the disk at the pressure maximum. As we will see from Figure \ref{fig:mdotw}, the `inflow equilibrium' has been reached at $\sim 30-40 r_g$ for the three models, i.e., the models have reached the steady state within this radius. A longer simulation time will help to extend the region of steady state, but this will be even more expensive. The relatively small inflow equilibrium radius will inevitably affect our evaluation of the wind properties at large radii such as the mass flux. This is because, e.g., the surface density of the accretion flow has not settled down to the steady value but is still subject to the initial condition. However, the extremely long simulation of SANE00 by \citet{White2020a}, which obtains a large radial range of inflow equilibrium, indicates that we may not be able to obtain a universal surface density within the inflow equilibrium radius. This suggests that it is possible that  beyond our inflow equilibrium radius the surface density of the accretion flow we obtain may still be realistic thus viable for our evaluation of wind properties. Our another ``defending'' argument is that the power-law indexes of wind properties we obtain, as we will show in the following subsections, are roughly same within and beyond the inflow equilibrium radius. In summary, we believe that our evaluations to wind should be roughly reliable. In fact, as we will see below, our results for various models are consistent with our physical expectations. But we emphasize that we should keep this caveat in mind and future works are required to examine this point.

Figure~\ref{fig:densityvelocity} shows the $t$- and $\varphi$-averaged two dimensional distribution of density and velocity in the $r-\theta$ plane for the three models  at the chosen time chunks of  $t=10{,}000\text{--}20{,}000$, $55{,}000\text{--}68{,}000$ and $8000\text{--}23{,}000$,  respectively. 
Figure \ref{fig:density} shows in a more quantitative way various quantities of the flow averaged over $\varphi$ and time as a function of $\theta$ at $40\,r_{\rm g}$. From the top to bottom panels we have density, the ratio of the gas pressure to magnetic pressure $\beta$, temperature, and $B^2$. 
From the figure we can see that the accretion flows of the two MAD models are geometrically thinner than that of the SANE model, consistent with  previous works \citep[e.g.,][]{Tchekhovskoy2011,McKinney2012}. The reason is because in the MAD model the magnetic field is so strong that it compresses the accretion flow vertically.  We also find that the velocity field  of the two MAD models are much more ordered than that of the SANE model.  This is because the magnetorotational instability (MRI) is suppressed in the MAD model so there is no turbulence, while MRI is still present in the SANE \citep[e.g.,][]{McKinney2012}.  

Figure~\ref{fig:mag} shows the $\varphi$- and $t$-averaged magnetic field lines of the three models. The field configuration of the two MAD models are very similar and both are very ordered. This is again because MRI is suppressed so there is no turbulence. For the SANE model, the field lines are less ordered due to the existence of turbulence. The plasma $\beta$ is also much smaller in the two MAD models than in the SANE model, as expected. The red line is the magnetic field line anchored at the ergosphere with $\theta=90^{\circ}$, i.e., the boundary between the BZ-jet and wind regions. From the figure,  we can  see that the BZ-jet region is strongly dominated by magnetic pressure, especially for the two MAD models. The shape of the red line is parabolic, i.e., $z\propto R^s$ in the cylindrical coordinates $(R, z)$ with $s>1$. There are many discussions about the shape of this magnetic field line or streamline. Readers can refer to, e.g., \citet{Nakamura2018} and \citet{Chen2021} for details. 

\subsection{Mass fluxes of wind and jet}
\label{massflux}
 To calculate the mass flux of the wind at a given time $t_0$, we first put  test particles at a given radius $r$ with different $\theta$ and $\varphi$ and obtain
their trajectories.  The mass flux of wind is calculated by summing up the corresponding mass flux carried by test particles whose trajectories belong to the real outflow (i.e., wind or jet) using the following formula:
\begin{equation}
    \dot{M}_{\rm wind (jet)}( r)=\sum_{i}\rho_i(r )u^r_i( r)\sqrt{-g}\delta\theta_i\delta\varphi_i.
	\label{eq:mdots}
\end{equation}
Here the subscript represent the different test particles, $\rho_i$ and $u^r_i$ are the mass density and four-velocity at the location where the test particle ``$i$'' originates, and $\delta\theta_i$ and $\delta\varphi_i$ are the ranges of $\theta$ and $\varphi$ the particles occupy.

\begin{figure}[h]
	\includegraphics[width=\columnwidth]{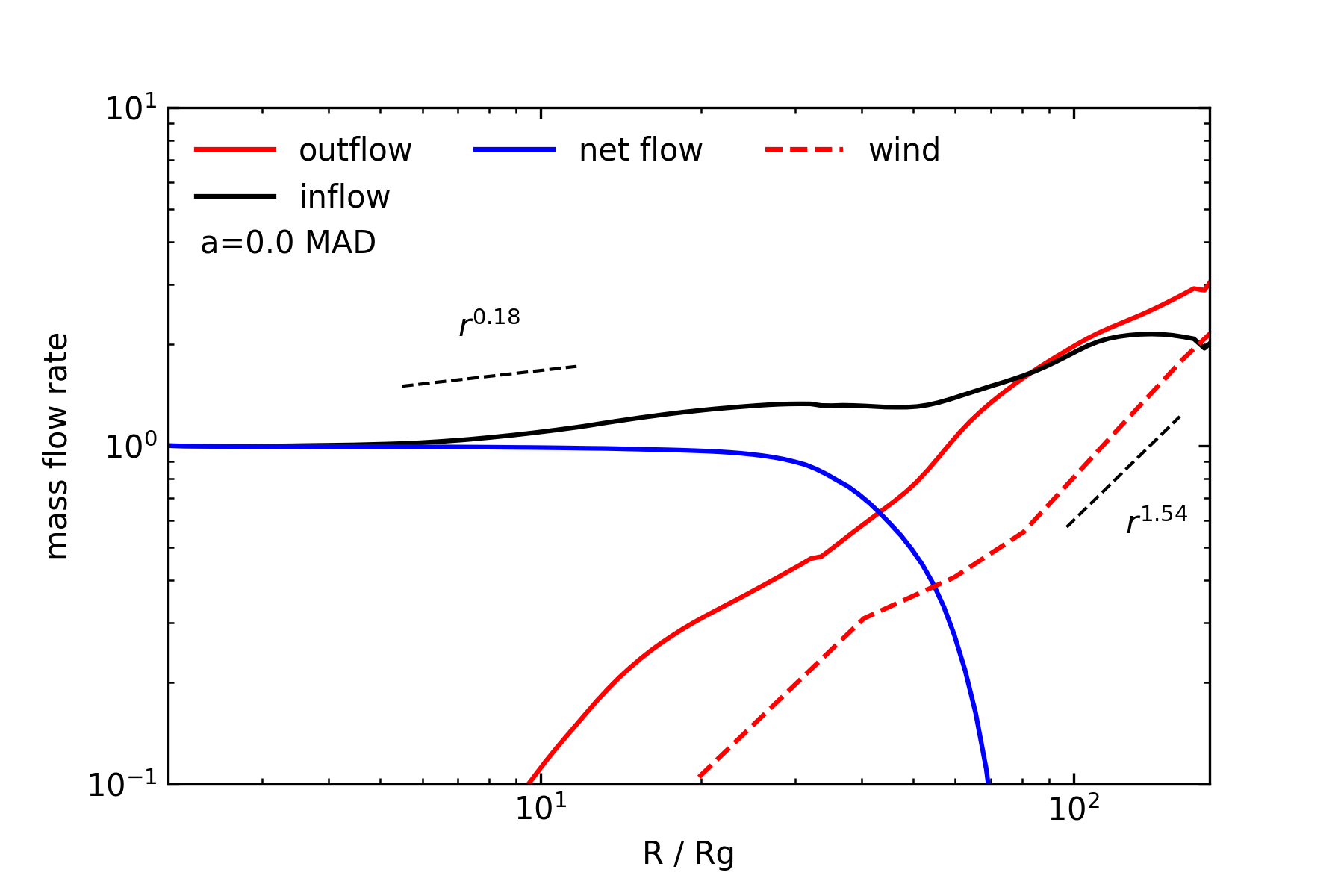}
    \includegraphics[width=\columnwidth]{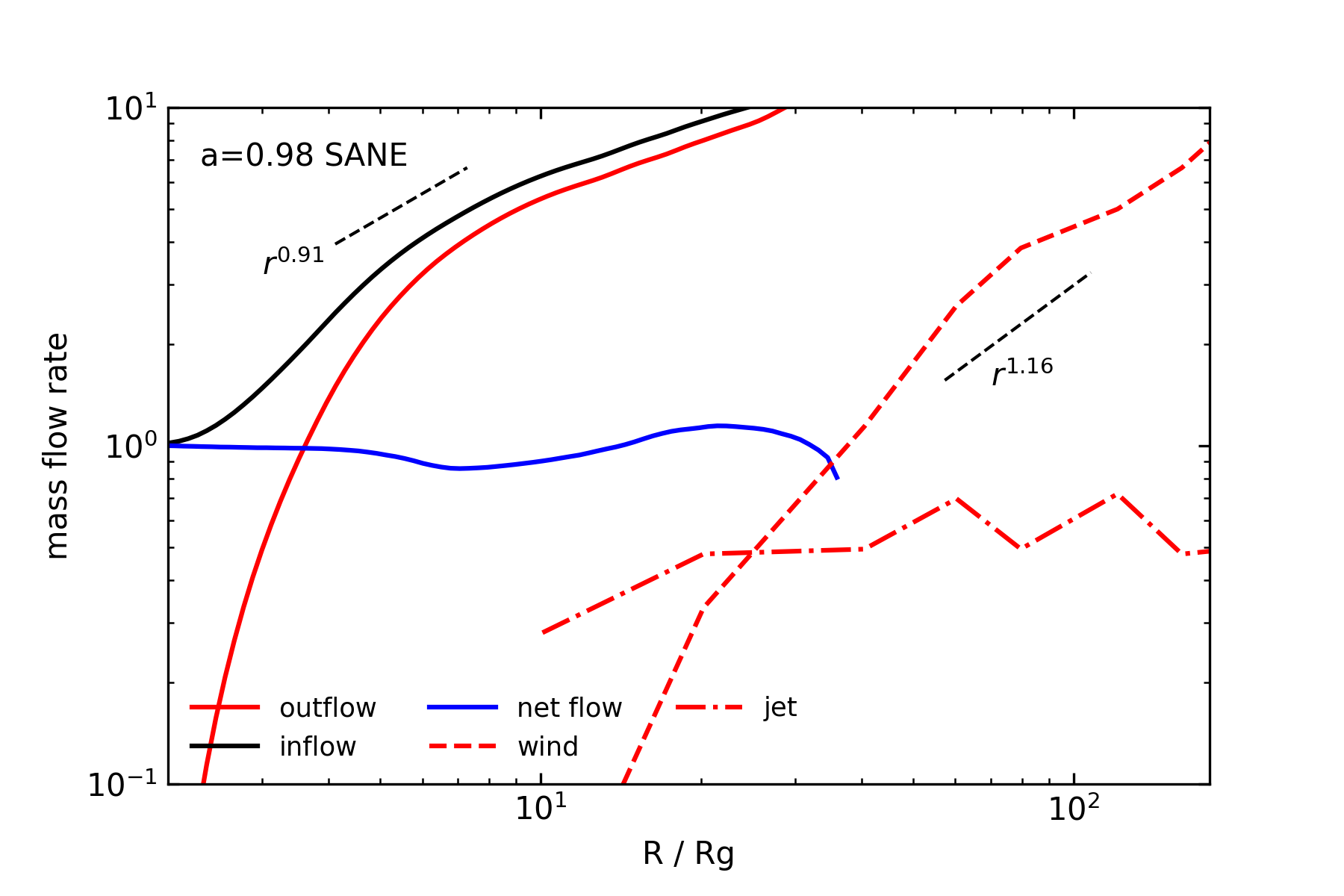}
	\includegraphics[width=\columnwidth]{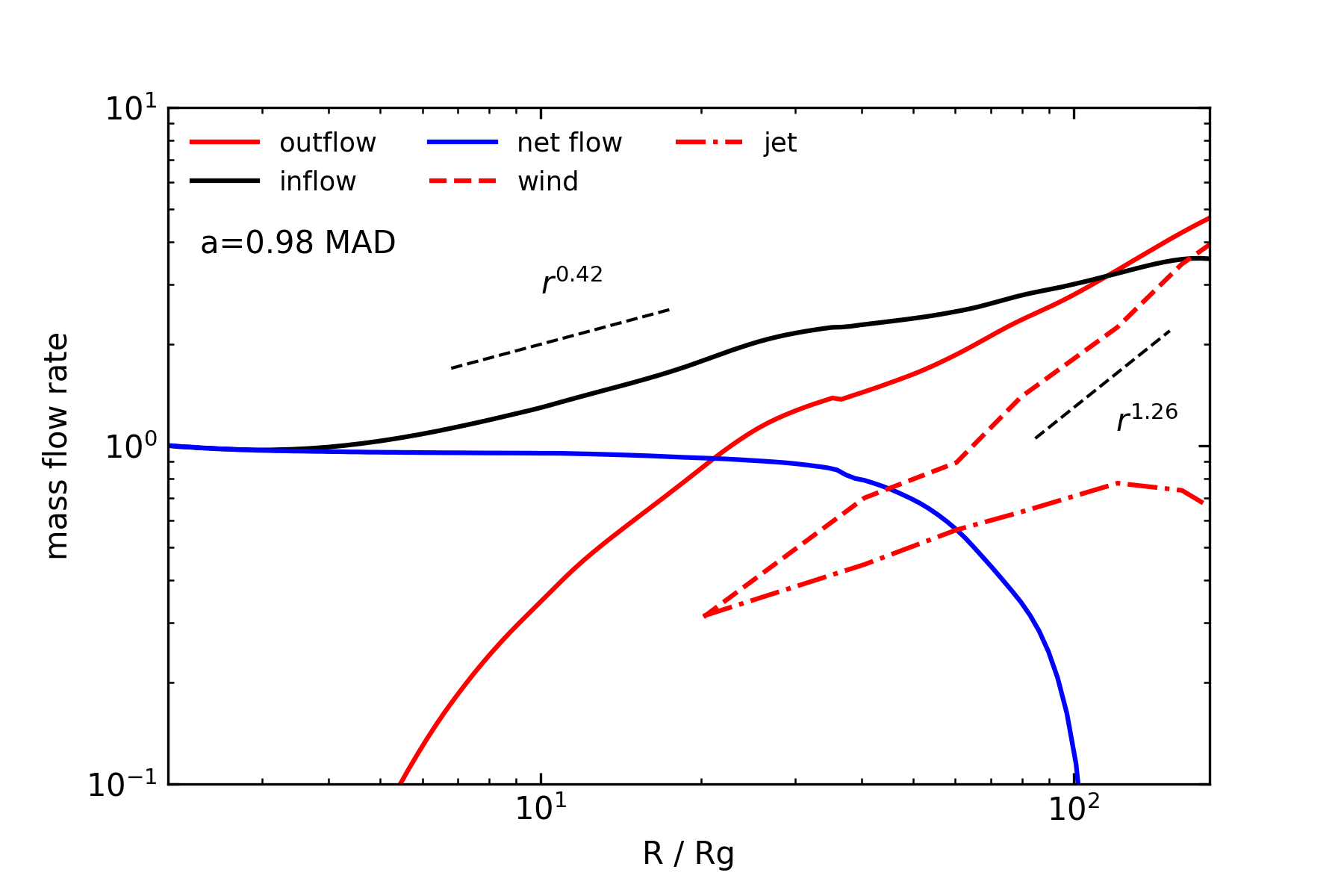}
    \caption{The mass fluxes of inflow, outflow, wind, and jet for MAD00 (top), SANE98 (middle) and MAD98 (bottom) respectively. The values have been normalized by the net rate $\dot{M}_{\rm BH}$. 
}
    \label{fig:mdotw}
\end{figure}

 In our calculations, we usually choose 10 different initial times $t_0$ to obtain the trajectories of these test particles and the mass flux corresponding to each choice of $t_0$. We then do time-average of these 10 groups to obtained the averaged mass flux.  These  10 initial times are uniformly selected after the simulation has reached steady state. Because the calculation of trajectory requires simulation data after time $t_0$, and it needs ``test particles'' to move away from their initial location far enough to distinguish the real outflow from turbulence, the initial time $t_0$ should not be too late. For MAD00, MAD98 and SANE98, the time intervals we choose for the time-average are $t_0=10{,}000$ to $20{,}000$, $t_0=8000$ to 23000, and $t_0=55{,}000$ to $68{,}000$, respectively. To calculate the wind properties (such as mass flux, velocity, and energy and momentum fluxes) as a function of radius, we have chosen a series of initial radii, namely $r=20\,r_{\rm g}$, $40\,r_{\rm g}$, $60\,r_{\rm g}$, $80\,r_{\rm g}$, $120\,r_{\rm g}$, $160\,r_{\rm g}$, $200\,r_{\rm g} $ for MAD98 and MAD00, and $r=10\,r_{\rm g}$, $20\,r_{\rm g}$, $40\,r_{\rm g}$, $60\,r_{\rm g}$, $80\,r_{\rm g}$, $120\,r_{\rm g}$, $160\,r_{\rm g}$, $200\,r_{\rm g} $ for SANE98. The initial position of the ``test particle'' in $\theta$ and $\varphi$ directions are consistent with our simulation grid, i.e., 128 grid points in the $\theta$ direction $[0,\pi]$ and 64 grid points in the $\varphi$ direction $[0, 2\pi]$. We  sum the mass flux of all real outflow as wind mass flux for MAD00 since there is no BZ-jet in this case; but for the $a=0.98$ model, i.e.,  SANE98 and MAD98, because of the existence of the BZ-jet, we calculate the mass flux of the wind and jet separately. 

\begin{figure*}[t]
\gridline{
\fig{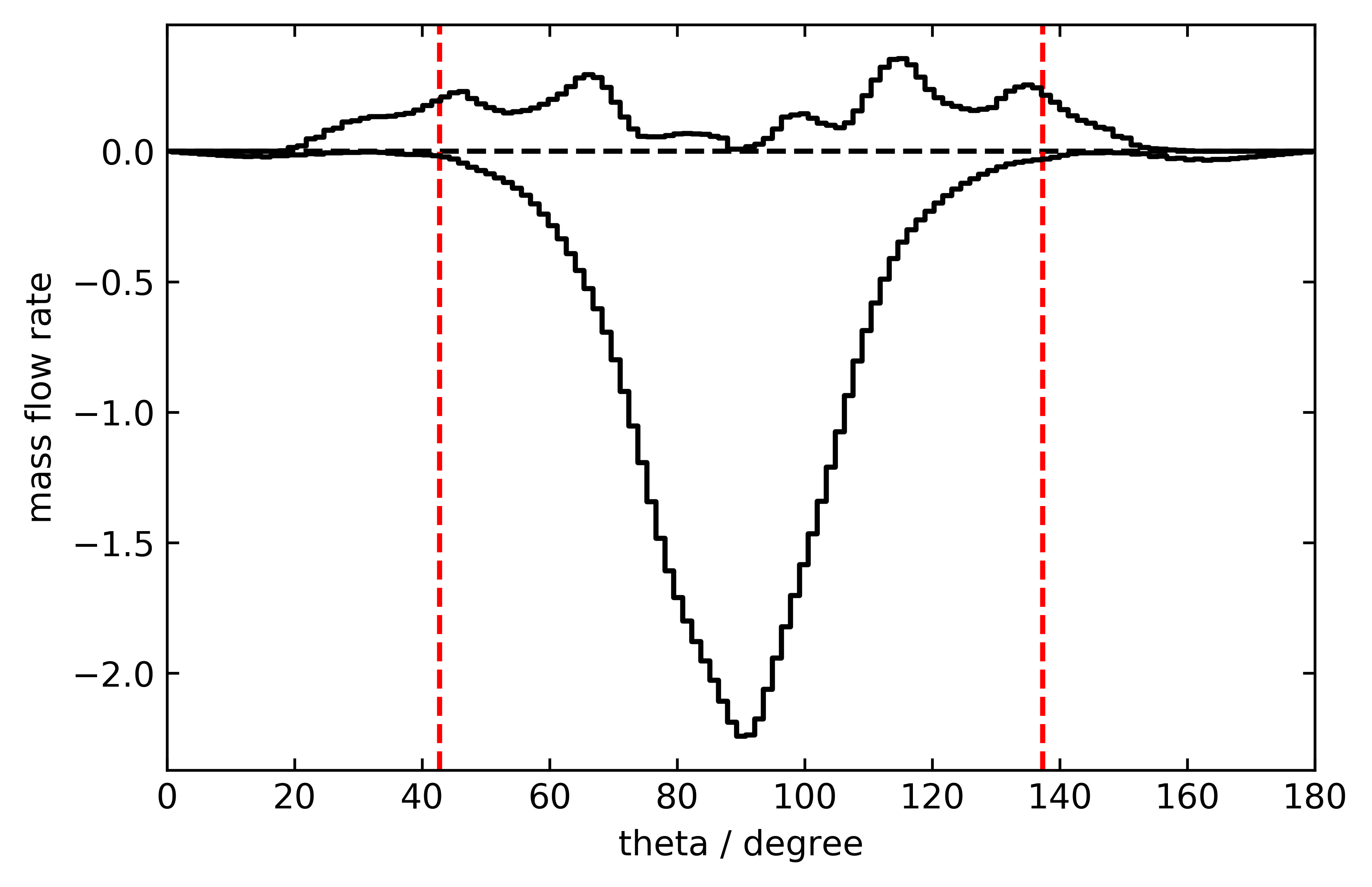}{0.33\textwidth}{}
\fig{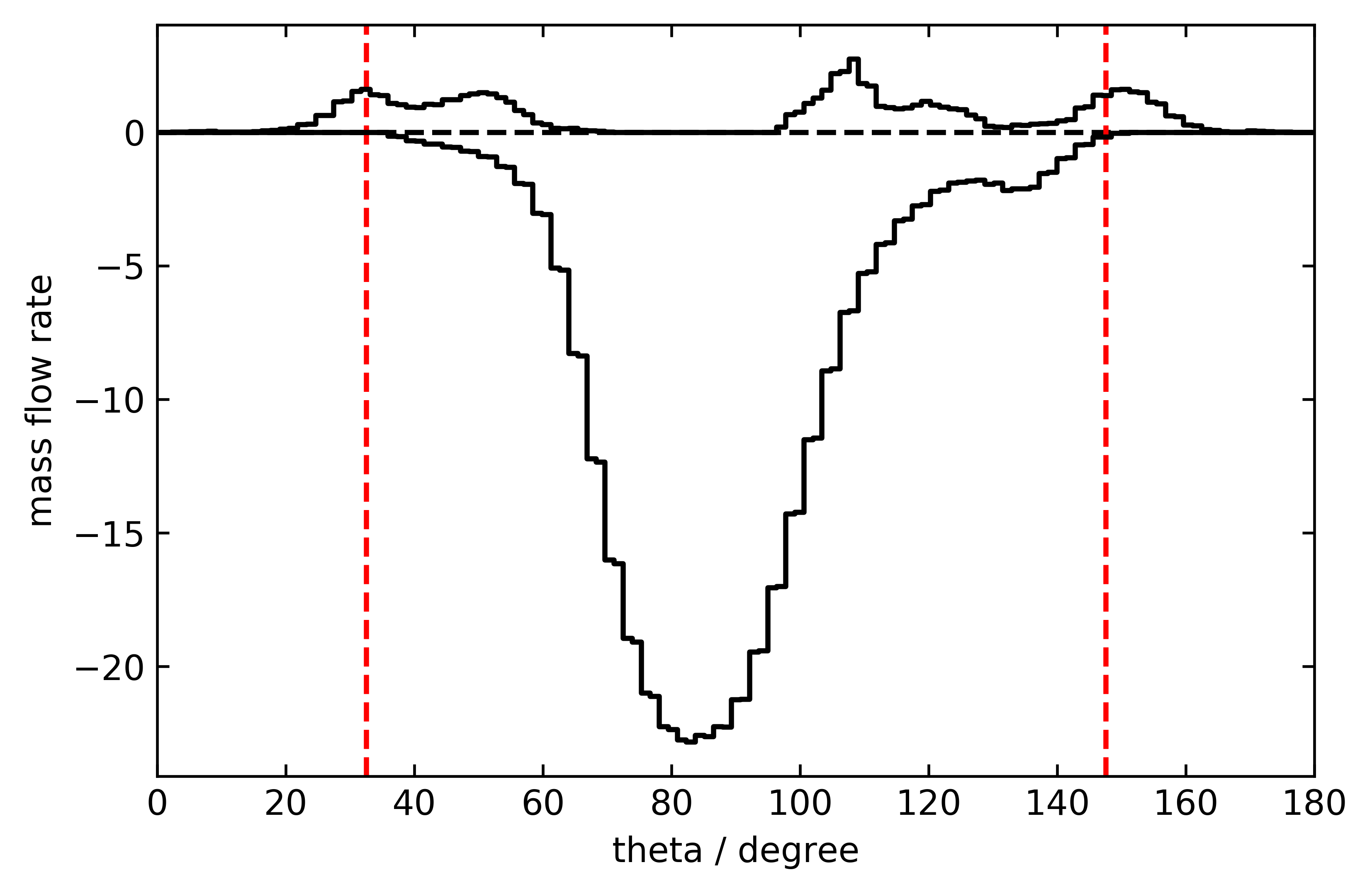}{0.33\textwidth}{}
\fig{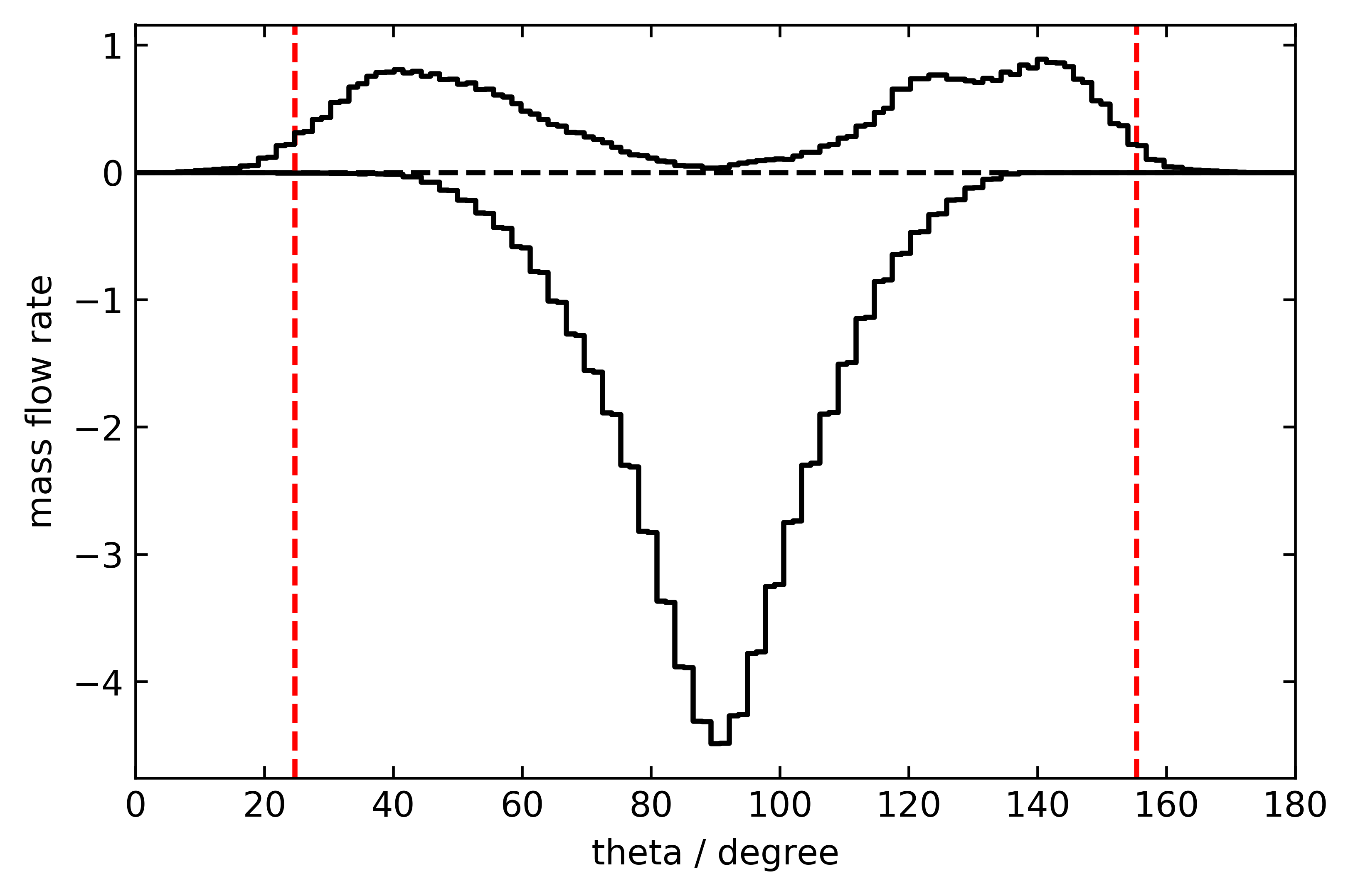}{0.33\textwidth}{}
}
\gridline{
\fig{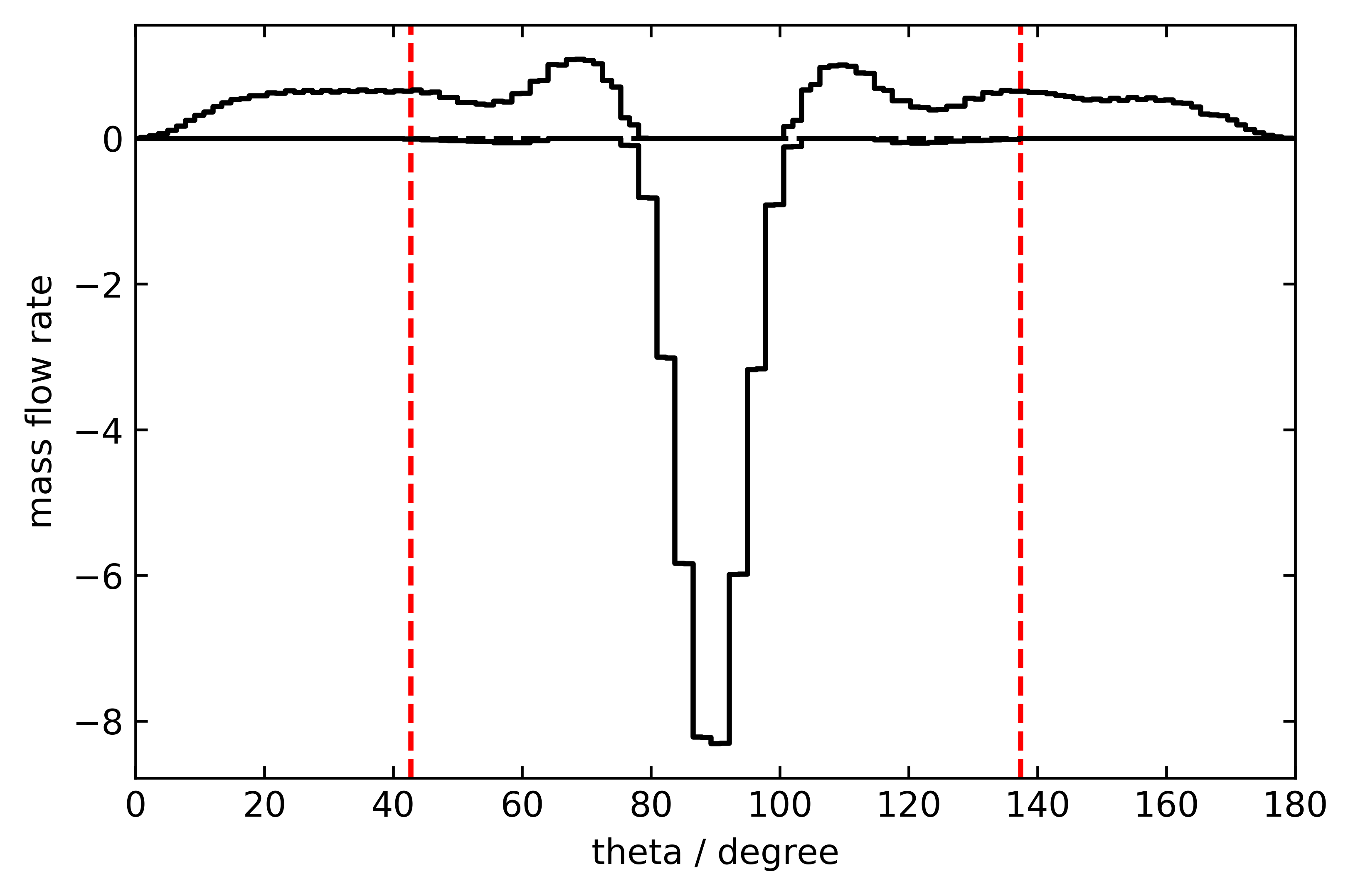}{0.33\textwidth}{}
\fig{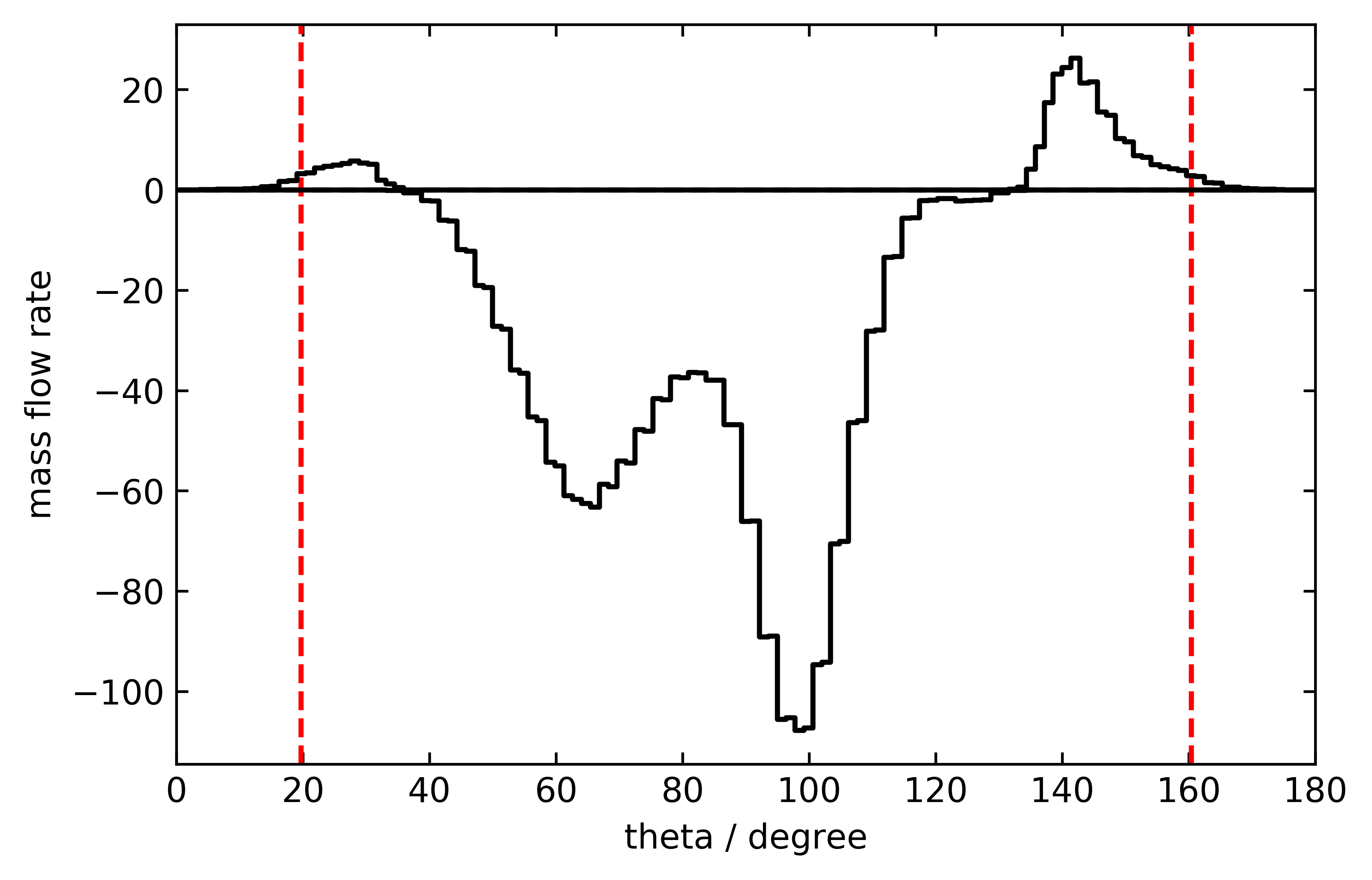}{0.33\textwidth}{}
\fig{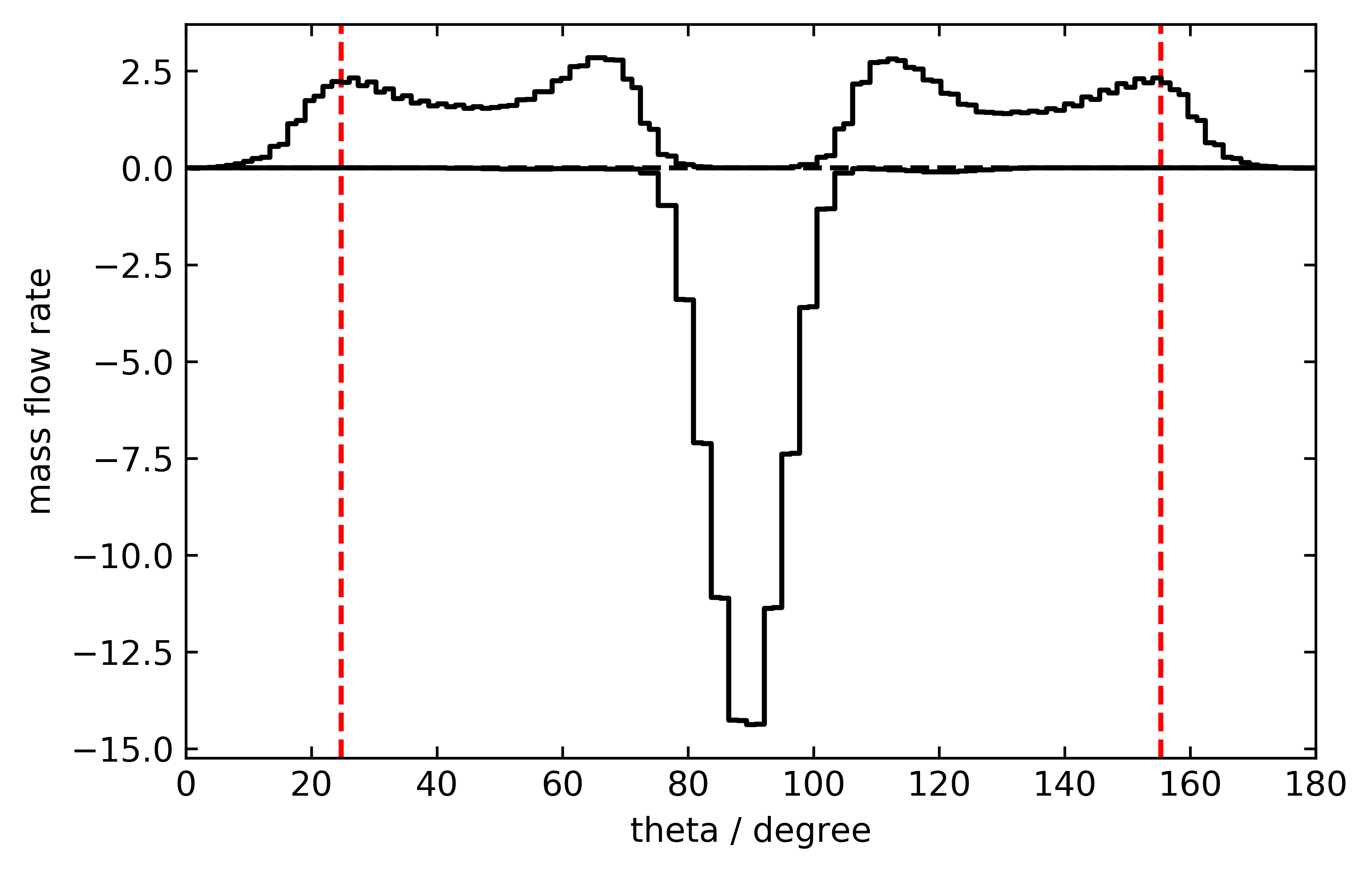}{0.33\textwidth}{}
}
\caption{The mass flux of wind per unit $\theta$ as a function of $\theta$ for  MAD00 (left), SANE98 (middle), and MAD98 (right), respectively. The values are integrated over all  $\varphi$ and averaged from time $55{,}000\text{--}68{,}000$, $10{,}000\text{--}20{,}000$ and $8000\text{--}23{,}000$ for SANE98, MAD00 and MAD98 respectively, and have been normalized by $\dot{M}_{\rm in}(r=2\, r_{\rm g})$ of the corresponding models. The top and bottom panels are for $r=40\,r_{\rm g}$ and $r=160\,r_{\rm g}$ respectively. The positive value denotes the wind flux, negative value denotes the total inflow $\dot{M}_{\rm in}$. The red dashed line represents the boundary between the BZ-jet and wind.  }
\label{fig:mdotall}
\end{figure*}

Figure~\ref{fig:mdotw} shows the results.  The black, red, and blue solid lines show the inflow,  outflow, and net rates, respectively. The inflow and outflow rates are calculated using the following equation,
\begin{equation}
    \dot{M}(r)=-\int_{0}^{2\pi}\int_{0}^{\pi}{\rho}u^{r}\sqrt{-g}d\theta d\varphi,
	\label{eq:M}
\end{equation}
where $u^r$ is the radial contravariant component of the 4-velocity, and the integral is over all $\theta, \varphi$ at fixed $r$. We add the negative sign so that the flux is positive when flow moves inward. The net rate is the difference between inflow and outflow rates. Note that here the outflow rate $\dot{M}_{\rm out}$ includes both real outflow (jet and wind) and the turbulent outflow. The red dashed and red dot-dashed lines denote the mass fluxes of  wind and jet respectively, obtained using our trajectory approach.  Combined with the result of SANE00 from \citet{Yuan2015}, beyond several $r_g$, the radial profiles of the inflow rate for SANE00, SANE98, MAD00 and MAD98 can be described by 
\begin{equation}
    \dot{M}_{\rm in}(r )=\left(\frac{r}{4.8\,r_{\rm s}}\right)^{0.54}\dot{M}_{\rm BH},
	\label{eq:im1}
\end{equation}
\begin{equation}
    \dot{M}_{\rm in}(r )=\left(\frac{r}{0.7\,r_{\rm s}}\right)^{0.91}\dot{M}_{\rm BH},
	\label{eq:im2}
\end{equation}
\begin{equation}
    \dot{M}_{\rm in}(r )=\left(\frac{r}{2.8\,r_{\rm s}}\right)^{0.18}\dot{M}_{\rm BH},
	\label{eq:im3}
\end{equation}
\begin{equation}
    \dot{M}_{\rm in}(r )=\left(\frac{r}{2.5\,r_{\rm s}}\right)^{0.42}\dot{M}_{\rm BH},
	\label{eq:im4}
\end{equation}
The radial profiles of the mass flux of wind for the four models are described by
\begin{equation}
    \dot{M}_{\rm w}(r )=\left(\frac{r}{20\,r_{\rm s}}\right)^{1.0}\dot{M}_{\rm BH},
	\label{eq:m1}
\end{equation}
\begin{equation}
    \dot{M}_{\rm w}(r )=\left(\frac{r}{15\,r_{\rm s}}\right)^{1.16}\dot{M}_{\rm BH},
	\label{eq:m2}
\end{equation}
\begin{equation}
    \dot{M}_{\rm w}(r )=\left(\frac{r}{55\,r_{\rm s}}\right)^{1.54}\dot{M}_{\rm BH},
	\label{eq:m3}
\end{equation}
\begin{equation}
    \dot{M}_{\rm w}(r )=\left(\frac{r}{30\,r_{\rm s}}\right)^{1.26}\dot{M}_{\rm BH},
	\label{eq:m4}
\end{equation}
respectively. Here $\dot{M}_{\rm BH}$ is the mass accretion rate at the black hole horizon. From these, we can find the following results:
\begin{itemize}
    \item By comparing SANE00 and MAD00, we can see that close to the black hole, the wind flux becomes weaker when the magnetic field becomes stronger. This is likely because turbulence helps the production of wind. The turbulence in SANE is caused by MRI; but in the case of MAD, the magnetic field is too strong close to the black hole thus MRI is suppressed and there is no longer turbulence \citep[e.g.,][]{Narayan2003,McKinney2012}.
    \item However, further away from the black hole, where the magnetic field is not too strong so that MRI is present in both SANE and MAD, the magnetic field helps the production of wind. This is why the power-law index for the MAD00 model is 1.54, larger than that of the SANE00, which is 1. Similar results can be found when we compare SANE98 and MAD98.
    \item By comparing SANE00 and SANE98, we can find that the wind flux become stronger with the increase of black hole spin if the radius is not too large. This is confirmed by comparing MAD00 and MAD98.  This indicates that the black hole spin helps the production of wind. 
   
\end{itemize}

For comparison purpose, we have also used the time-averaged approach calculating the mass flux of wind for the three models. As in \citet{Yuan2015}, we find that the mass flux obtained by this approach is smaller than that obtained by the trajectory approach, and the discrepancy depends on the radius and time interval of average. For example, at 50\,$r_{\rm g}$ and 200\,$r_{\rm g}$, for the time interval we use, the former is found to be smaller than the latter by a factor of $2-4$. The difference for SANE98 is slightly larger than MAD00 and MAD98, which is likely because turbulence in MAD is weaker. 
Such a discrepancy is smaller than that found in \citet{Yuan2015}. We speculate that this is because  different simulation codes and different time interval are used.

Note that, however, the values of the power-law index of eqs. \ref{eq:im1}--\ref{eq:im4} are smaller than those of eqs. \ref{eq:m1}--\ref{eq:m4}. This implies that beyond a certain radius, which are $214\,r_{\rm g}$, $2\times 10^{6}r_{\rm g}$, $164\,r_{\rm g}$, and $210\,r_{\rm g}$ for SANE00, SANE98, MAD00, and MAD98 respectively, the mass flux of wind described by eqs. \ref{eq:m1}--\ref{eq:m4} would be larger than those of the inflow rate, which is unphysical. Therefore, eqs. \ref{eq:m1}--\ref{eq:m4} are valid only within these radii\footnote{The outer boundary of a hot accretion flow, i.e, the Bondi radius, is usually about $10^5\,r_{\rm g}$ for typical temperature of the gas in the galactic center region. So eq. \ref{eq:m2} is applicable only within the Bondi radius.}. Beyond these radii, the mass fluxes of wind in the four models are roughly equal to the inflow rates, which are described by eqs. \ref{eq:im1}-\ref{eq:im4}.

Now let us discuss the mass flux of jet.  From Figure~\ref{fig:mdotw}, we can see that for SANE98, the jet is produced at small radii and gradually increases until it saturates at $r \approx 20\,r_{\rm g}$.  For MAD98, the jet mass flux increases faster, and saturates at a larger radius of $r \approx 100\, r_{\rm g}$. The saturated mass fluxes of the BZ-jet in SANE98 and MAD98 are
\begin{equation}
    \dot{M}_{\rm jet}\approx 0.5\dot{M}_{\rm BH},
    \label{sane98jetmassflux}
\end{equation}
\begin{equation}
    \dot{M}_{\rm jet}\approx 0.8\dot{M}_{\rm BH},
    \label{mad98jetmassflux}
\end{equation}
respectively. Again, $\dot{M}_{\rm BH}$ denotes the mass accretion rate at the black hole horizon. The mass flux of the jet in MAD98 is larger than that of the SANE98, implying that magnetic field helps the formation of jet. 

Figure~\ref{fig:mdotall} shows the mass flux of wind per unit $\theta$ integrated over all $\varphi$ and averaged from $t=55{,}000\text{--}68{,}000$, $10{,}000\text{--}20{,}000$ and $8000\text{--}23{,}000$ for SANE98, MAD00 and MAD98, respectively.  The positive value denotes real outflow while negative value denotes the inflow. The figure shows that for most cases,  the mass flux of inflow is concentrated on the equatorial plane. But at $r=160\,r_{\rm g}$ of the SANE98 model, the inflow flux has some bimodal distribution, the accretion is mainly via the coronal region. Such a result has been discussed and termed ``coronal accretion'' in \citet{Zhu2018}, and is also similar to the accretion model of Quadrupole Topology QDPa in \citet{Beckwith2008}. 

\begin{figure*}
\gridline{\fig{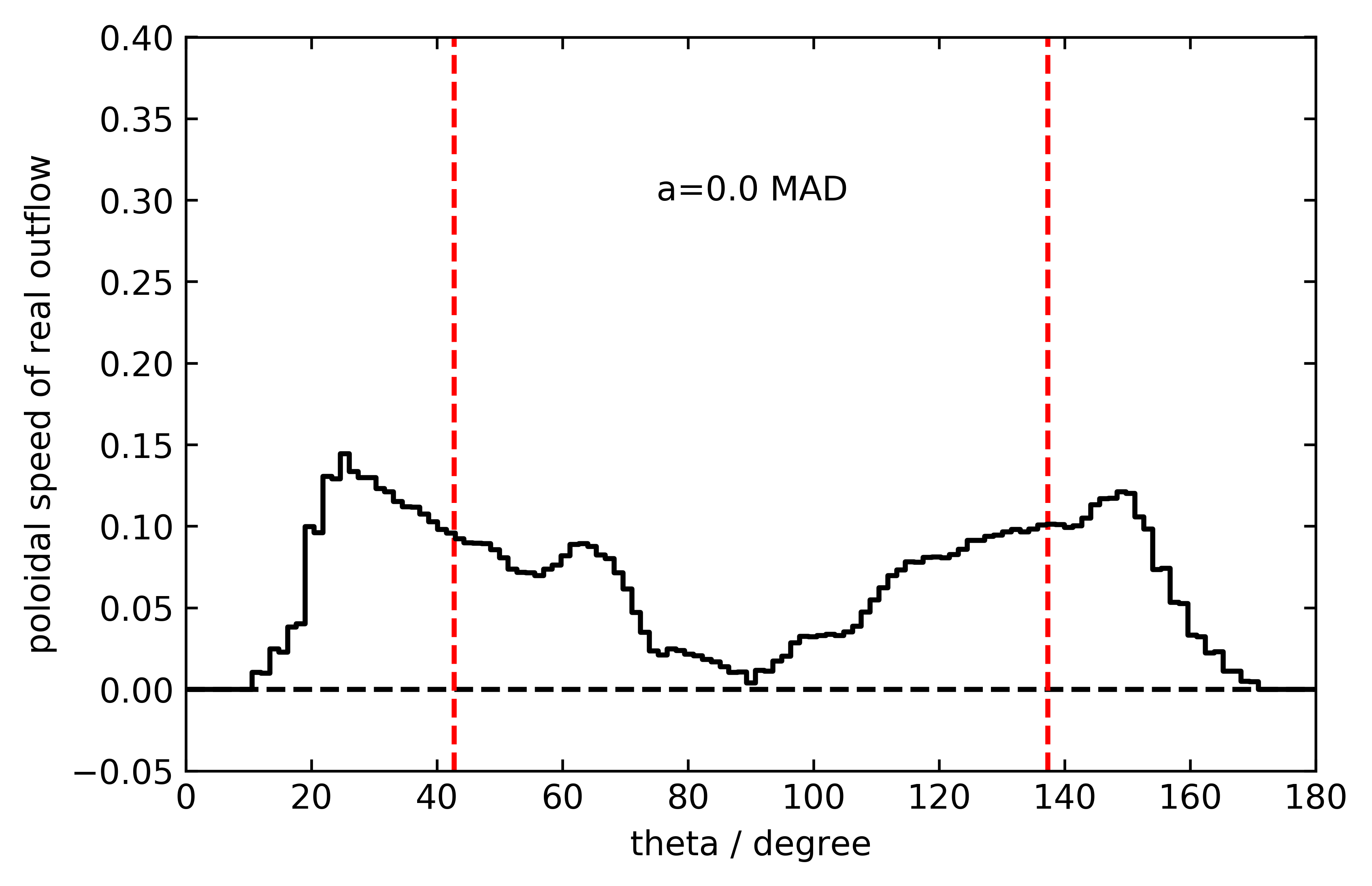}{0.33\textwidth}{}
\fig{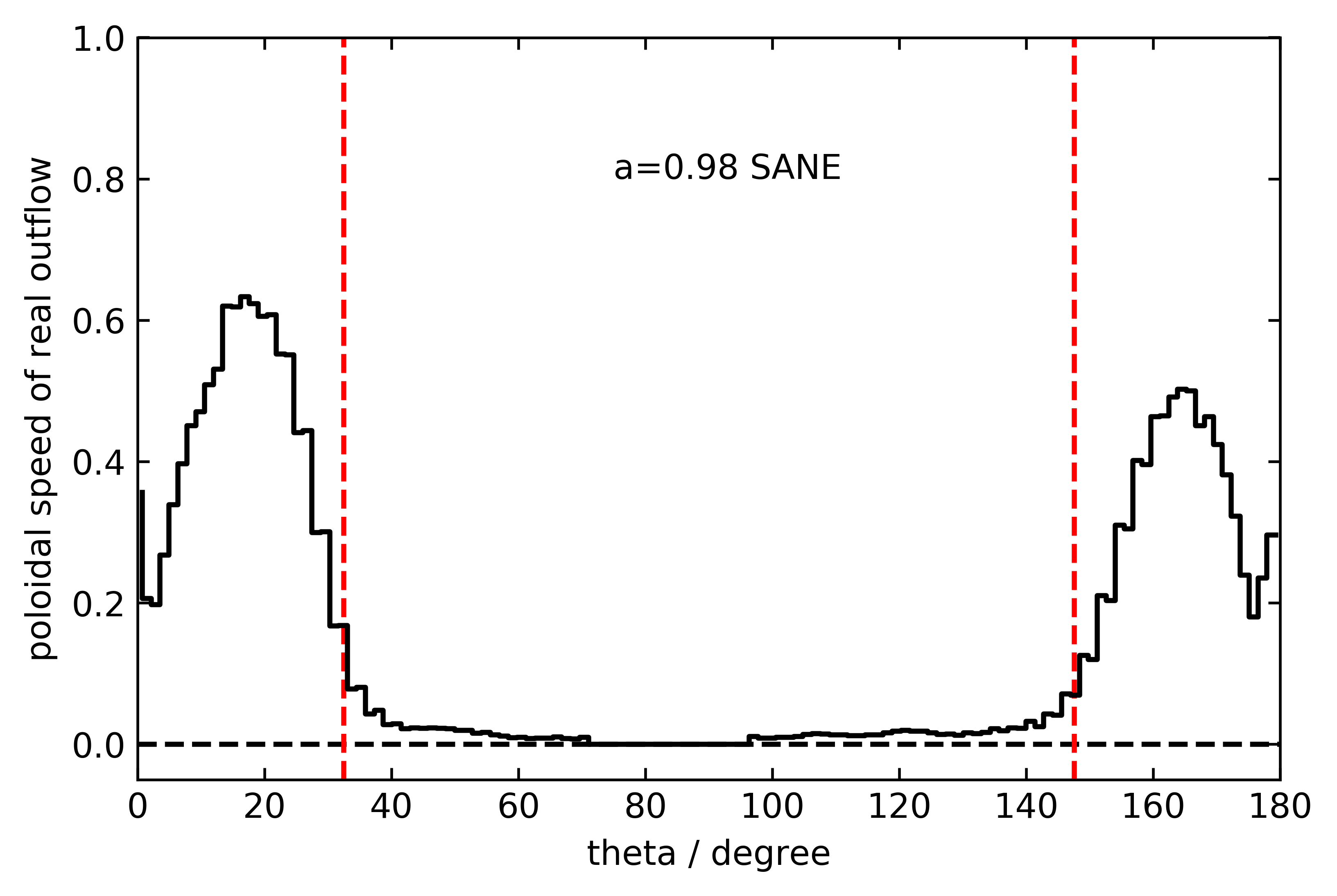}{0.33\textwidth}{}
\fig{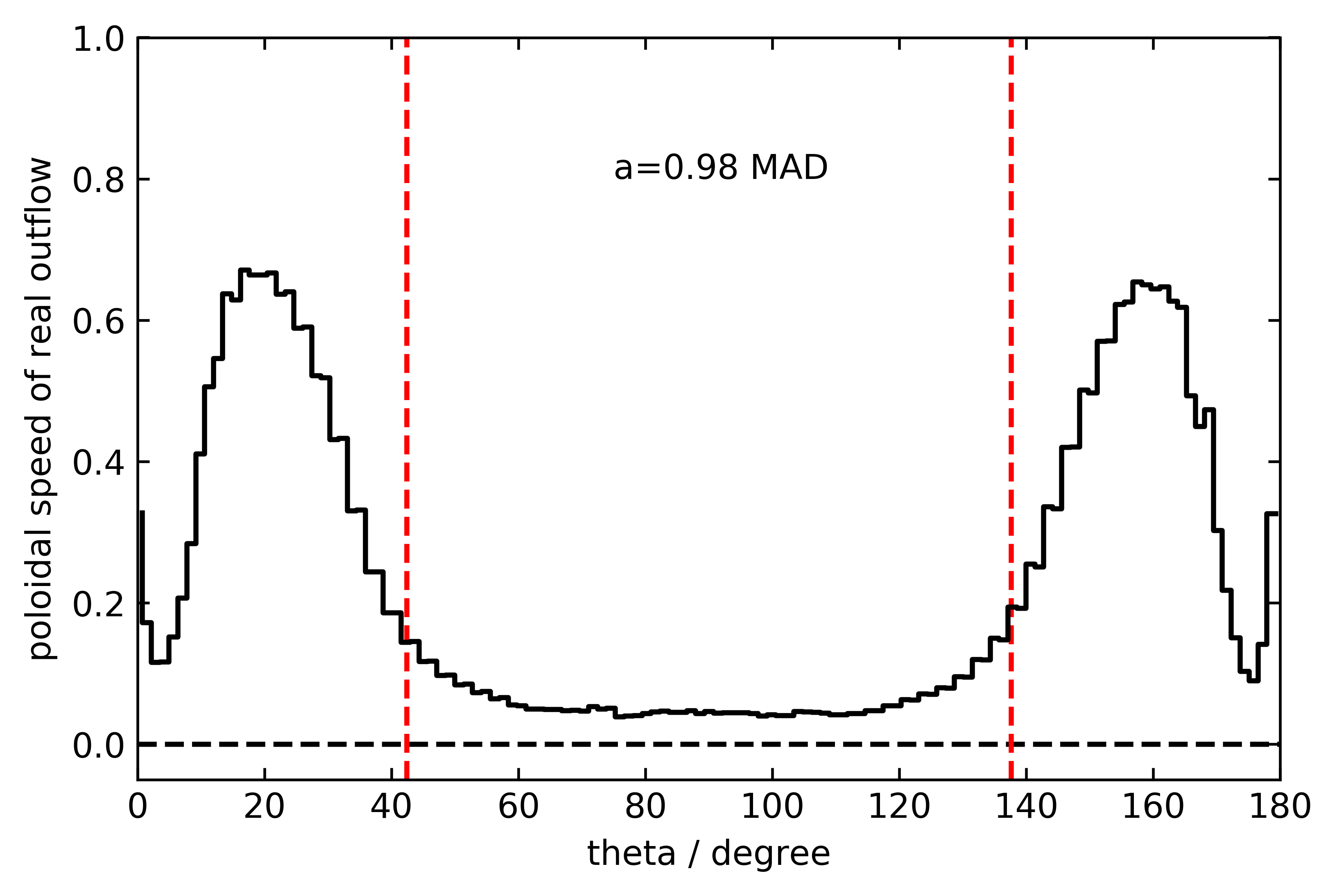}{0.33\textwidth}{}
}
\gridline{\fig
{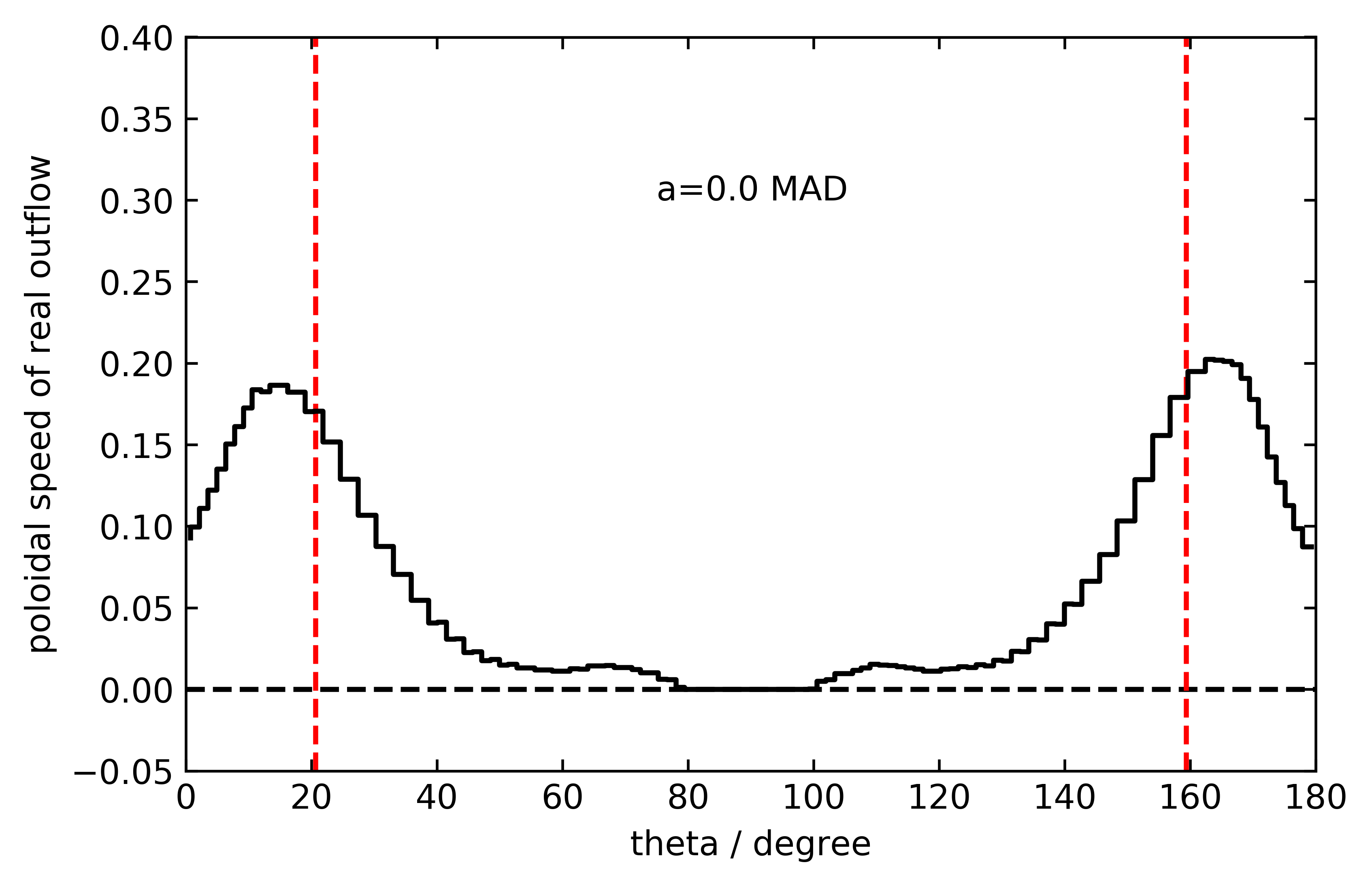}{0.33\textwidth}{}
\fig{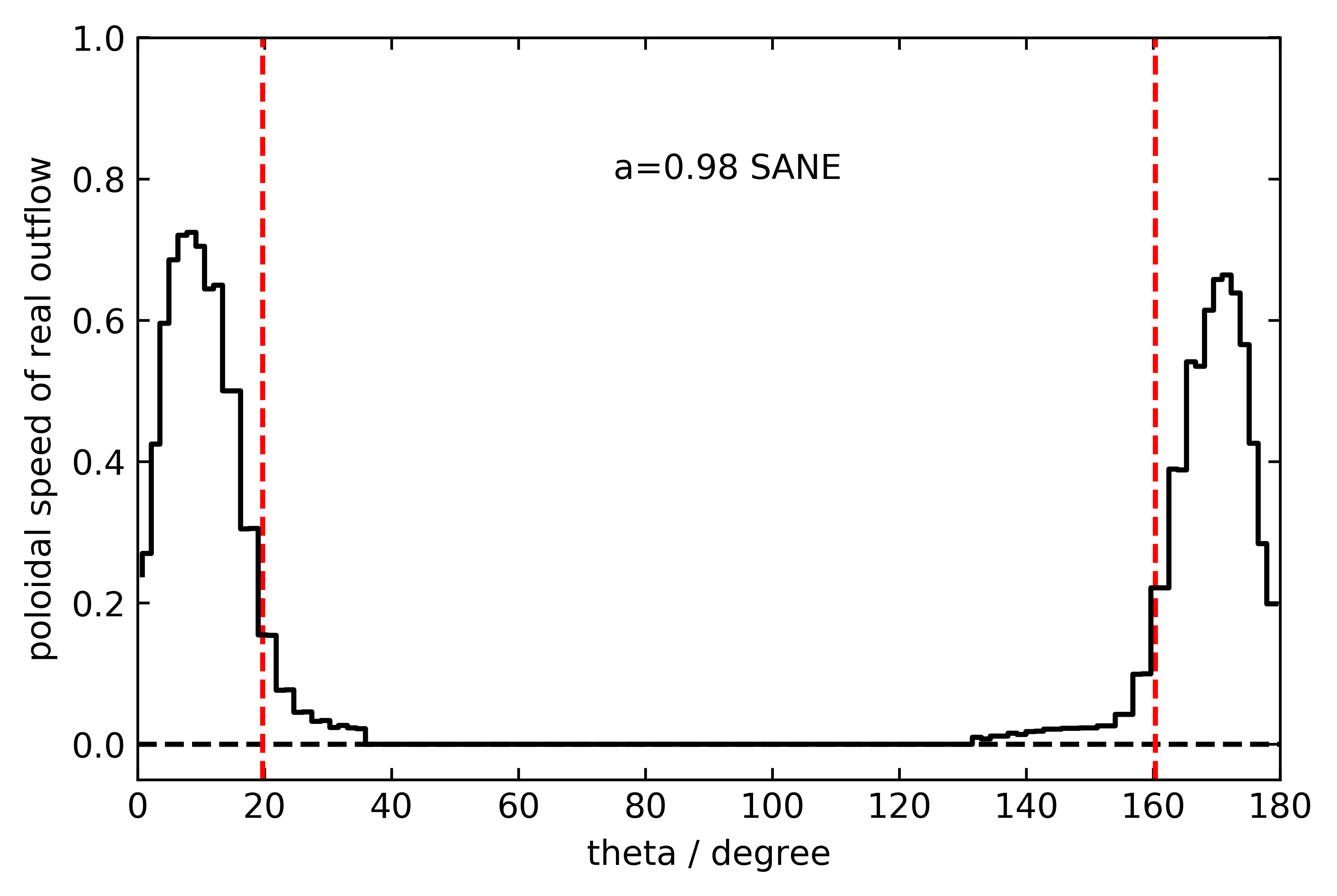}{0.33\textwidth}{}
\fig{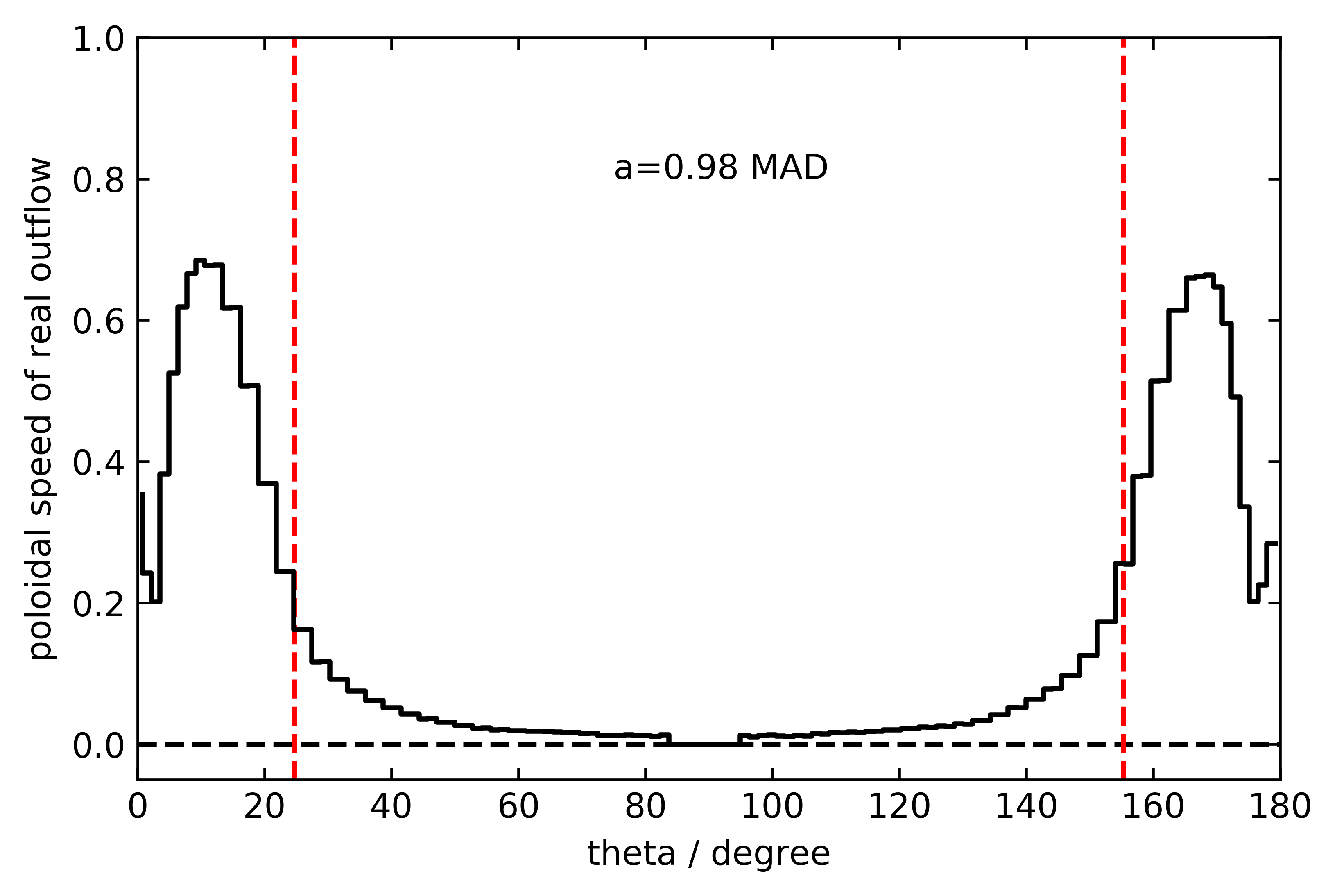}{0.33\textwidth}{}
}
\caption{The poloidal speed of wind and jet  in unit of speed of light averaged over $\varphi$ and time from $10{,}000\text{--}20{,}000$, $55{,}000\text{--}68{,}000$ and $8000\text{--}23{,}000$ for MAD00 (left), SANE98 (middle) and MAD98 (right), respectively. The top and bottom panels are for $r=40\,r_{\rm g}$ and $r=160\,r_{\rm g}$. The red dashed line represents the boundary between jet and wind. }
\label{fig:pv}
\end{figure*}

\subsection{The poloidal speed of wind and jet}
\label{windvelocity}

\begin{figure*}[t]

\gridline{\fig{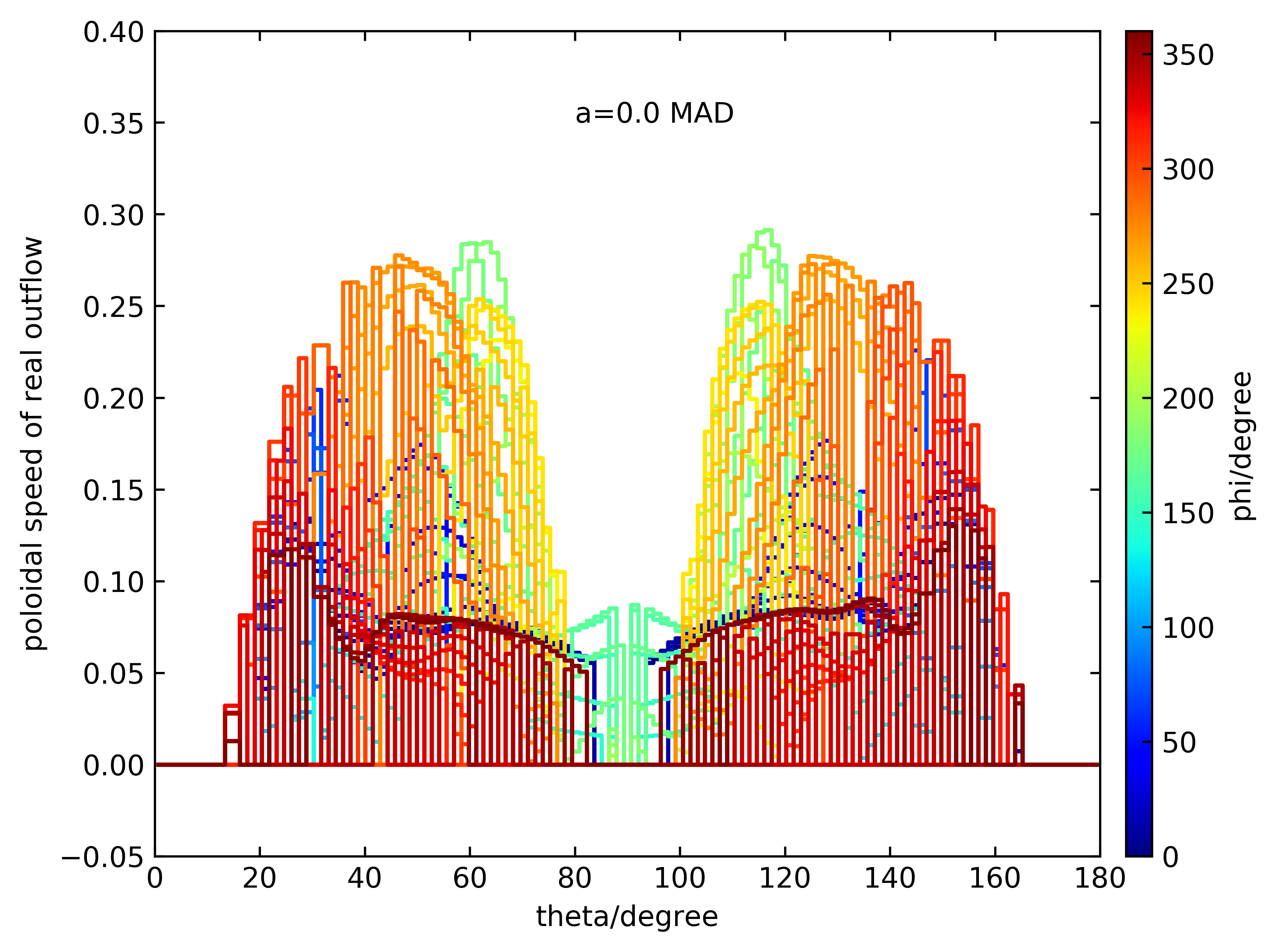}{0.33\textwidth}{}
\fig{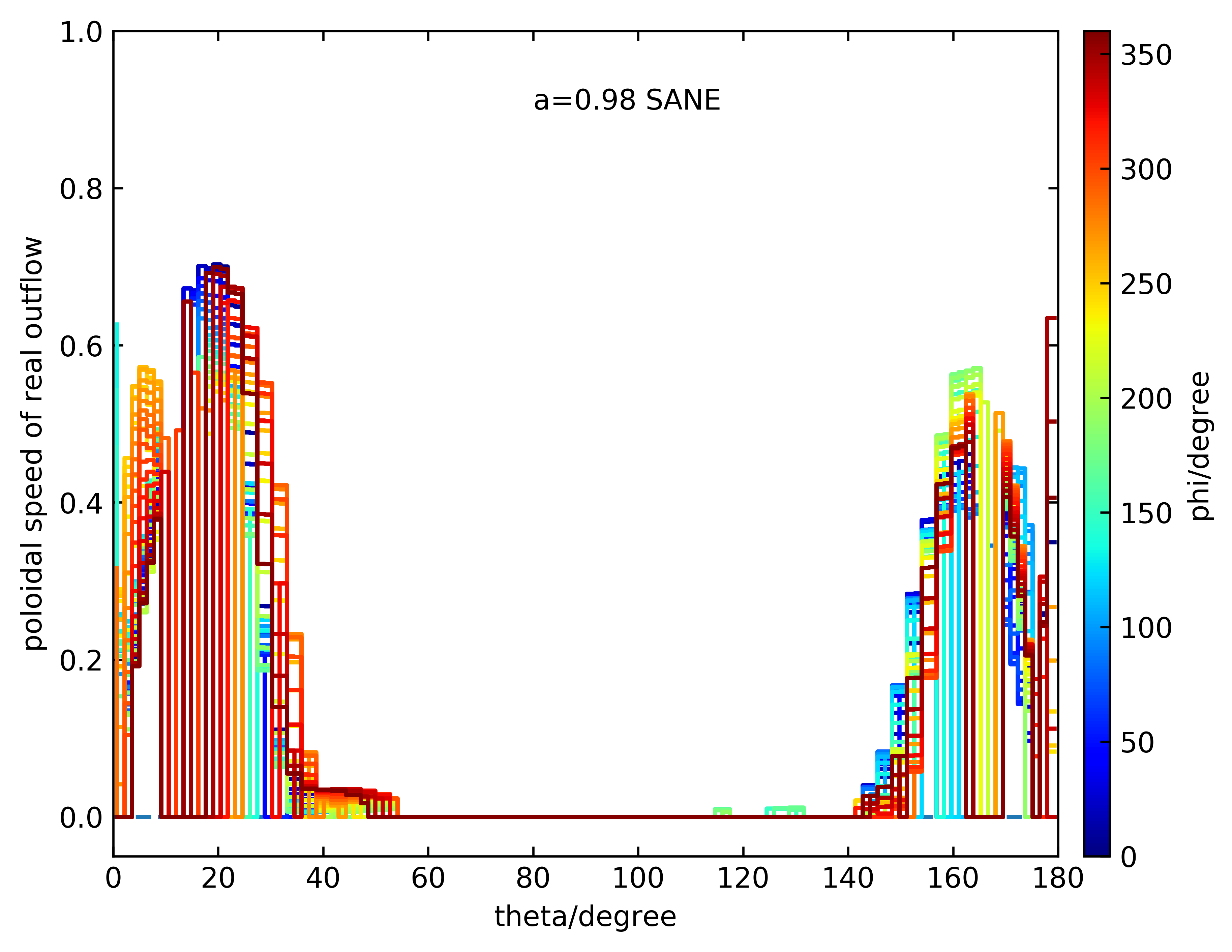}{0.33\textwidth}{}
\fig{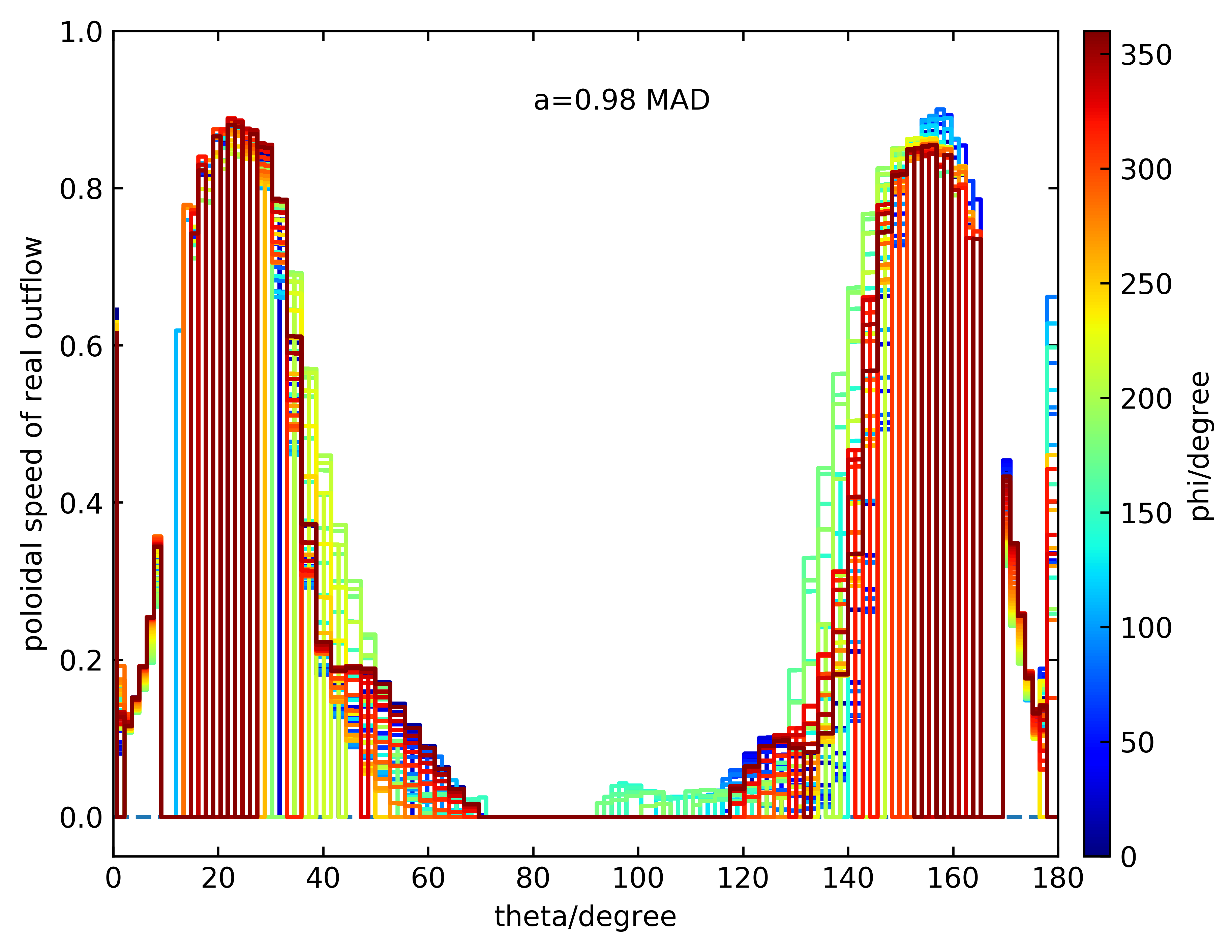}{0.33\textwidth}{}
}
\gridline{\fig
{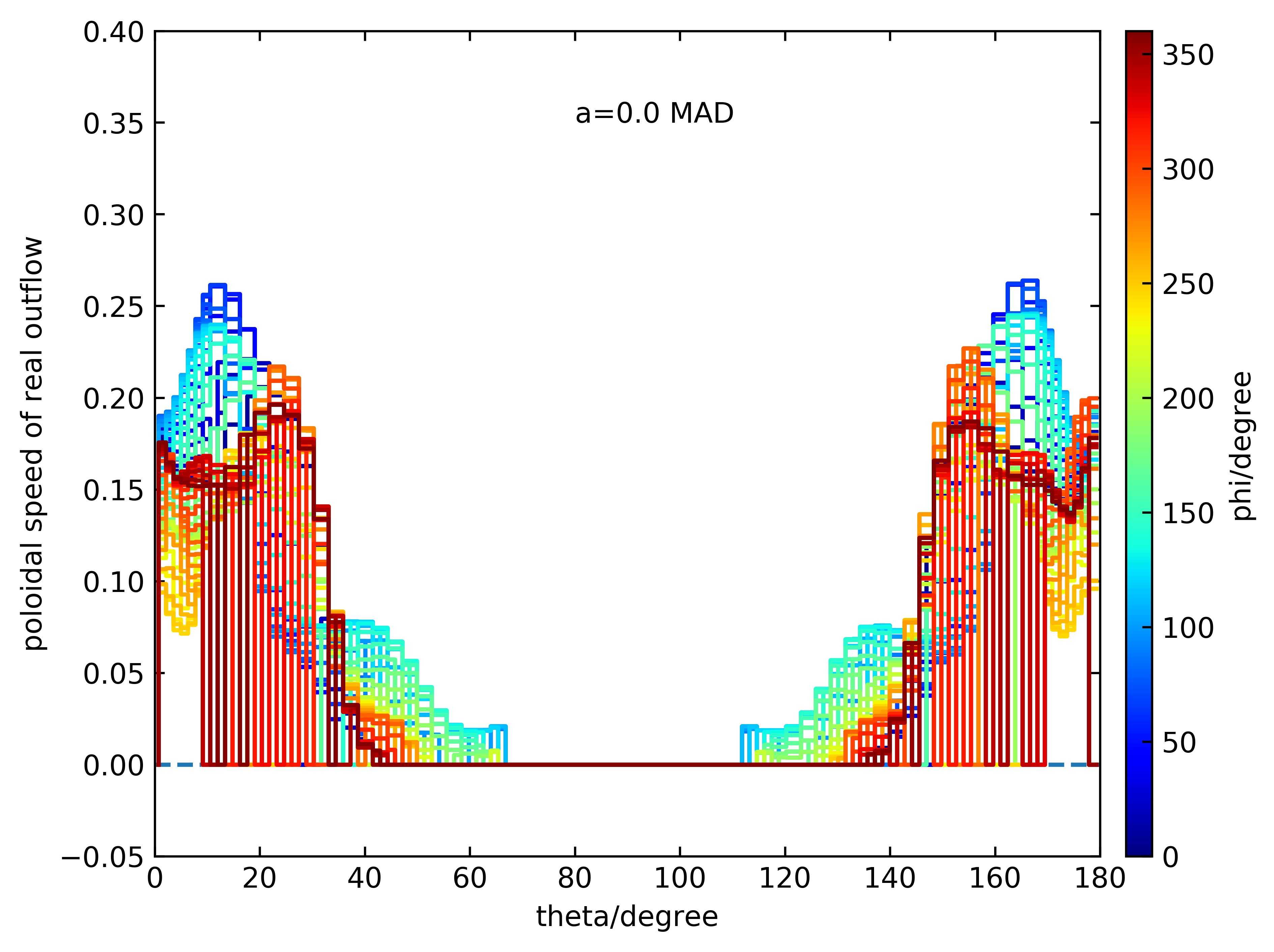}{0.33\textwidth}{}
\fig{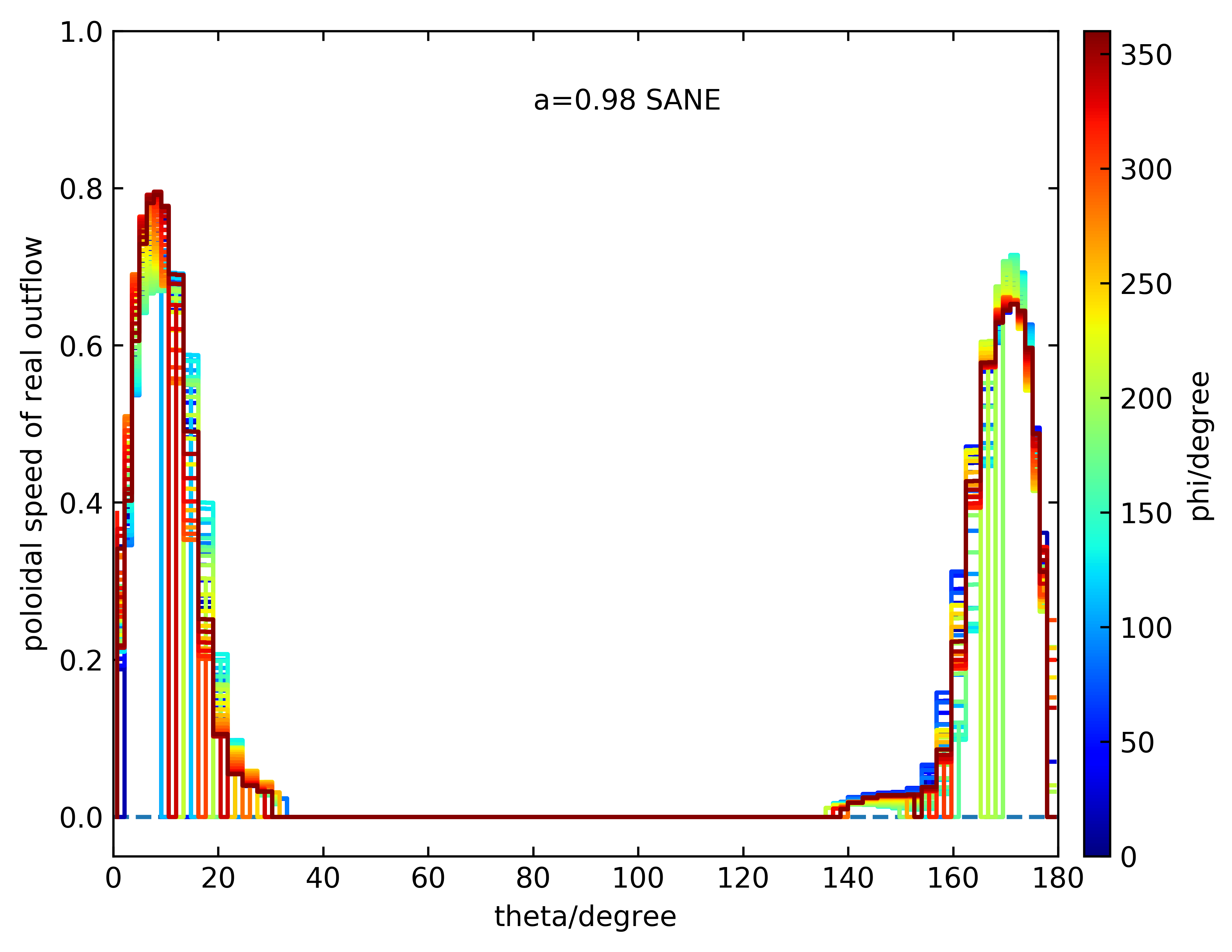}{0.33\textwidth}{}
\fig{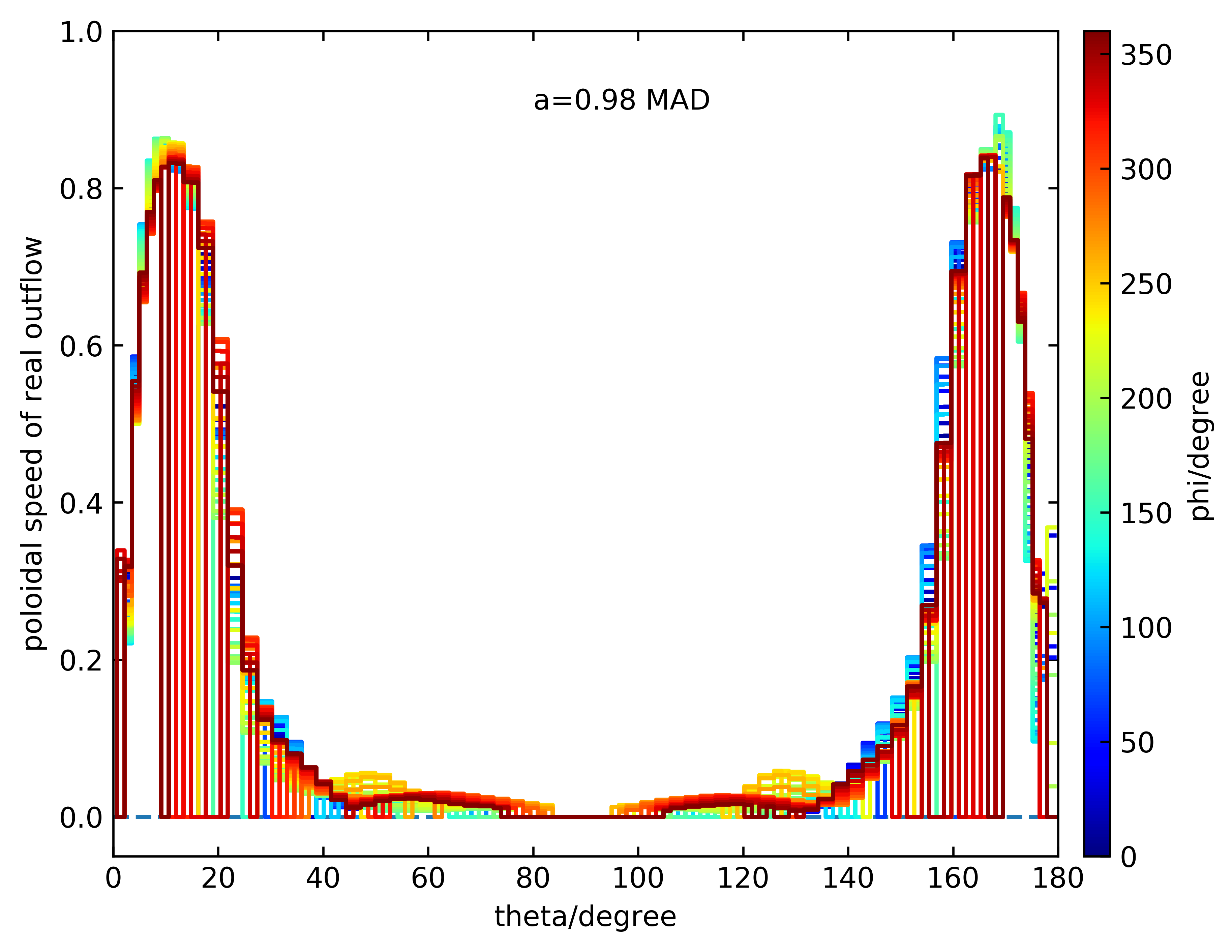}{0.33\textwidth}{}
}
\caption{The poloidal speed of wind and jet as a function of $\theta$ at a given time of for MAD00 (left; $t=19{,}000$), SANE98 (middle; $t=61{,}500$), and MAD98 (right; $t=15{,}500$) respectively. Different color denotes different $\varphi$. The top and bottom panels are for 40\,$r_{\rm g}$ and 160\,$r_{\rm g}$. }
\label{fig:pvt}
\end{figure*}

In this section, we discuss the poloidal speed of wind and jet. This is the dominant component of the outflow velocity at large radii where the magnetic field becomes weak \citep[e.g.,][]{Cui2020a,Cui2020b}. Figure~\ref{fig:pv} shows the poloidal speed of wind and jet averaged over $\varphi$ and time from $55{,}000$ to $68{,}000$, $10{,}000$ to $20{,}000$ and $8000$ to $23{,}000$ for SANE98, MAD00, and MAD98 respectively. 
We can see the same with the SANE00 model presented in \citet{Yuan2015},  for the three models the velocity is higher in the polar direction but smaller closer to the equatorial plane. By comparing MAD98 and SANE98, we can see that  the largest  velocity of the jet of the two models are similar. However, the velocity shown in this figure is time-averaged. We find that the instantaneous velocity of the two models are quiet different when examining the snapshot result. The velocity  of jet in SANE98 varies little with time, while that in MAD98 varies greatly. This means that the velocity fluctuation in MAD98 is stronger than SANE98. For illustration purposes, Figure~\ref{fig:pvt} shows the poloidal speed of wind and jet at a certain time and radius as functions of $\theta$ and $\varphi$. For MAD00, SANE98, and MAD98, the chosen time is $t=19{,}000$, $61{,}500$, and $15{,}500$ respectively. We can see that the peak of the poloidal speed of the real outflow is about $0.27\,c, 0.8\,c, 0.88\,c$ for MAD00, SANE98, and MAD98, respectively. 

\begin{figure}[t]
	\includegraphics[width=\columnwidth]{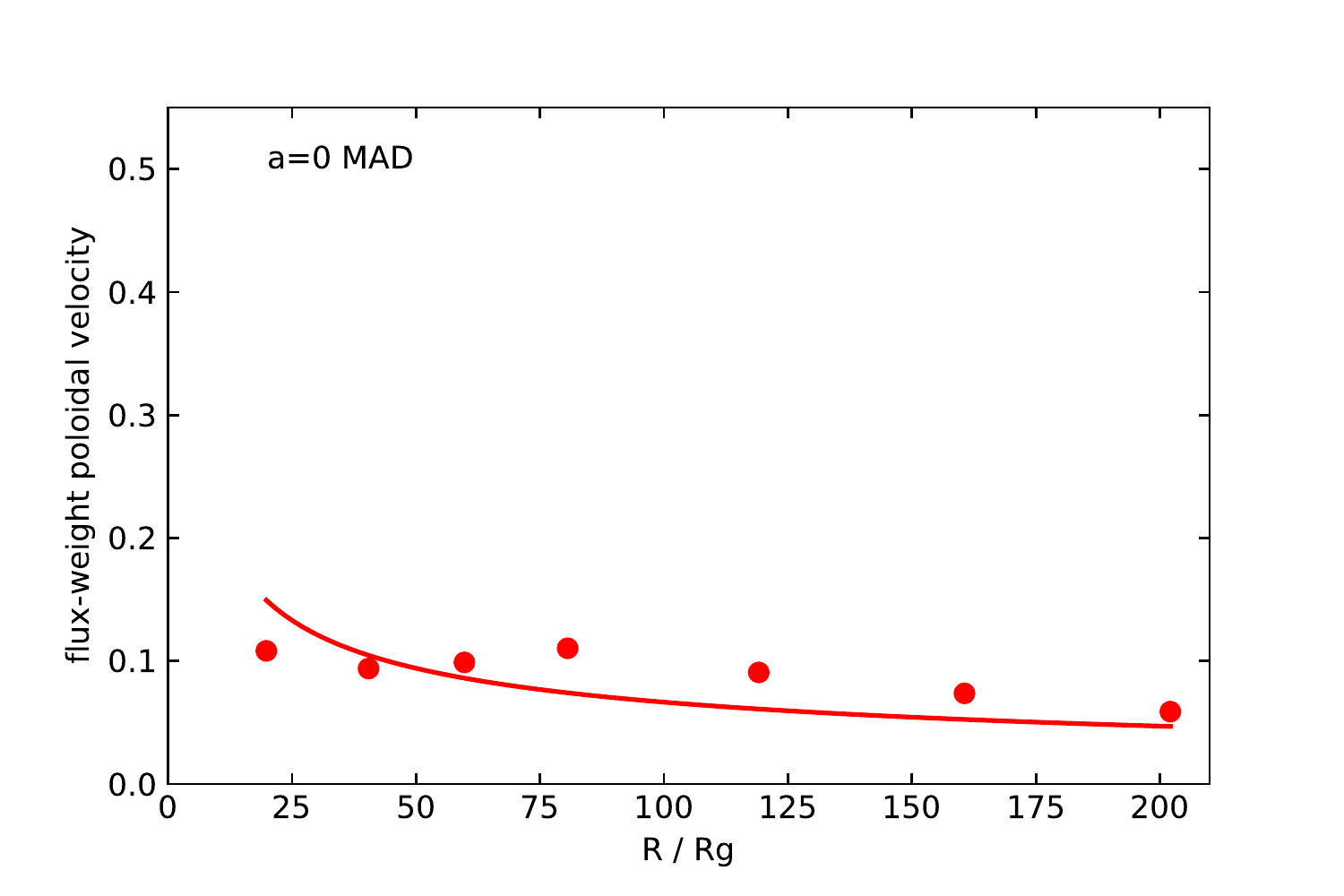}
    \includegraphics[width=\columnwidth]{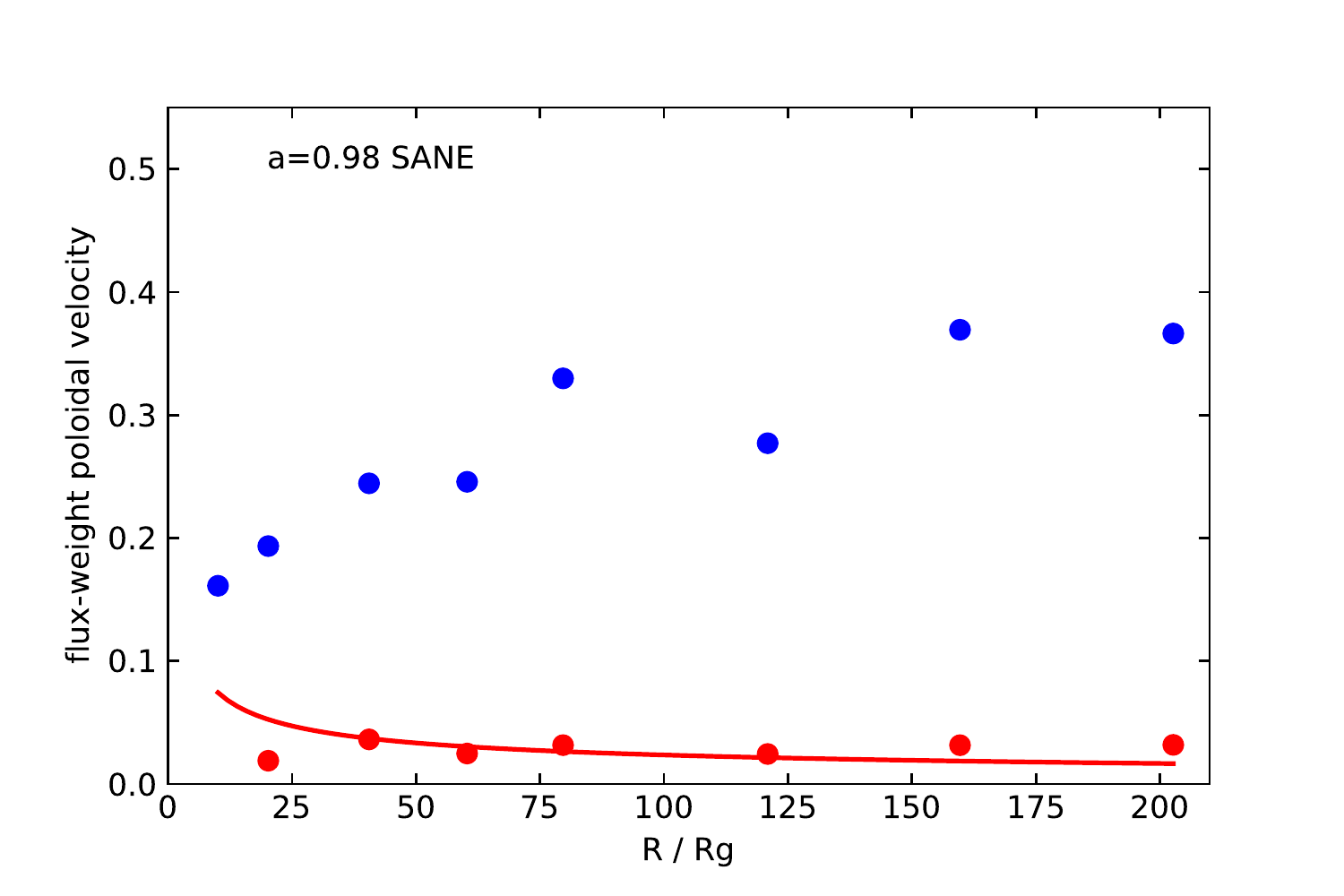}
	\includegraphics[width=\columnwidth]{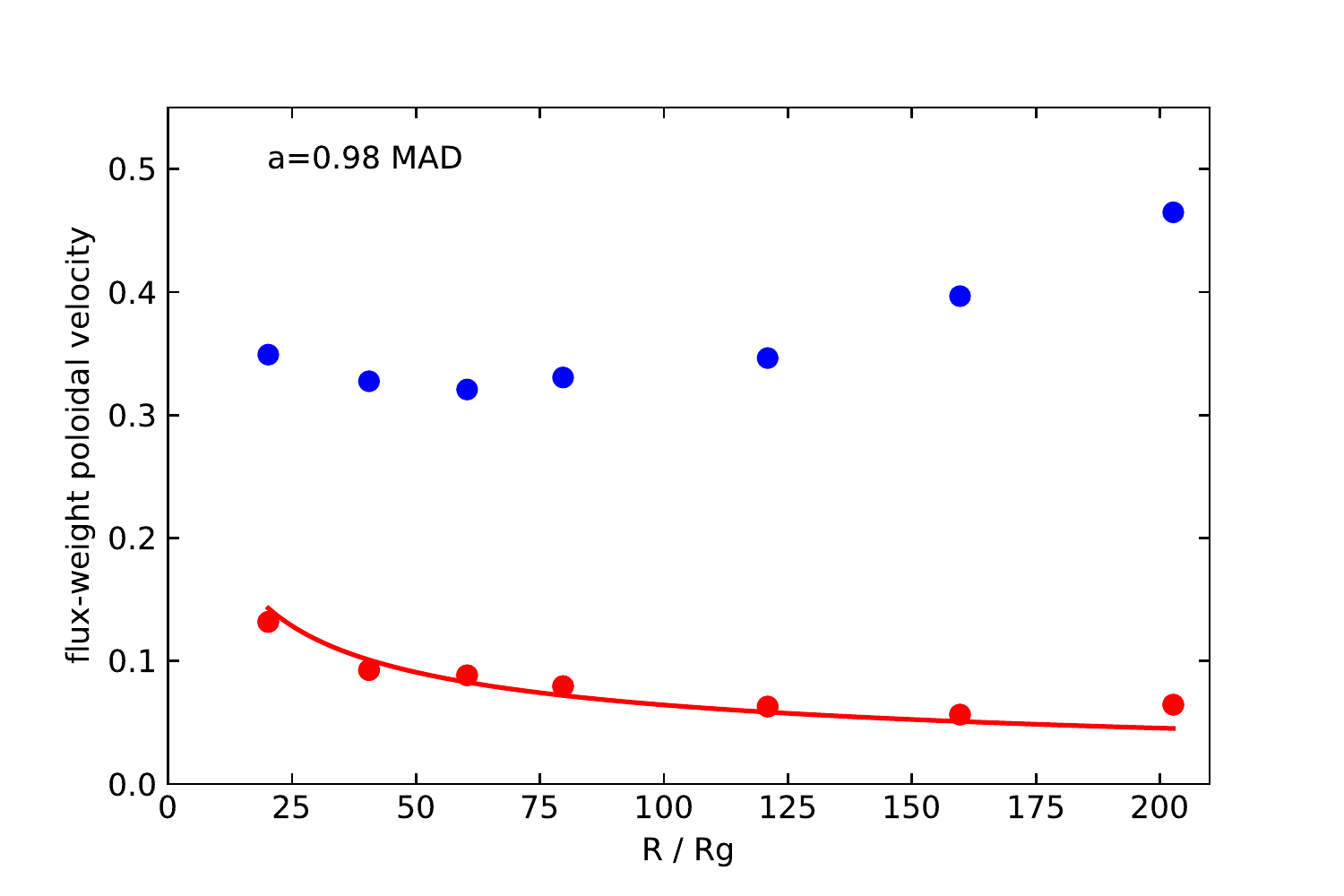}
    \caption{Radial profiles of mass-flux-weighted and time averaged poloidal velocity of wind (red dots) and jet (blue dots) for  MAD00, SANE98, and MAD98.  }
    \label{fig:pvk}
\end{figure}

 Figure~\ref{fig:pvk} shows the mass-flux-weighted poloidal velocity of wind and jet as a function of radius.  The blue and red dots are for jet and wind respectively, while the red line is the fitting curve for the wind. The poloidal velocity of wind, combined with the result for SANE00 taken from \citet{Yuan2015}, are described by: 
\begin{equation}
    v_{\rm p}(r )=0.21v_{\rm k}(r),
	\label{eq:v0}
\end{equation}
\begin{equation}
    v_{\rm p}(r )=0.66v_{\rm k}(r),
	\label{eq:v1}
\end{equation}
\begin{equation}
    v_{\rm p}(r)=0.24v_{\rm k}(r),
	\label{eq:v2}
\end{equation}
\begin{equation}
    v_{\rm p}(r)=0.64v_{\rm k}(r)
	\label{eq:v3}
\end{equation}
for SANE00, MAD00, SANE98 and MAD98, respectively. Here $v_{\rm k}(r)$ is the Keplerian speed at radius $r$. We can find that the poloidal speed of wind of MAD models is larger than that of SANE models. This is also the case for jet. As we can see from the figure, the mass-flux-weighted velocities of jet at $200\, r_{\rm g}$ are
\begin{equation}
    v_{\rm p,jet}\approx 0.4~c,
	\label{eq:sanejetv}
\end{equation}
\begin{equation}
    v_{\rm p,jet}\approx 0.5~c,
	\label{eq:madjetv}
\end{equation}
for SANE98 and MAD98, respectively. This is because Lorentz force is one of the main acceleration forces of wind \citep{Yuan2015}\footnote{In this sense, the acceleration mechanism is similar to the \citet{BP82}.}. In other words, the Poynting flux of outflow is converted into the kinetic energy flux, as we will discuss in \S\ref{sec:BZ-jet}. Thus when the magnetic field is stronger, as in the MAD model, the poloidal speed will be larger.  Another result we can find is that the black hole spin has little effect on the velocity of wind, although the black hole spin can strongly strengthen the BZ-jet. This is because, unlike the jet, wind is produced at relatively large radii where the effect of general relativity is weak. 

In AGN feedback study, a useful quantity is the power of wind. The mass flux weighted poloidal velocity $v_p$ is not suitable for this aim; we should instead use the mass-flux-weighted $v_p^2$. We find that they are described by,
\begin{equation}
    v_{\rm p}^2(r)=0.053v_{\rm k}^2(r),
	\label{eq:v02}
\end{equation}
\begin{equation}
    v_{\rm p}^2(r )=0.57v_{\rm k}^2(r),
	\label{eq:v12}
\end{equation}
\begin{equation}
    v_{\rm p}^2(r)=0.11v_{\rm k}^2(r),
	\label{eq:v22}
\end{equation}
\begin{equation}
    v_{\rm p}^2(r)=0.58v_{\rm k}^2(r)
	\label{eq:v32}
\end{equation}
for SANE00, MAD00, SANE98 and MAD98, respectively.

The mass-flux-weighted poloidal speed of wind decreases with increasing radius. But we note that this does not mean that if we follow the trajectory of a given ``wind particle'', its polodal speed will decrease when propagating outward. The study in \citet{Yuan2015} has indicated that the poloidal speed actually increases outward or at least keeps constant. This is because of the acceleration due to the gradient of magnetic and thermal pressure of the wind. They find that the speed of wind launched from radius $r$ can be described by $v_{\rm p}(r)\approx (0.2-0.4) v_{\rm k}(r)$ for SANE00 \citep{Yuan2015}. The outward decrease of the mass-flux-weighted wind speed shown in eqs. \ref{eq:v1}-\ref{eq:v3} is because, when the wind propagates outward, more and more wind material whose velocity is becoming smaller will join in. In fact, we can see in Figure~\ref{fig:mdotw} that the wind mass flux increases rapidly with radius. Therefore, the mass-flux-weighted wind velocity $v(r)$ mainly reflects wind launched close to radius $r$.

Unlike the wind, the blue dots of Figure~\ref{fig:pvk} show that the mass-flux-weighted poloidal velocity of the jet increases with radius. On the one hand, this is because of the strong acceleration of jet matter during its outward motion. On the other hand, unlike the case of wind, there is little low-speed matter joining into the jet when it propagates outward.

\subsection{The energy and momentum fluxes of wind and jet}
\label{momentumflux}

\begin{figure}[t]
	\includegraphics[width=\columnwidth]{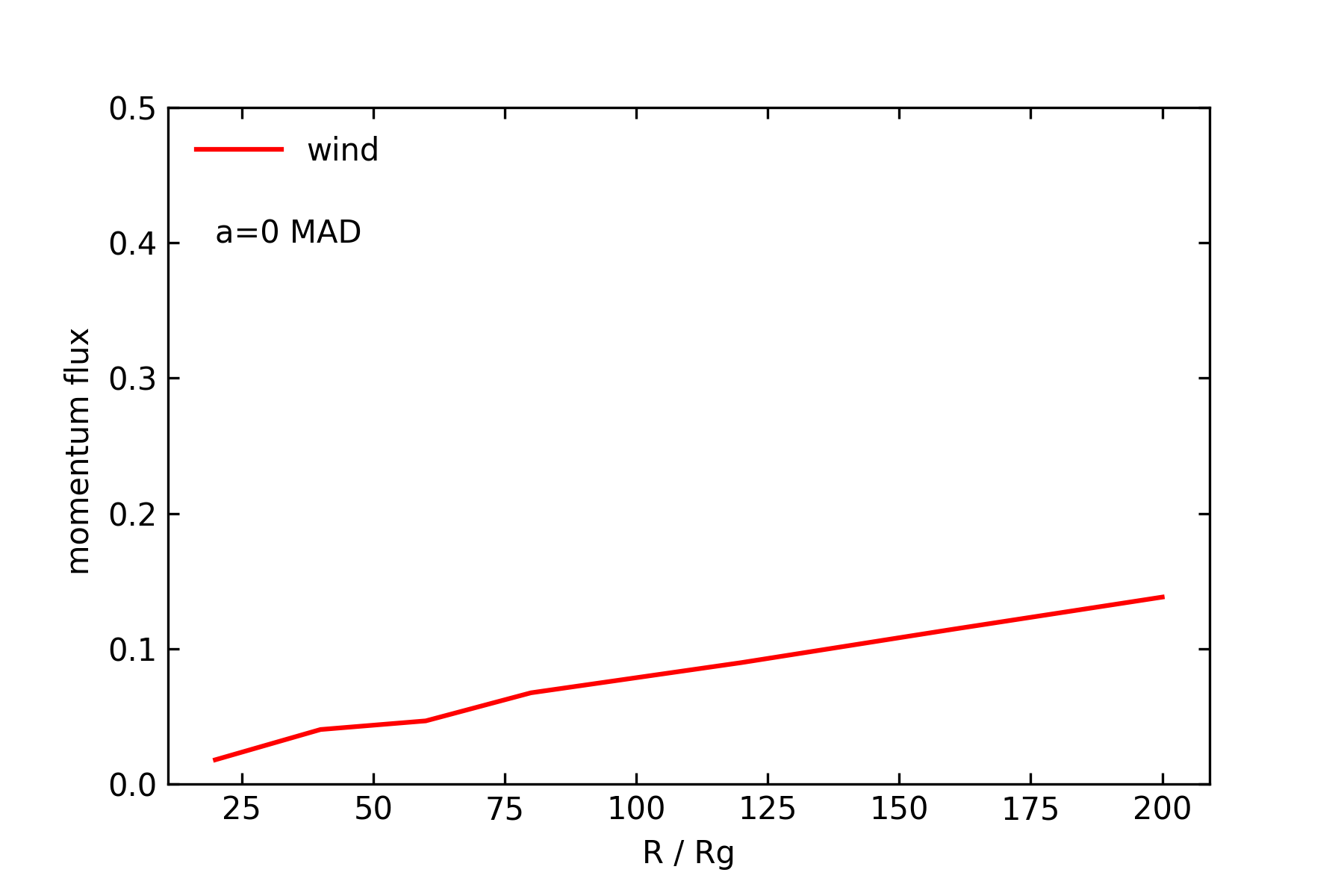}
    \includegraphics[width=\columnwidth]{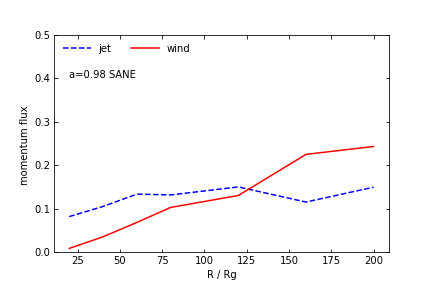}
	\includegraphics[width=\columnwidth]{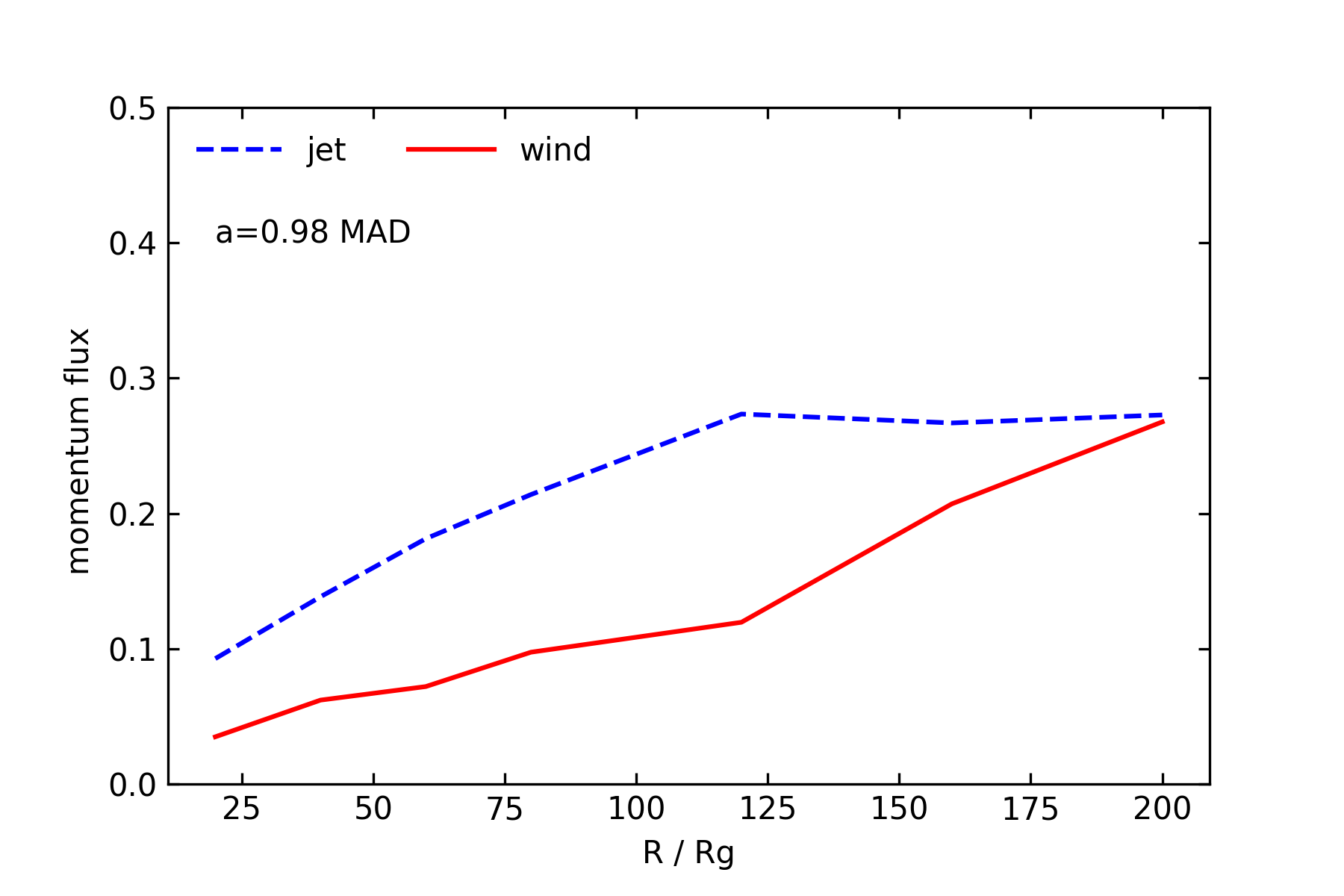}
    \caption{The radial profiles of the momentum fluxes of  wind and jet. The values have been normalized by $\dot{M}(r=2\, r_{\rm g})c$ of MAD00, SANE98 and MAD98 respectively.}
    \label{fig:mn}
\end{figure}

\begin{figure}[t]
	\includegraphics[width=\columnwidth]{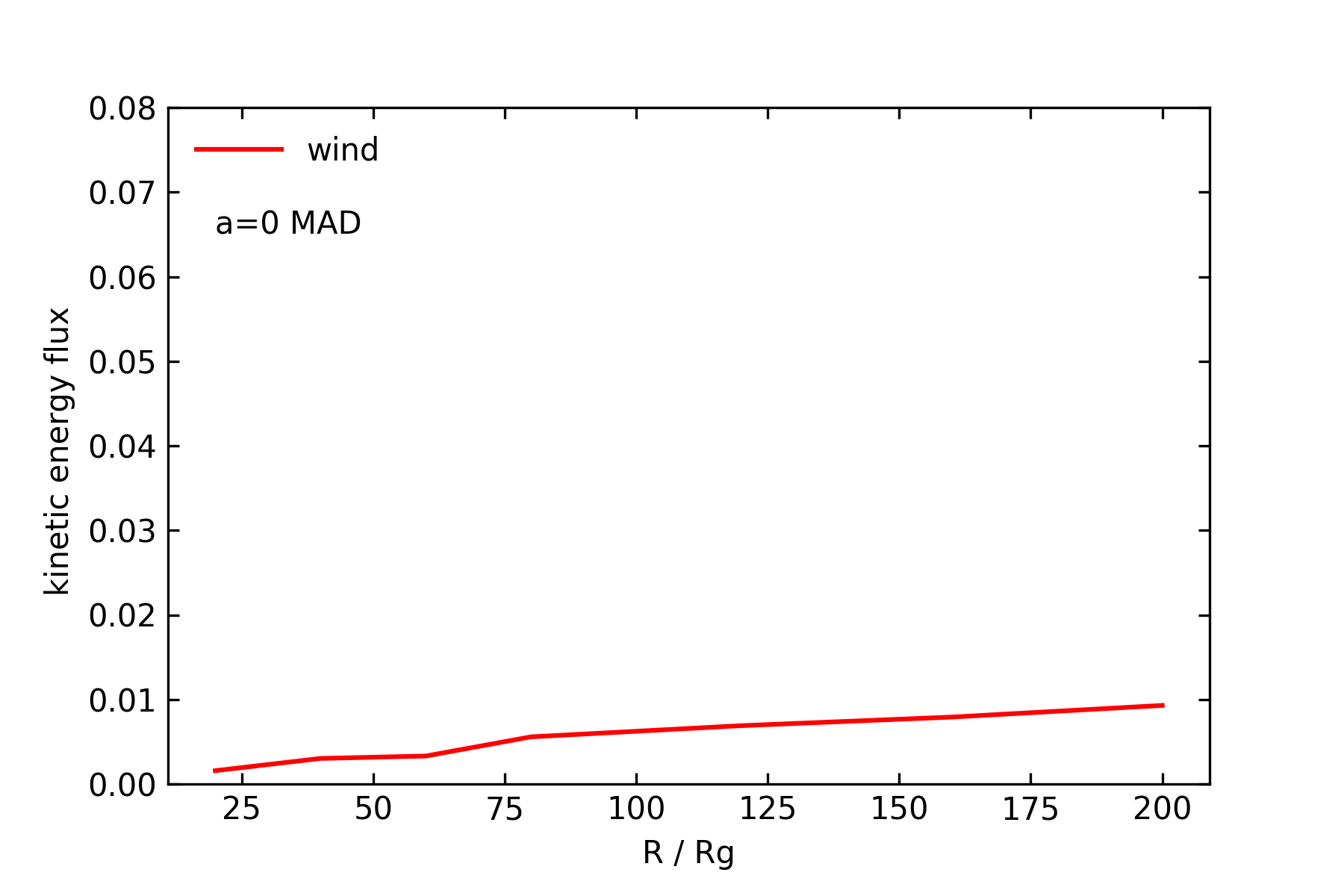}
    \includegraphics[width=\columnwidth]{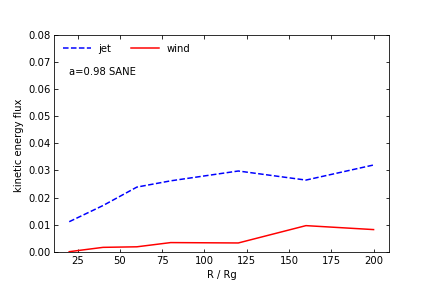}
	\includegraphics[width=\columnwidth]{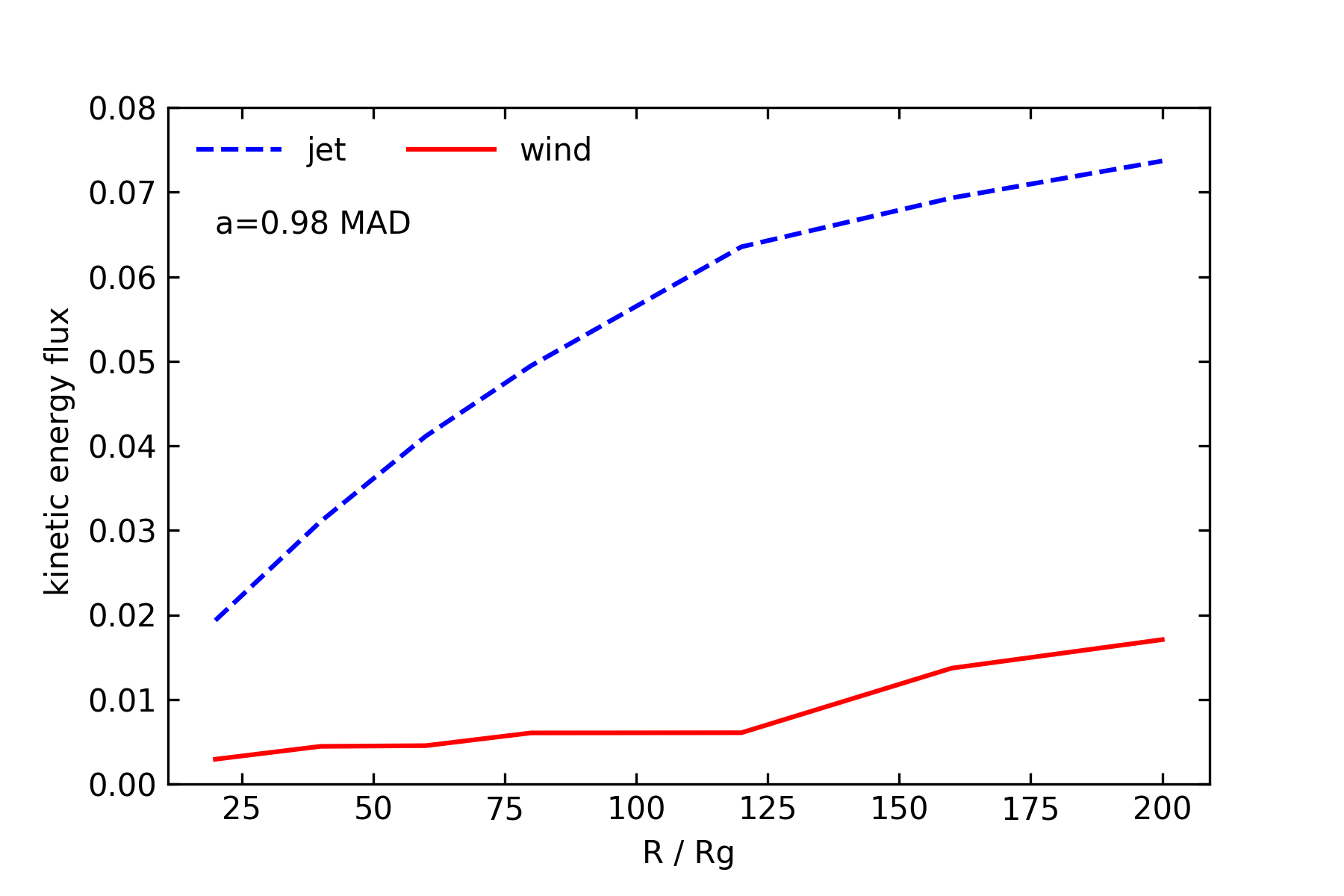}
    \caption{The radial profiles of the kinetic energy fluxes of wind and jet. Those values have been normalized by $\dot{M}(r=2\, r_{\rm g})c^2$ of the three models respectively. }
    \label{fig:ke}
\end{figure}

The kinetic energy and momentum fluxes of wind and jet due to their poloidal velocity (dominated by radial velocity) are calculated by the following equations,
\begin{equation}
\begin{split}
    \dot{E}_{\rm jet(wind)}(r)=\frac{1}{2}\int{\gamma}\rho(r, {\theta}, \varphi)v^{3}_{r}(r, \theta, \varphi)\\(r^2+a^2\cos^2\theta)\sin\theta d\theta d\varphi,
\end{split}
	\label{eq:en}
\end{equation}

\begin{equation}
\begin{split}
    \dot{P}_{\rm jet(wind)}(r)=\int{\gamma}\rho(r, {\theta}, \varphi)v^{2}_{r}(r, \theta, \varphi)\\ (r^2+a^2\cos^2\theta)\sin\theta d\theta d\varphi,
\end{split}
	\label{eq:mn}
\end{equation}
where $\gamma={1}/{\sqrt{1-v_r^2}}$ and $v_r$ is the radial 3-velocity component in the locally non-rotating frame defined in the appendix. Figures~\ref{fig:mn} \& \ref{fig:ke} show the radial profiles of the momentum and kinetic energy fluxes of wind and jet. From these results, combined with the results of SANE00 shown in \citet{Yuan2015}, we find that both momentum and kinetic energy fluxes of wind increase with the black hole spin. This is because the mass flux of outflow increases with spin (\S\ref{massflux}) although the poloidal velocity of wind is almost independent of spin (\S\ref{windvelocity}).  As for the effect of magnetic field, we find that, at $200\, r_{\rm g}$, the momentum flux of outflow (i.e., wind and jet) of MAD00/MAD98 is slightly larger than that of SANE00/SANE98, while the kinetic energy flux of MAD00/MAD98 is much larger than that of SANE00/SANE98. This is because, at $200\,r_{\rm g}$ the mass flux of wind of MAD00/MAD98 is less than that of SANE00/SANE98 (eqs. \ref{eq:m1}-\ref{eq:m4} in \S\ref{massflux}) by about a factor of 1.5, while the velocity of the wind of MAD00/MAD98 is greater than SANE00/SANE98 by a factor of about 3 (eqs. \ref{eq:v0}-\ref{eq:v3} in \S\ref{windvelocity}). For jet, we can easily understand the results by combining eqs. \ref{sane98jetmassflux}, \ref{mad98jetmassflux}, \ref{eq:sanejetv}, \& \ref{eq:madjetv}.

When the black hole spin is large, both BZ-jet and wind will be present. It is interesting in this case to compare the momentum and energy fluxes of jet and wind. Such information is needed when we compare the roles of wind and jet in AGN feedback. The results can be obtained from Figures \ref{fig:mn} \& \ref{fig:ke}. For SANE98, the comparison  of momentum and kinetic energy fluxes between jet and wind at $200\,r_{\rm g}$ are,
\begin{equation}
  \dot{P}_{\rm wind}\approx 1.5 \dot{P}_{\rm jet},
  \label{sane98momentumratio}
  \end{equation}
  \begin{equation}
  \dot{E}_{\rm jet}\approx 3.5 \dot{E}_{\rm wind}.
  \label{sane98kineticratio}
\end{equation} 
 For MAD98, the results are,
\begin{equation}
  \dot{P}_{\rm wind}\approx \dot{P}_{\rm jet},
  \label{mad98momentumratio}
  \end{equation}
  \begin{equation}
  \dot{E}_{\rm jet}\approx 4 \dot{E}_{\rm wind}.
  \label{mad98kineticratio}
\end{equation}
From these results, we can see that even in the case of rapidly spinning black hole where the jet is supposed to be the strongest, no matter the accretion flow is SANE or MAD, the power of jet is only a factor of $\sim 4$ larger than that of wind, while the momentum of jet is even smaller or at most similar to that of wind. This comparison only considers the kinetic energy flux. When the total energy is considered, i.e., the Poynting flux and enthalpy are also included, as we will discuss in \S\ref{sec:BZ-jet}, the jet power will be larger than that of wind by a factor of $\sim 7$ and 10 for SANE98 and MAD98, respectively (refer to eqs.\ \ref{sanejetwindcomp} and \ref{madjetwindcomp}). 

In the case of a slowly rotating black hole, the wind becomes relatively more important compared to jet. In this case, the jet is mainly powered by the rotation of the accretion flow.  \citet{Yuan2015} show that  the momentum and energy fluxes of wind are much larger than that of jet in the case of SANE00 (refer to their Figures 11 \& 12): 
\begin{equation}
  \dot{P}_{\rm wind}\approx 15 \dot{P}_{\rm jet},
  \label{sane00momentumratio}
  \end{equation}
  \begin{equation}
  \dot{E}_{\rm wind}\approx 3 \dot{E}_{\rm jet}.
  \label{sane00kineticratio}
\end{equation}

These results have important implications in the study of AGN feedback, indicating the importance of wind compared to jet, which are often neglected in many AGN feedback studies. We will discuss this issue further in \S\ref{discussion}.

So far we only focus on discussing the kinetic energy flux of wind due to its poloidal velocity component. 
Other components of the energy flux include the kinetic energy due to rotational velocity, enthalpy energy flux, and Poynting flux. The mass-flux-weighted rotational velocity $v_{\varphi}^2$ are found to be:
\begin{equation}
    v_{\rm \varphi}^2(r)=0.98v_{\rm k}^2(r),
	\label{eq:vphi0}
\end{equation}
\begin{equation}
    v_{\rm \varphi}^2(r )=0.70v_{\rm k}^2(r),
	\label{eq:vphi1}
\end{equation}
\begin{equation}
    v_{\rm \varphi}^2(r)=0.94v_{\rm k}^2(r),
	\label{eq:vphi2}
\end{equation}
\begin{equation}
    v_{\rm \varphi}^2(r)=0.52v_{\rm k}^2(r)
	\label{eq:vphi3}
\end{equation}
for SANE00, MAD00, SANE98, and MAD98, respectively. Note that for MAD the rotational velocity is significantly less than the Kepler velocity, which is because of the strong outward magnetic pressure gradient force in MAD. We can see that the rotational velocity is larger than the poloidal velocity at these relatively small radii. 

\begin{figure*}[t]
\gridline{
\fig{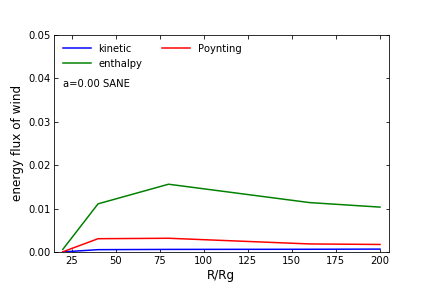}{0.48\textwidth}{}
\fig{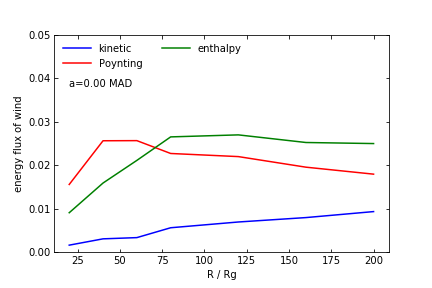}{0.48\textwidth}{}
}
\gridline{
\fig{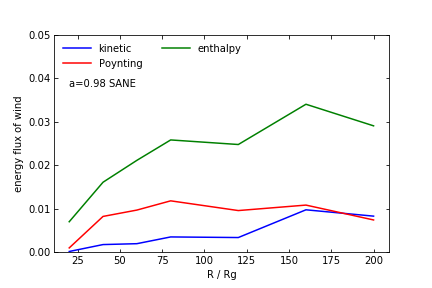}{0.48\textwidth}{}
\fig{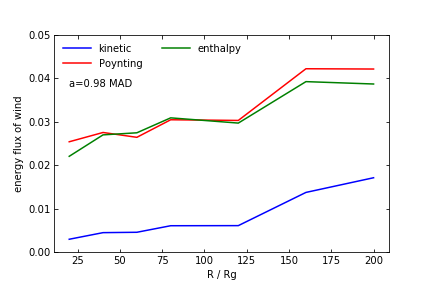}{0.48\textwidth}{}
}
    \caption{The radial profiles of various components of the energy flux of wind. These values have been normalized by $\dot{M}(r=2\, r_{\rm g})c^2$ of the four models respectively. The kinetic energy flux only includes the component due to the poloidal velocity.}
    \label{fig:enwind}
\end{figure*}

The Poynting flux and enthalpy flux are calculated by:
\begin{equation}
    F=\int-T^{r}_{t}\sqrt{-g} d\theta d\varphi.
	\label{eq:QEn}
\end{equation}
For the Poynting flux:
\begin{equation}
    T^{r}_{t} (EM)=b^{2}u^{r}u_{t}-b^{r}b_{t}.
	\label{eq:Qmag}
\end{equation}
For the enthalpy flux:
\begin{equation}
    T^{r}_{t} (ENT)=(u+p)u^{r}u_{t}.
	\label{eq:Qenth}
\end{equation}
Here $u^{r}$, $u_{t}$, $b^{r}$, $b_{t}$ are the quantities in the locally non-rotating frame, defined in the appendix. The various energy fluxes are shown in Figure~\ref{fig:enwind}. The energy flux due to the rotational velocity is not shown in the figure because its magnitude relative to the kinetic energy due to the poloidal velocity is easy to estimate by comparing eqs. \ref{eq:v02}--\ref{eq:v32} and eqs. \ref{eq:vphi0}--\ref{eq:vphi3}.
We can see that for SANE, the enthalpy flux of the gas is several times larger than the kinetic and Poynting fluxes; while for MAD, the enthalpy and Poynting fluxes are comparable and both of them are larger than the kinetic energy flux. 

All these results are for wind at relatively small radii, when they are still within the accretion flow region. \citet{Cui2020a} and \citet{Cui2020b} have studied the large-scale dynamics of wind after they are launched from the accretion disk scale, with and without taking into account the effect of magnetic field. The boundary conditions of wind are taken from the numerical simulations of wind launching from accretion flows. Both the black hole and galaxy potentials are included. During the outward
propagation, roughly speaking, it is found that the enthalpy and rotational energy and magnetic field energy compensate for the increase of gravitational potential. As a result, the wind can travel for a long distance with roughly constant poloidal velocity. So at large distance, the kinetic energy flux due to poloidal velocity will be the dominant component.

\subsection{The Poynting flux of BZ-jet }
\label{sec:BZ-jet}

\begin{figure}

  \includegraphics[width=\columnwidth]{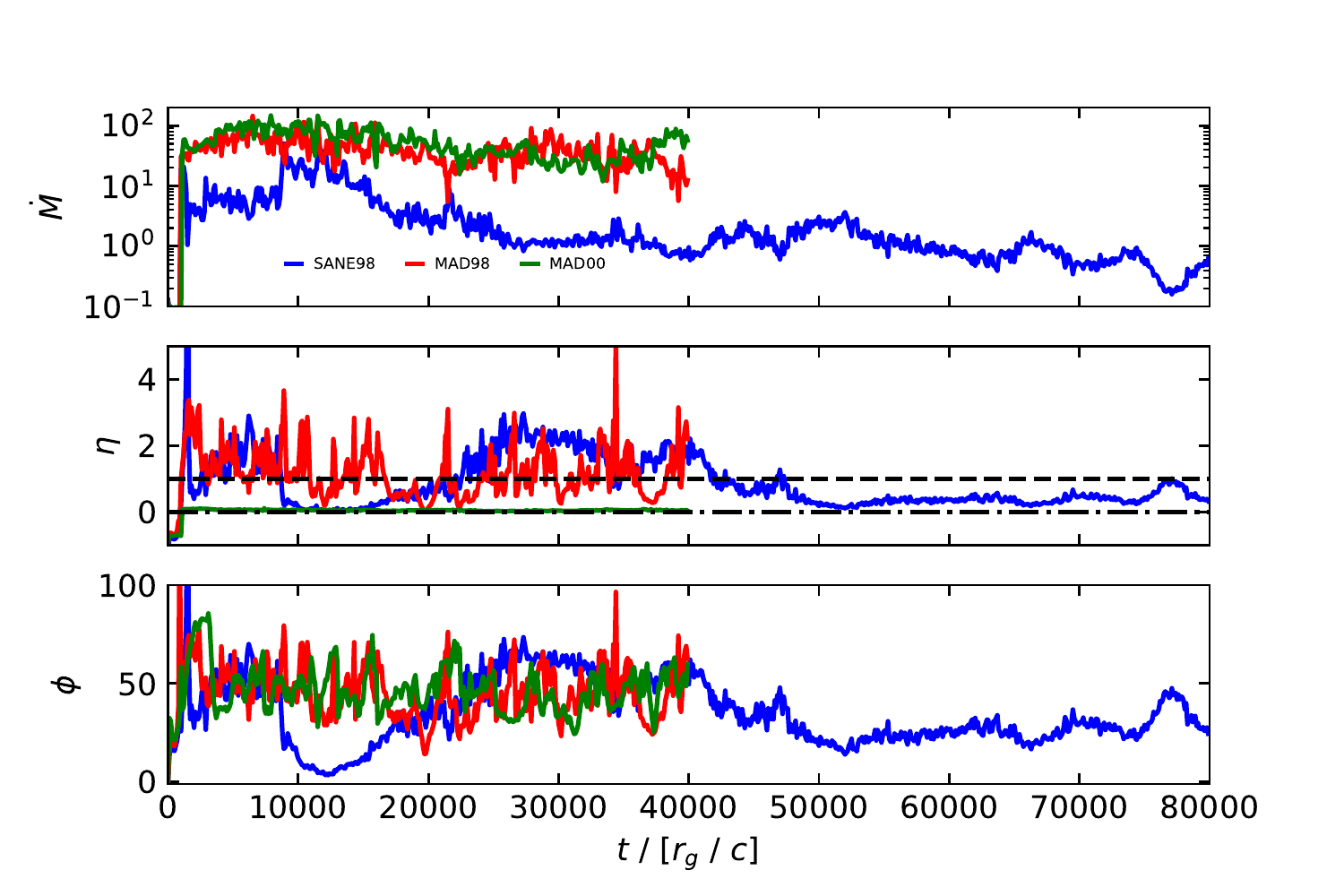}
    \caption{Time evolution of mass flux, energy extraction efficiency, and dimensionless magnetic flux for MAD00 (green),  SANE98 (blue), and MAD98 (red), respectively.  }
    \label{fig:eta}
\end{figure}

\begin{figure}[htbp]
    \includegraphics[width=\columnwidth]{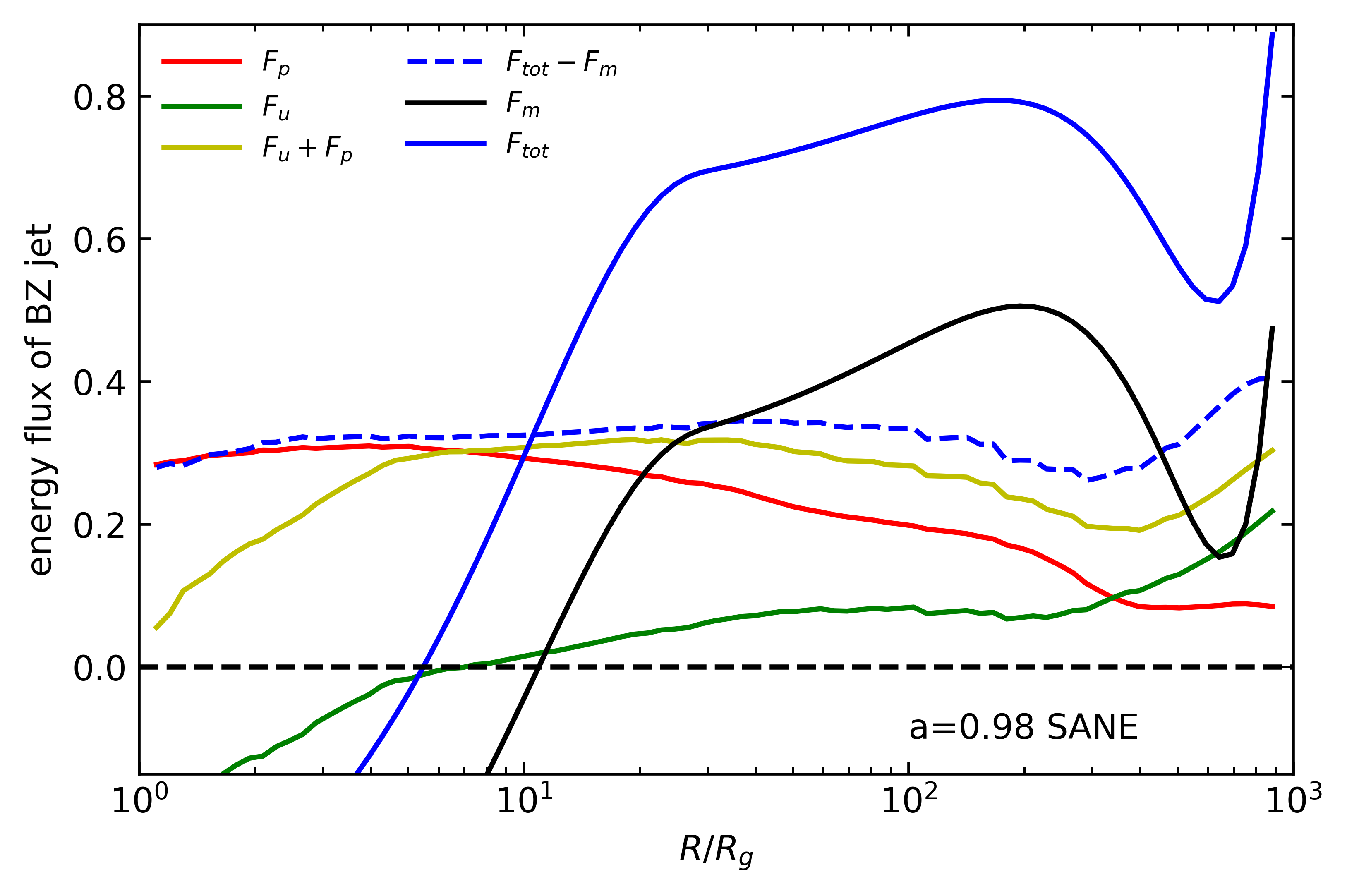}
	\includegraphics[width=\columnwidth]{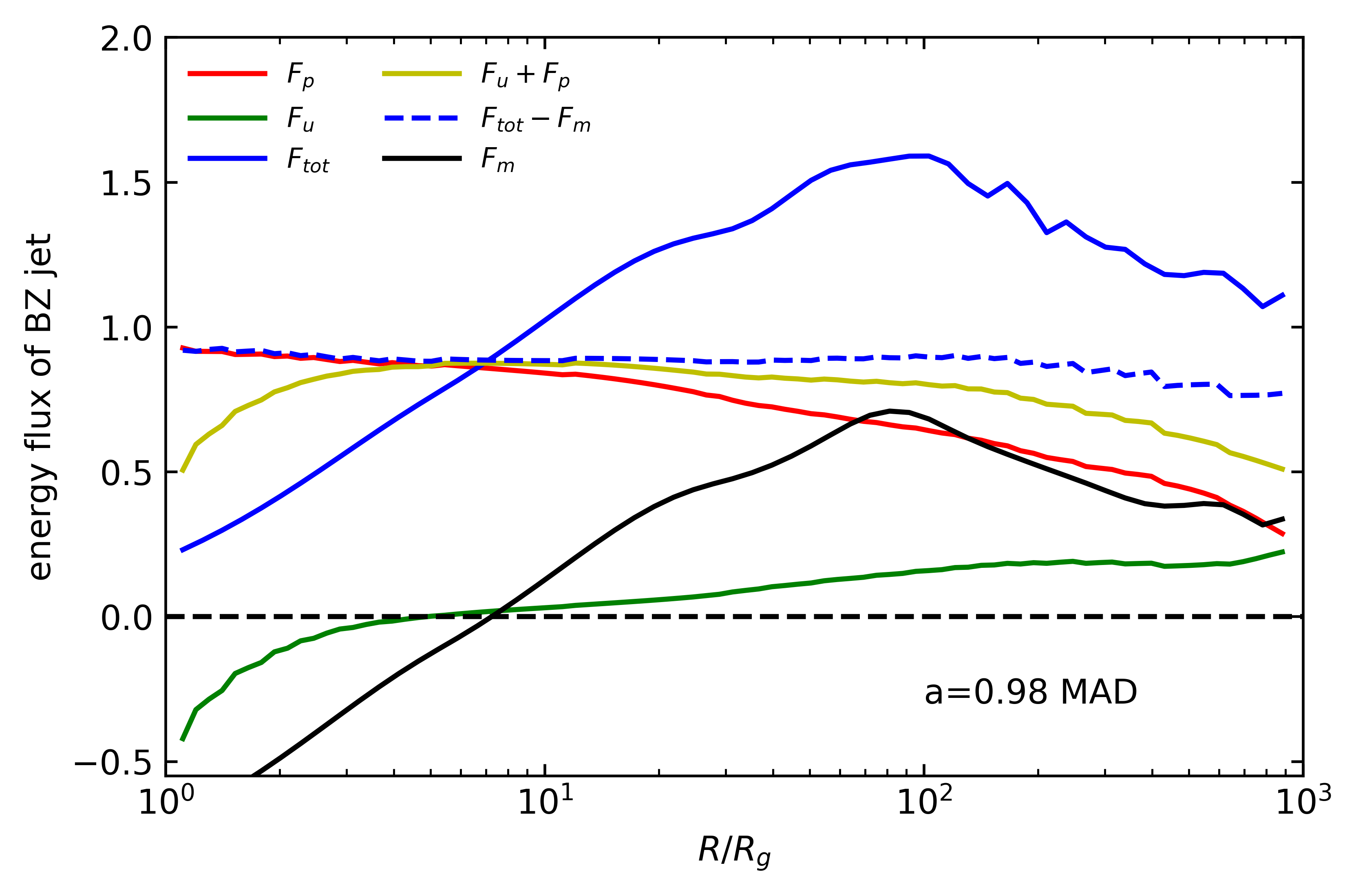}
    \caption{The radial profiles of various components of the energy flux of the BZ-jet for SANE98 (upper panel) and MAD98 (bottom panel).   They are the total energy flux $F_{\rm tot}$, the Poynting flux $F_{\rm p}$, the rest mass energy subtracted energy flux $F_{\rm tot}-F_{\rm m}$, the flux of enthalpy $F_{\rm u}$,  the sum of the enthalpy and Poynting fluxes $F_{\rm u}+F_{\rm p}$, and the rest mass energy flux  $F_{\rm m}=\rho u^r$. Those values have been normalized by $\dot{M}c^2$ at the black hole horizon of the respective models. }
    \label{fig:bzflux}
\end{figure}

Following \citet{Tchekhovskoy2011} and \citet{White2019}, the total energy fluxes (as measured at infinity) of the BZ-jet is calculated as follows,

\begin{equation}
    \dot{E}_{\rm tot}(r)=\int_{0}^{2\pi}\int_{0}^{\pi}T^{r}_{t} ( r^2+a^2\cos^2\theta )\sin\theta d\theta d\varphi
	\label{eq:Etot}
\end{equation}
here $T^{r}_{t}$ is a component of the stress-energy tensor $ T^{\mu}_{\nu}$ describing the radial flux of energy:
\begin{equation}
    T^{r}_{t}=(b^2+u+p+\rho)u^ru_t-b^rb_t.
	\label{eq:trt}
\end{equation}
Note that $T^r_t$ represents the total energy transported by the
fluid and the magnetic field, including the rest mass energy of the gas. The positive sign of $\dot{E}_{\rm tot}$ means the energy flux is inward. Given that rest mass energy doesn't play a role in AGN feedback, following \citet{Sadowski2013}, we define the following ``energy flux'' by eliminating the rest mass energy,
\begin{equation}
    \dot{e}(r)=-T^{r}_{t}-\rho u^{r}.
	\label{eq:etot}
\end{equation}
Positive values of  $\dot{e}(r)$ mean energy is lost from the system. The magnetic flux that threads  the hemisphere of black hole horizon, $\Phi_{\rm BH}$, is calculated as follows,
\begin{equation}
    \Phi(r)=\frac{1}{2}\int_{0}^{2\pi}\int_{0}^{\pi}\sqrt{4\pi}|B^{r}|(r^2+a^2\cos^2\theta)\sin\theta d\theta d\varphi.
	\label{eq:bflux}
\end{equation}
A dimensionless magnetic flux that is normalized with the mass accretion rate is
\begin{equation}
    \phi(r)=\frac{\Phi}{\sqrt{\dot{M}c}r_{\rm g}}.
	\label{eq:phi}
\end{equation}
The energy extraction efficiency, or the energy outflow efficiency, $\eta$, is defined as the energy return rate
to infinity divided by the time-average rest-mass accretion power:
\begin{equation}
    \eta=\frac{\dot{M}(r_{\mathrm{ H}})c^2-\dot{E}_{\rm tot}(r_{\mathrm{H}} )}{\langle\dot{M}(r_{\mathrm{H}})c^2\rangle} \times 100\,{\rm per\,cent}.
	\label{eq:eta}
\end{equation}
Here $\dot{M} (r_{\rm H})\equiv \dot{M}_{\rm BH}$ is the mass flux at the black hole horizon $r_{\rm H}$ calculated by eq.~\ref{eq:M}. Figure~\ref{fig:eta} shows the time evolution of mass flux $\dot{M}(r_{\rm H})$ (top panel),  energy outflow efficiency (middle panel), and dimensionless magnetic flux (bottom panel) for MAD00, SANE98, and MAD98.  They all strongly fluctuate with time, especially the two MAD models. For example, the energy outflow  efficiency of MAD98 ranges from  3 to 0. The time-averaged values of some physical quantities for SANE98 and MAD98 are given in Table~\ref{tab:BZ}. The dimensionless magnetic fluxes $\phi$ are $\sim 24$ and 40. The energy outflow efficiency of MAD00 is zero, as expected. The time-averaged energy outflow efficiency $\eta$ of  SANE98 and MAD98 models are $35\%$ and $110\%$. We can see that for MAD98, the variability of $\eta$ and $\phi$ are synchronous, and both of them are larger  compared to SANE98, which is consistent with our expectation  \citep[e.g.,][]{Narayan2003,Tchekhovskoy2011}. The values of $\eta$ and $\phi$ for MAD98 are also roughly quantitatively consistent with the result of \citet{Tchekhovskoy2011}. While SANE98 has a larger value of $\phi$ than is often seen in SANE simulations in the literature \citep[usually less than  $7$, see][]{Porth2019}, the lack of short-term variability in this quantity clearly distinguishes it from a true MAD state. Such large values of $\phi$ occur when initial field configurations like ours are run beyond the common stopping point of $t = 10{,}000$ or $20{,}000$ \citep{White2020a}.
 
\begin{deluxetable}{lcccccr}[th]
	\centering
	\tablecaption{Time-averaged quantities of BZ-jet\label{tab:BZ}}
	\tablenum{4}
	\tablehead{		
		\colhead{Model} &
		\colhead{ $\dot{M}$} & 
		\colhead{$\phi$} &
		\colhead{ ${P}_{\rm P}$} &
		\colhead{  $ {P}^{\rm BZ}$}&
		\colhead{ $ \eta$}
		}
		\startdata
		 SANE98 & 0.91& 23.81 &0.275& 0.29 &35.3\%\\
		MAD98 & 47.27&40.53 & 42.60& 44.22 &110\%\\
		\hline
		\enddata	
\end{deluxetable}

\citet[][see also BZ77]{Tchekhovskoy2010} give the BZ-jet power as,
\begin{equation}
    P^{\rm BZ}=\frac{\kappa}{4\pi}{\Omega}^2_{\rm H}{\Phi}^2_{\rm BH}f(\Omega_{\rm H}),
	\label{eq:Pbzn}
\end{equation}
where $\kappa$ is a numerical constant that depends on the geometry of the magnetic field, we adopt $\kappa=0.044$ in this paper. $\Omega_{\rm H}=ac/2r_{\rm H}$ is the angular velocity of the black hole horizon, $\Phi_{\rm BH}$ is the magnetic flux threading the hemisphere of black hole horizon, and $f(\Omega_{\rm H})$ is a modifying factor for high spin $a$, which is $f(\Omega_{\rm H})\approx1+1.38({\Omega_{\rm H}r_{\rm g}}/{c})^2-9.2({\Omega_{\rm H}r_{\rm g}}/{c})^4$\citep{Tchekhovskoy2010}.
Using this formula, combined with the magnetic flux from our simulations, we obtain $P^{\rm BZ}=\,0.29, 44.22$ for SANE98 and MAD98 respectively. To compare these values with our simulation results, we have also calculated the Poynting flux of the BZ jet based on the following equation,
\begin{equation}
    P_{P}=\int_{0}^{2\pi}\int_{0}^{\theta_{\rm BZ}}-T^{r}_{t}({\rm EM})(r^2+a^2\cos^2\theta)\sin\theta d\theta d\varphi,
	\label{eq:Ebz}
\end{equation}
where $T^{r}_{t}({\rm EM})$ is the electromagnetic component of stress-energy tensor  describing the radial flux of the Poynting flux described by eq. \ref{eq:Qmag}.
We obtain $P_{\rm P}=\,0.275, 42.60$ for SANE98 and MAD98 respectively. These values are in good agreement with that predicted by eq.\ (\ref{eq:Pbzn}), only slightly smaller. We have also calculated the ``Poynting flux jet efficiency'' $\eta_{\rm P}$ defined by
\begin{equation}
    \eta_{\rm P}=\frac{\langle P_{\rm P}\rangle}{\langle\dot{M}_{\rm BH}c^2\rangle}
\end{equation} For  MAD98, we obtain $\eta_{\rm P}=90.1\%$, while the total energy efficiency $\eta=99.2\%$\footnote{This value is slightly different from that given in Table \ref{tab:BZ}. This is because, for simplicity, here the time interval used when we do the time-average is shorter than that used in Table \ref{tab:BZ}.}, so the Poynting flux of jet accounts for $\sim 91\%$ of the total energy flux. For SANE98, $\eta_{\rm P}=32.3\%$ while $\eta=35.3\%$, so the Poynting flux of jet accounts for $\sim 92\%$ of the total energy flux. Thus the time-averaged  Poynting powers of jet for SANE98 and MAD98 are,
\begin{equation}
\dot{E}_{\rm jet}=0.32\dot{M}_{\rm BH}c^2,
\label{sanejetpower}
\end{equation}
\begin{equation}
 \dot{E}_{\rm jet}=0.9\dot{M}_{\rm BH}c^2,
 \label{madjetpower}
\end{equation}
respectively.

To understand the conversion among various components of  the energy fluxes in the jet, we have calculated the radial profiles of these components. The results are shown in Figure~\ref{fig:bzflux}. They are the total energy flux (blue line) $F_{\rm tot} (\equiv -\left[(b^2+u+p+\rho)u^ru_t-b^rb_t\right])$,  the Poynting flux (red line) $F_{\rm p}(\equiv -(b^2u^ru_t-b^rb_t))$, the rest mass energy-subtracted energy flux (blue dotted line)  $F_{\rm tot}-F_{\rm m} (\equiv -\left[(b^2+u+p+\rho)u^ru_t-b^rb_t\right]-{\rho}u^r$), the enthalpy flux (green line) $F_{\rm u} (\equiv -(u+p)u^ru_t)$, the sum of the enthalpy and Poynting flux (yellow line) $F_{\rm u}+F_{\rm p}$, and the rest-mass energy flux (black line) $F_{\rm m}(\equiv {\rho}u^r)$. A positive value of flux means that the direction of the flux is outward. The following results can be found from Figure~\ref{fig:bzflux}: 
\begin{itemize}
    \item The general ``shape'' of the radial profile of the total energy flux $F_{\rm tot}$ is very similar to that of the rest mass energy flux $F_{\rm m}$, implying that the change of the total energy is mainly due to the change of the rest mass energy.  
    \item The rest mass energy-subtracted total energy flux $F_{\rm tot}-F_{\rm m}$ roughly remains constant within $200\,r_{\rm g}$. 
    \item Both the enthalpy and rest-mass energy fluxes ($F_{\rm u}$ and $F_{\rm m}$) are pointing inward within the stagnation radius \citep{Broderick2015,Nakamura2018} of the black hole, while the Poynting flux $F_{\rm p}$ is always pointing outward. 
    \item When the jet is initially launched from the black hole, its energy is dominated by the Poynting flux $F_{\rm p}$. With the increase of radius, $F_{\rm p}$ decreases while the enthalpy flux $F_{\rm u}$ increases.  Their sum, $F_{\rm u} + F_{\rm p}$, decreases with  radius, and such a decrease is faster than that of $F_{\rm tot}-F_{\rm m}$.  This means that the reduced electromagnetic energy is not only converted into  enthalpy of the jet, but  also converted into  kinetic energy.  This is consistent with the increase of jet kinetic energy with radius shown in Figure~\ref{fig:ke}.  The  difference between the blue dashed line and yellow line is the magnitude of the increased kinetic energy. 
    \item The increase of enthalpy flux with radius is larger than that of the kinetic energy flux.
\end{itemize}

\begin{figure}
\plotone{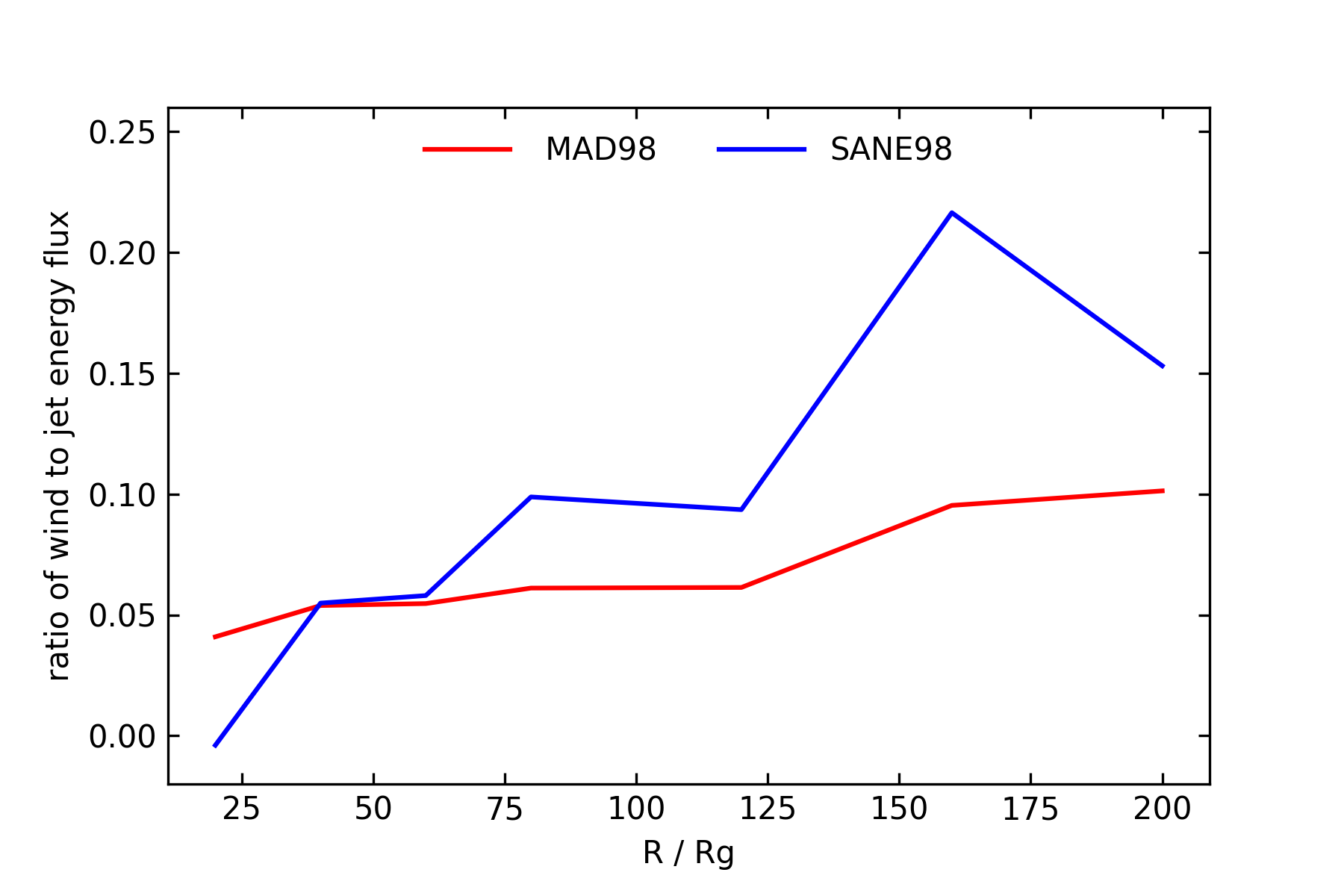}
\caption{The ratio of wind to jet energy flux for SANE98 and MAD98.}
\label{fig:windvsjet}
\end{figure}

To compare the energy flux between wind and jet, we have calculated the ratio of the wind to jet total energy fluxes. The results for SANE98 and MAD98 are shown in Figure~\ref{fig:windvsjet}. We find that in general, the ratio increases with increasing radius, implying that the wind energy flux increases faster than that of jet. This is because, unlike the jet, which is produced only at small radii, wind is produced in  a large range of radius from small to large radii. Thus the total energy flux of jet roughly remains constant but the total energy flux of wind increases with radius. At $\sim 200\,r_{\rm g}$, the ratio of wind to jet total energy fluxes reaches
\begin{equation}
   F_{\rm tot,wind}/F_{\rm tot,jet} \approx 15\% 
   \label{sanejetwindcomp}
\end{equation}
and
\begin{equation}
  F_{\rm tot,wind}/F_{\rm tot,jet} \approx 10\%  
  \label{madjetwindcomp}
\end{equation}
for SANE98 and MAD98, respectively. Referring to eqs.\ (\ref{sane98kineticratio}) \& (\ref{mad98kineticratio}), we can see that the ratio of the wind to jet total energy flux is smaller compared to the ratio of their kinetic energy fluxes. Such a result implies that the relative contribution of kinetic energy to the total energy in jet is smaller than that in wind.  

\section{Summary}
\label{sec:sum}
In a previous work \citep{Yuan2015}, based on three dimensional general relativity MHD (3D GRMHD) simulations of black hole hot accretion flows, we have investigated the wind launched from the accretion flow. A ``virtual particle trajectory'' approach has been adopted in that work, which can faithfully reflect the motion of fluid elements and thus discriminate turbulence and real outflows. Compared to the time-averaged approach often adopted in the literature, this approach can much more precisely calculate the properties of wind, especially the mass flux. Due to the instantaneous nature of wind, the time-average approach strongly underestimates the mass flux of wind by an order of magnitude or so. The \citet{Yuan2015} paper only focuses on the SANE and non-spinning black holes (i.e., SANE00). In the present work, we extend the \citet{Yuan2015} work by considering wind and jet in both SANE and MAD and rapidly spinning black holes. Since the magnetic field lines are quite ordered close to the rotation axis, we define jet as the region occupied by field lines connected to the ergosphere of the rotating black hole; while outflow beyond the jet region is called wind. The boundary between jet and wind is shown by, e.g., the red line in Figure~\ref{fig:mag}. Three models have been considered, namely  SANE98, MAD00, and MAD98, which denotes SANE accretion flow around a black hole with spin $a=0.98$, MAD accretion flow around a black hole with $a=0$ and $a=0.98$, respectively. Our  main results can be summarized as follows. 
\begin{itemize}
    \item The radial profiles of mass flux of wind and jet in various models are presented in Figure~\ref{fig:mdotw} and in eqs. (\ref{eq:m1}-\ref{eq:m4}). Close to the black hole, the wind becomes weaker in MAD compared to SANE, which is due to the suppression of turbulence when the magnetic field becomes too strong. Further away from the black hole, the wind mass flux becomes stronger in MAD compared to SANE, which is because at large radii turbulence is present and in general the magnetic field helps the formation of wind. 
    \item When the radius is not too large, the wind becomes stronger when the black hole spin becomes larger. 
    \item For rapidly spinning black holes,  the mass flux of jet for SANE98 and MAD98 can be described in units of the black hole accretion rate by eqs. \ref{sane98jetmassflux} \& \ref{mad98jetmassflux}. 
    \item The time and $\varphi-$averaged poloidal velocity of wind and jet as a function of $\theta$ are presented in Figure~\ref{fig:pv}. It is much larger in the jet region than in the wind region, as expected. The mass-flux-weighted poloidal velocity of wind and jet are shown in Figure~\ref{fig:pvk}. The quantitative results are given by eqs. \ref{eq:v1}-\ref{eq:v3} and eqs. \ref{eq:sanejetv}-\ref{eq:madjetv} for wind and jet, respectively. Magnetic field significantly enhances the poloidal velocity of wind and jet, as expected, but the effect of black hole spin on the wind velocity is weak. 
    \item The radial profiles of the momentum and kinetic energy fluxes of wind and jet are shown in Figures \ref{fig:mn} \& \ref{fig:ke}. Both the momentum and kinetic energy fluxes of wind increase with the black hole spin and the strength of magnetic field. The values of momentum and kinetic energy fluxes of wind can be easily calculated by combining the above-mentioned mass flux and poloidal velocity results. 
    \item The comparisons of the fluxes of momentum and kinetic energy  between wind and jet for SANE98 and MAD98 are described by eqs.\ \ref{sane98momentumratio}--\ref{mad98kineticratio}. Even in these two cases with extremely rapidly rotating black hole where the jet is supposed to the strongest, the momentum flux of wind is larger or similar to that of jet; the kinetic energy flux of jet is only stronger than that of wind by a factor of $\sim 4$. In the case of a non-spinning black hole such as SANE00, both the momentum and kinetic fluxes of wind will be significantly larger than that of jet (eqs.\ \ref{sane00momentumratio}--\ref{sane00kineticratio}). When other energy components such as enthalpy and Poynting flux are included, the jet power will be larger than that of wind by a factor of $\sim7$ and 10 for SANE98 and MAD98. These results have important implications for the study of AGN feedback, when we consider the respective role of wind and jet. This issue will be further discussed in \S\ref{discussion}.
    \item  The time-averaged Poynting power of jet for SANE98 and MAD98 is described by eqs.\ \ref{sanejetpower} and \ref{madjetpower}. As shown by Figure \ref{fig:bzflux}, the power of jet is dominated by Ponyting flux close to the black hole and is gradually converted into enthalpy and kinetic energy when the jet propagates outward.  
\end{itemize}

\section{Discussion: the relative importance of wind and jet in AGN feedback}
\label{discussion}

In many papers studying the effect of hot mode (also radio or jet or maintenance mode) AGN feedback in galaxy evolution, the authors only take into account jets but neglect wind. Our results suggest that wind should also be included, and they may even be more important than jet. 

First, the momentum of wind is in general larger than that of jet.  In the study of AGN feedback, some important aspects, including the growth of the black hole mass, the determination of the AGN mass accretion rate, and the degree of suppression of star formation in the central region of the galaxy, are mainly controlled by the momentum rather than energy feedback \citep[e.g.,][]{Ostriker2010}.  
Even in the case that energy feedback is more relevant than momentum feedback, since the power of jet is larger than wind only by a factor of $4\text{--}10$ even in the extreme case of a very rapidly spinning black hole, and since the opening angle of wind is much larger than that of jet, which makes the energy deposition efficiency of wind to the interstellar medium potentially much larger than that of jet, the role of wind  must be included. 

On the other hand, in the extreme case of rapidly spinning black holes, since the jet power is much larger than wind and the jet is well collimated, it can easily penetrate through the galaxy and propagate into much further distance than wind. In this way, the jet is able to heat the circum-galactic or intergalactic medium, which is hard for wind to do so. 

In the above discussions, one caveat is that the comparisons of momentum and power of wind and jet are conducted at $r=200\,r_{\rm g}$ due to limitation of our GRMHD simulations. It is not clear whether the relation roughly remains correct or not at much larger radii when the jet  propagates deeply into the galaxy. In the case of momentum, we can see from Figure \ref{fig:mn} that the flux of wind seems to keep increasing with increasing radius while the flux of jet seems to saturate. This will make the role of wind relatively more important. 

\section*{Acknowledgements}
We thank the referee for his/her constructive report. HY thanks Dr.\ Z.\ Gan for his help in the initial setup of the simulations. Drs. Xuening Bai and Jerry Ostriker are acknowledged for their useful comments and suggestions. HY and FY are supported in part by the  National Key Research and Development Program of China (grant
2016YFA0400704), the Natural Science Foundation of China (grant
11633006), and the Key Research Program of Frontier Sciences
of CAS (No. QYZDJSSW-SYS008). YFY is supported by the National Natural Science
Foundation of China (Grant No. 11725312, 11421303). This work has used the High
Performance Computing Resource in the Core Facility for Advanced
Research Computing at Shanghai Astronomical Observatory.
\appendix

\section{Kerr--Schild coordinates}

Since our simulation output data is in Kerr--Schild coordinates, we need to know the tetrad carried by the locally non-rotating frame (LNRF) observer in Kerr--Schild coordinates. We can then use the tetrad to convert the simulation data into physical quantities in LNRF coordinates.
We know that in Boyer--Lindquist coordinates the Kerr metric form is
\begin{equation}
\begin{split}
     d{s}^{2}=&-\left(1-\frac{2r}{\Sigma}\right)dt^{2}+\frac{\Sigma}{\Delta}dr^2+{\Sigma}d{\theta}^2+\frac{A\sin^2\theta}{\Sigma}d\varphi^2\\&-\frac{4ar\sin^2\theta}{\Sigma}d{\varphi}dt
	\label{eq:ks1}
\end{split}
\end{equation}
with the definitions
$\Sigma=r^2+a^2\cos^2\theta$,  $\Delta=r^2-2r+a^2$,  $A=(r^2+a^2)^2-a^2{\Delta}\sin^2\theta$
Both $r$ and $a$ have units of the black hole mass $M$.
In contravariant form we have
\begin{equation}
\begin{split}
     \left(\frac{\partial}{{\partial}s}\right)^2=&-\frac{A}{\Sigma\Delta} \left(\frac{\partial}{{\partial}t}\right)^2-\frac{4ar}{\Sigma\Delta}\left(\frac{\partial}{{\partial}t}\right)\left(\frac{\partial}{{\partial}\varphi}\right)+\frac{\Delta}{\Sigma}\left(\frac{\partial}{{\partial}r}\right)^2\\&
+\frac{1}{\Sigma}\left(\frac{\partial}{{\partial}\theta}\right)^2+\frac{\Delta-a^2\sin^2\theta}{\Sigma\Delta{\sin}^2\theta}\left(\frac{\partial}{\partial{\varphi}}\right)^2
\end{split}
	\label{eq:ks2}
\end{equation}

The frame carried by the LNRF observer has basis vectors:
\begin{equation}
e^{\mu}_{(t)}=\sqrt{\frac{A}{\Sigma\Delta}}\left(1, 0, 0, \frac{2ar}{A}\right)
	\label{eq:ks3}
\end{equation}
\begin{equation}
e^{\mu}_{(r)}=\sqrt{\frac{\Delta}{\Sigma}}(0, 1, 0, 0)
	\label{eq:ks4}
\end{equation}
\begin{equation}
e^{\mu}_{(\theta)}=\sqrt{\frac{1}{\Sigma}}(0, 0, 1, 0)
	\label{eq:ks5}
\end{equation}
\begin{equation}
e^{\mu}_{(\varphi)}=\sqrt{\frac{\Sigma}{A}}\frac{1}{\sin\theta}(0, 0, 0, 1)
	\label{eq:ks6}
\end{equation}
in contravariant form and
\begin{equation}
e^{(t)}_{\mu}=\sqrt{\frac{\Sigma\Delta}{A}}(1, 0, 0, 0)
	\label{eq:ks7}
\end{equation}
\begin{equation}
e^{(r)}_{\mu}=\sqrt{\frac{\Sigma}{\Delta}}\left(0, 1, 0, 0\right)
	\label{eq:ks8}
\end{equation}
\begin{equation}
e^{(\theta)}_{\mu}=\sqrt{\Sigma}(0, 0, 1, 0)
	\label{eq:ks9}
\end{equation}
\begin{equation}
e^{(\varphi)}_{\mu}=\sqrt{\frac{A}{\Sigma}}{\sin\theta}\left(-\frac{2ar}{A}, 0, 0, 1\right)
	\label{eq:ks10}
\end{equation}
in covariant form.

In Kerr--Schild coordinates, the line element of the Kerr space time is
\begin{equation}
\begin{split}
ds^2=&-\left(1-\frac{2r}{\Sigma}\right)dt^2+\left(\frac{4r}{\Sigma}\right)drdt+\left(1+\frac{2r}{\Sigma}\right)dr^2\\&
+{\Sigma}d{\theta}^2+\sin^2\theta\left[\Sigma+a^2\left(1+\frac{2r}{\Sigma}\right)\sin^2\theta\right]d{\varphi}^2\\&
-\left(\frac{4ar\sin^2\theta}{\Sigma}\right)d{\varphi}dt-2a\left(1+\frac{2r}{\Sigma}\right)\sin^2{\theta}d{\varphi}dr
\end{split}
	\label{eq:ks11}
\end{equation}
In contravariant form, it is
\begin{equation}
\begin{split}
 \left(\frac{\partial}{{\partial}s}\right)^2=&-\left(1-\frac{2r}{\Sigma} \right)\left(\frac{\partial}{{\partial}t}\right)^2+\frac{4r}{\Sigma}\left(\frac{\partial}{{\partial}t}\right)\left(\frac{\partial}{{\partial}r}\right)+\frac{\Delta}{\Sigma}\left(\frac{\partial}{{\partial}r}\right)^2\\&
+\frac{2a}{\Sigma}\left(\frac{\partial}{{\partial}t}\right)\left(\frac{\partial}{{\partial}\varphi}\right)+\frac{1}{\Sigma}\left(\frac{\partial}{\partial{\theta}}\right)^2+\frac{1}{{\Sigma}\sin^2\theta}\left(\frac{\partial}{\partial{\varphi}}\right)^2
\end{split}
	\label{eq:ks12}
\end{equation}

The transformation matrix from Boyer--Lindquist coordinates to Kerr--Schild coordinates is:
\begin{equation}
     \frac{\partial{x}^{\mu}(\rm KS)}{\partial{x}^{\nu}(\rm BL)}=
\left(\begin{array}{cccc}
 1 &  \frac{2r}{\Delta}&  0 & 0 \\ 
    0 &    1 & 0&     0 \\ 
   0 &     0 & 1 &    0\\
   0 &     \frac{a}{\Delta}&  0 & 1
\end{array}\right)
	\label{eq:ks13}
\end{equation}

The transformation matrix from Kerr--Schild coordinates coordinates to Boyer--Lindquist coordinates is:
\begin{equation}
     \frac{{\partial}x^{\mu}(\rm BL)}{\partial{x}^{\nu}(\rm KS)}=
\left(\begin{array}{cccc}
 1 &  -\frac{2r}{\Delta}&  0 & 0 \\ 
    0 &    1 & 0&     0 \\ 
   0 &     0 & 1 &    0\\
   0 &     -\frac{a}{\Delta} & 0 & 1
\end{array}\right)
	\label{eq:ks14}
\end{equation}
Through the vector transformation formula:
\begin{equation}
     e^{\mu}_{(\alpha)}(\rm KS)=\frac{{\partial}x^{\mu}(\rm KS)}{{\partial}x^{\nu}(\rm BL)}e^{\nu}_{(\alpha)}(\rm BL)
	\label{eq:ks15}
\end{equation}
and
\begin{equation}
     e^{(\alpha)}_{\mu}(\rm KS)=\frac{{\partial}x^{\nu}(\rm BL)}{{\partial}x^{\mu}(\rm KS)}e^{(\alpha)}_{\nu}(\rm BL)
	\label{eq:ks16}
\end{equation}
It can be obtained that in Kerr--Schild coordinates, the Locally Non-Rotating Frame observers carry the following frames respectively:
In contravariant form,
\begin{equation}
e^{\mu}_{(t)}=\sqrt{\frac{A}{\Sigma\Delta}}\left(1, 0, 0, \frac{2ar}{A}\right)
	\label{eq:ks17}
\end{equation}
\begin{equation}
e^{\mu}_{(r)}=\sqrt{\frac{\Delta}{\Sigma}}\left(\frac{2r}{\Delta}, 1, 0, \frac{a}{\Delta}\right)
	\label{eq:ks18}
\end{equation}
\begin{equation}
e^{\mu}_{(\theta)}=\sqrt{\frac{1}{\Sigma}}(0, 0, 1, 0)
	\label{eq:ks19}
\end{equation}
\begin{equation}
e^{\mu}_{(\varphi)}=\sqrt{\frac{\Sigma}{A}}\frac{1}{\sin\theta}(0, 0, 0, 1);
	\label{eq:ks20}
\end{equation}
in covariant form,
\begin{equation}
e^{(t)}_{\mu}=\sqrt{\frac{\Sigma\Delta}{A}}\left(1, \frac{-2r}{\Delta}, 0, 0\right)
	\label{eq:ks21}
\end{equation}
\begin{equation}
e^{(r)}_{\mu}=\sqrt{\frac{\Sigma}{\Delta}}(0, 1, 0, 0)
	\label{eq:ks22}
\end{equation}
\begin{equation}
e^{(\theta)}_{\mu}=\sqrt{\Sigma}(0, 0, 1, 0)
	\label{eq:ks23}
\end{equation}
\begin{equation}
e^{(\varphi)}_{\mu}=\sqrt{\frac{A}{\Sigma}}{\sin\theta}\left(-\frac{2ar}{A}, \frac{4ar^2}{A\Delta}-\frac{a}{\Delta}, 0, 1\right).
	\label{eq:ks24}
\end{equation}

We can use this frame to convert the simulation data into LNRF coordinates. For instance, we can use the below formulas to have the velocity vector in the LNRF coordinates.
Athena++ outputs velocity components $\nu^i=(\nu_1, \nu_2, \nu_3)$ in Kerr--Schild-like coordinates where the time basis vector is orthogonal to hypersurfaces of constant time and has unit length. Then  $\nu_j=g_{ij}\nu^i$ for $i, j=1, 2, 3$;
$\gamma=\sqrt{1+\nu^i\nu_i}$, $\alpha=\sqrt{{-1}/{g^{00}}}$; 
$u^{t}(\rm KS)={\gamma}/{\alpha}$; $u^{r}(\mathrm{ KS})=\nu_{1}-\gamma{\alpha} g^{01}(\rm KS)$; $u^{\theta}(\mathrm {KS})=\nu_{2}-\gamma{\alpha}g^{02}(\rm KS)$; $u^{\varphi}(\mathrm{ KS})=\nu_{3}-\gamma{\alpha}{g}^{03}(\rm KS)$;
\begin{equation}
 u^{a}(\mathrm{LNRF})=e^{( a )}_{\mu}(\mathrm {KS})u^{\mu}(\rm {KS});
	\label{eq:ks25}
\end{equation}
$v_{r}=\frac{u^{r}(\rm LNRF)}{u^{t}(\rm LNRF)}$; $v_{\theta}=\frac{u^{\theta}(\rm LNRF)}{u^{t}(\rm LNRF)}$; and $v_{\varphi}=\frac{u^{\varphi}(\rm LNRF)}{u^{t}(\rm LNRF)}$.
Similarly, we can get the magnetic field vector in the LNRF coordinates. Athena++ outputs primitive magnetic field vector is $B^{i}=(B^1, \,B^{2}, \,B^{3})$, then we could have four-vector $b^{\mu}$ with $b^{t}=g_{i\mu}({\rm KS} )B^iu^{\mu}(\rm{KS})$ and $b^{i}=[B^i+b^tu^{i}({\rm KS})]/u^t({\rm KS})$. 

\begin{equation}
 b^{a}(\mathrm{LNRF})=e^{( a )}_{\mu}(\mathrm {KS})b^{\mu}(\rm {KS});
	\label{eq:ks26}
\end{equation}

$B_{r}=b^{r}(\rm LNRF)$; $B_{\theta}=b^{\theta}(\rm LNRF)$; and $B_{\varphi}=b^{\varphi}(\rm LNRF)$.


\bibliography{yang.bib}
\bibliographystyle{aasjournal}
\listofchanges

\end{document}